\documentclass[11pt, bibliography=numbered]{scrartcl} %

\usepackage{lmodern}
\usepackage{ai-usage-card}
\usepackage{svg}
\usepackage{annotation}
\usepackage{longtable}
\usepackage{geometry}
\usepackage{tablefootnote}
\usepackage[nameinlink]{cleveref}
\usepackage[nolist]{acronym}
\usepackage[nopatch=footnote]{microtype}
\usepackage{array}
\usepackage{booktabs}

\usepackage[skip = 7pt plus 3pt, tocskip = -0.6mm]{parskip}

\usepackage{xcolor} %
    \definecolor{UBlau}{HTML}{153268}
    \definecolor{LBlau}{HTML}{005f9b}
    \definecolor{LMBlau}{HTML}{0091c8}
    \definecolor{LHBlau}{HTML}{50a5d2}
    \definecolor{WiWi}{HTML}{2b7ab3}
    \definecolor{Grau60}{HTML}{878786}
    
\usepackage{listings} %
\lstnewenvironment{configpython}[2][] %
{
    \lstset{
        commentstyle=\color{UBlau},
        keywordstyle=\color{WiWi},
        numberstyle=\tiny\color{Grau60},
        basicstyle=\ttfamily\footnotesize,
        language=python,
        numbers=none,
        frame=tblr,           %
        tabsize=4,
        captionpos=b,         %
        caption={#1},
        label={#2},
        breaklines=true,      %
        breakatwhitespace=false,
        breakautoindent=true,
        prebreak=\mbox{\textcolor{red}{$\hookrightarrow$}} %
    }
}{}

\lstnewenvironment{configjson}[2][] %
{
    \lstset{
        commentstyle=\color{UBlau},
        keywordstyle=\color{WiWi},
        numberstyle=\tiny\color{Grau60},
        basicstyle=\ttfamily\footnotesize,
        numbers=none,
        frame=tblr,           %
        tabsize=4,
        captionpos=b,         %
        caption={#1},
        label={#2},
        breaklines=true,      %
        breakatwhitespace=false,
        breakautoindent=true,
        prebreak=\mbox{\textcolor{red}{$\hookrightarrow$}} %
    }
}{}

\usepackage{tcolorbox} %
\usepackage{lipsum}    %

\definecolor{systemcolor}{rgb}{0.9,0.9,1} %
\definecolor{usercolor}{rgb}{1,0.9,0.9} %

\newtcolorbox{combinedprompt}{
    colframe=black,
    colback=white,
    boxrule=0.6mm,
    width=\linewidth,
    fonttitle=\bfseries,
    rounded corners,
    coltitle=black
}
    
\usepackage[backend=biber, style=ieee, sorting=nyt, maxcitenames=2, mincitenames=1]{biblatex} %
    \DeclareNameAlias{default}{family-given} %
    
\usepackage{csquotes} %
\usepackage{amssymb} %
\usepackage{mathtools} %
\usepackage{setspace} %
\usepackage{enumitem} %
\usepackage{wrapfig} %
\usepackage{float}
    
\usepackage{abstract} %
\usepackage[bf]{caption} %
\usepackage{subcaption}

\usepackage[automark, autooneside, headsepline]{scrlayer-scrpage} %
    \clearpairofpagestyles
    \addtokomafont{pagehead}{\footnotesize\scshape}
    \ihead{Decision Protocols in Multi-Agent Large Language Model Conversations}
    \ohead{\headmark}
\usepackage{hyperref} %
    \hypersetup{colorlinks,
            linkcolor={UBlau},
            citecolor={LHBlau},
            urlcolor={LMBlau},
            pdftitle={Master Thesis Lars Kaesberg},
            pdfauthor={Lars Benedikt Kaesberg}}

\begin{document}

\begin{titlepage}
\newgeometry{top=2.2cm,bottom=4.2cm,right=0.8in,left=0.8in}
\pagenumbering{gobble}
\centering \thispagestyle{empty}

\begin{figure}[!h]
    \centering
    {\includegraphics[width=0.85\textwidth]{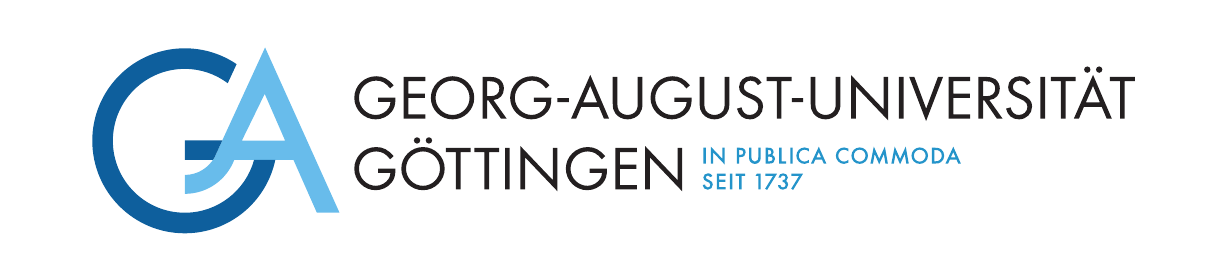}}
\end{figure}

\vspace{2cm}

\textsc{Master's Thesis
\vspace{5pt}
\begin{spacing}{2}
    {\LARGE Decision Protocols in Multi-Agent Large Language Model Conversations}
\end{spacing}
\vspace{5pt}
Lars Benedikt Kaesberg}

\vfill

In Fulfillment of the Requirements\\
for the Degree of\\
\vspace{5pt}
\textit{Master of Science}

\vspace{1cm}

\begin{tabular}{@{}ll}
    Main Examiner: & Prof. Dr. Bela Gipp\\
    Second Examiner: & Dr. Terry Lima Ruas\\
    Supervisor: & Jan Philip Wahle
\end{tabular}

\vspace{1cm}

GippLab\\
Georg August University of Göttingen\\
Göttingen, December 2024

\restoregeometry
\end{titlepage}

\newpage
\thispagestyle{empty}
\setcounter{tocdepth}{2}
\tableofcontents
\vfill

\hrule
\vspace{2mm}
I hereby declare that I have written this thesis independently without any help from others and
without the use of documents or aids other than those stated. I have mentioned all used sources
and cited them correctly according to established academic citation rules. \\
\vspace{2mm}

Helsinki, December 31th, 2024

\begin{acronym}
  \acro{CoT}{Chain-of-Thought}
  \acro{fpc}{finite population correction}
  \acro{LLM}{Large Language Model}
  \acro{MALLM}{Multi-Agent LLM}
  \acro{MAS}{Multi-Agent Systems}
  \acro{regex}{regular expression}
  \acro{SOM}{society of mind}
  \acro{SOP}{Standard Operating Procedure}
\end{acronym}

\newpage
\pagestyle{scrheadings}
\pagenumbering{arabic}

\renewcommand{\abstracttextfont}{\normalfont\normalsize}
\begin{abstract}
    Improving the task performance of \acp{LLM} is essential, yet scaling these models faces significant challenges such as diminishing returns and high costs. \ac{MAS} offer a promising solution by distributing tasks among specialized agents to improve the overall task performance. This can reduce training costs at the expense of increased test time due to the discussion and decision-making process. The decision protocol is a critical component of \ac{MAS} because it specifies how multiple agents collaborate to create a final solution. This thesis introduces the \ac{MALLM} framework, which implements and evaluates various decision protocols, namely voting, consensus, and judge decision mechanisms, to simulate multi-agent discussions for conversational task solving. Unlike previous work that used a single decision protocol or tested them on limited datasets, this study systematically examines their impact on a diverse set of tasks, ranging from knowledge-based datasets (MMLU, MMLU-Pro, GPQA) and logic-based datasets (StrategyQA, MuSR, Math-lvl-5, SQuAD 2.0). The results indicate that consensus protocols excel in knowledge-intensive domains while voting and judge protocols are more effective for logic-based tasks. Increasing response diversity through independent solution generation improves decision quality, while changes in information access during the decision process have minimal impact.

\end{abstract}

\section{Introduction} \label{sec:intro}
Recent years have shown a trend to scale the size of \acp{LLM} to improve task performance \cite{openai_gpt-4_2024}. It is unclear how far this trend can be pushed, as a further increase in the model size could bring diminishing returns in task performance. \textcite{sutskever2024neurips} recently stated that “Pre-training as we know it will unquestionably end” highlighting potential upcoming limitations of scaling. Additionally, the scalability of LLMs presents substantial challenges, with only major corporations like OpenAI, Meta, and Anthropic possessing the necessary talent, computing resources, and financial resources to train such expansive models \cite{abdalla_elephant_2023, naveed_comprehensive_2024}. As the benefits of training larger \acp{LLM} remain unclear, and the costs and environmental impact continue to rise, many researchers are looking for alternative methods to improve task performance instead of scaling the model size \cite{liu_green_2024}. This revealed that not only increasing compute power for training larger \acp{LLM} but also for inference (i.e., generating answers), can significantly enhance task performance \cite{snell_scaling_2024}. Consequently, researchers are exploring a range of strategies beyond scaling, such as prompting techniques \cite{wei_chain--thought_2023} and collaborative frameworks \cite{hong_metagpt_2024,wu_autogen_2023}, to further boost task performance without the need for continually larger models.

\begin{figure}[H]
    \centering
    \includegraphics[width=\textwidth]{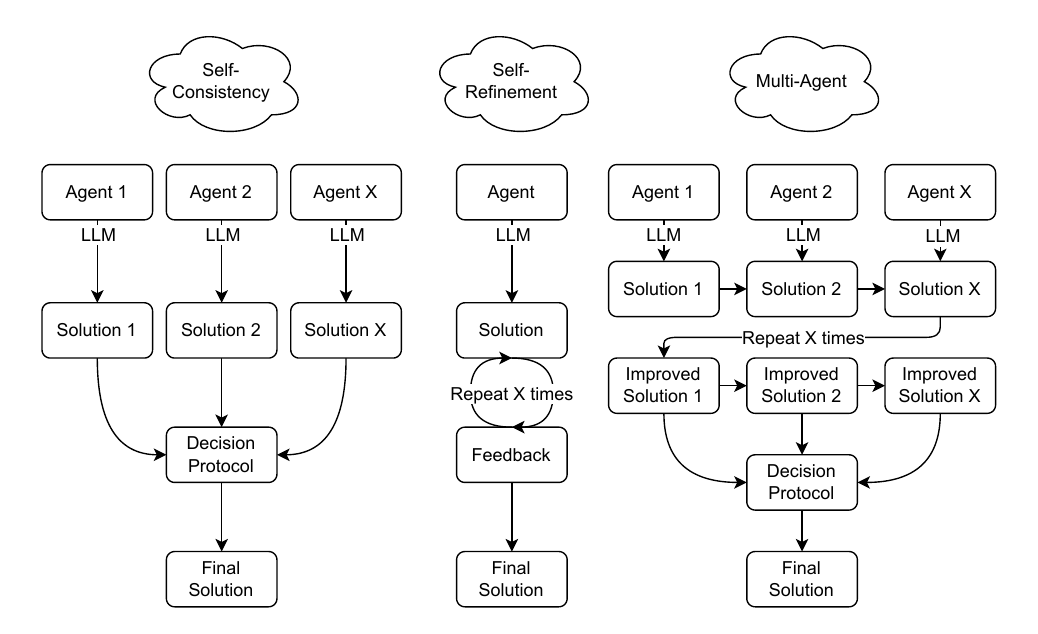}
    \caption{Illustration of solution generation process for self-consistency, self-refinement, and multi-agent discussions.}
    \label{fig:mas_sc_sr}
\end{figure}

Many methods can be used to achieve this new kind of scaling. One of the first approaches is the so-called \ac{CoT} from \textcite{wei_chain--thought_2023}. It allows models to provide reasoning steps before the final answer generation. This method is still widely used for simplicity, but many more sophisticated approaches have been developed. For example, OpenAI refined this approach with its new o1-preview model\footnote{\href{https://openai.com/index/introducing-openai-o1-preview/}{openai.com/index/introducing-openai-o1-preview}}. This relies on further training of a model on these correct reasoning steps to give \acp{LLM} better reasoning capabilities \cite{winter_system_2024}. Other approaches without additional training can be roughly divided into three categories: self-consistency \cite{wang_self-consistency_2022}, which involves generating multiple reasoning paths and selecting the most consistent answer; self-refinement \cite{madaan_self-refine_2023}, which allows the model to iteratively improve its own reasoning and answers; and multi-agent discussions. An overview of how these methods work can be found in \Cref{fig:mas_sc_sr}.

In multi-agent discussions, each agent generates a solution, and these solutions are refined and improved through iteration during the discussion. The method is effective because it combines the strengths of self-consistency and self-refinement, mirroring how humans solve complex problems by bringing together experts from different fields to exchange ideas and enhance the outcome. The downside of using multiple agents is that each one creates its own solution. This can also be seen as an opportunity, as having multiple answers and using an agent to select the correct one may be easier than coming up with a new solution. This can be compared to how multiple-choice questions tend to be easier than free-text questions in exams. Still, a method is needed to combine these solutions to produce a final solution.

This is often done using a majority metric to resolve these conflicts \cite{wang_self-consistency_2022, chen_reconcile_2024}. These simple approaches have been shown to yield good results, but it needs to be tested whether more sophisticated approaches can further improve the results \cite{wang_self-consistency_2022}. Decision protocols can have a significant impact on task performance as they only need to select the correct solution from a pool of possible answers \cite{yang_llm_2024, chen_reconcile_2024, wang_self-consistency_2022}. Some results from \textcite{yang_llm_2024} already demonstrate that incorporating a voting step to determine the final solution can lead to improved outcomes. Further testing is necessary to evaluate whether integrating additional information or confidence values can enhance this voting process. In addition, more decision protocols, such as consensus or judge decision protocols, need to be compared. To address these considerations, the following research questions are posed:

\begin{itemize}
    \item[1.] \textbf{How do decision-making mechanisms influence LLM performance in conversational tasks?}
    \vspace{-0.5em}
    \begin{itemize}
    \itemsep-0.5em 
        \item[1.1] Is there a decision protocol that consistently performs best?
        \item[1.2] How effectively can these mechanisms adapt to different tasks?
        \item[1.3] Does the round in which voting starts affect task performance?
        \item[1.4] Does the number of agents proposing solutions matter?
    \end{itemize}

    \item[2.] \textbf{How does the diversity of responses affect the performance of decision protocols?}
    \vspace{-2em}
    \begin{itemize}
    \itemsep-0.5em 
        \item[2.1] Is it more efficient to start with one solution and iterate on it, or should each agent propose its own initial solution?
        \item[2.2] Can different discussion protocols or prompting help increase the answer diversity?
    \end{itemize}
    
    \item[3.] \textbf{How do small variations in information provided during the decision phase impact decision protocols in multi-agent LLM systems?}
    \vspace{-0.5em}
    \begin{itemize}
    \itemsep-0.5em 
        \item[3.1] How do small changes in the amount of information provided during the decision phase affect task performance and decision stability? 
        \item[3.3] How does providing agents with new information (e.g., confidence scores or additional facts) after the discussion phase influence their voting decisions and final results?
    \end{itemize}
    
    \item[4.] \textbf{How do language agents react when challenged in a multi-agent setting?}
    \vspace{-0.5em}
    \begin{itemize}
    \itemsep-0.5em 
        \item[4.1] How often do language agents change their decisions when challenged?
        \item[4.2] Can challenging an answer improve task performance?
    \end{itemize}
\end{itemize}

This work introduces \ac{MALLM}\footnote{\href{https://github.com/Multi-Agent-LLMs/mallm}{github.com/Multi-Agent-LLMs/mallm}}, a new framework designed to simulate multi-agent discussions for solving conversational tasks. \ac{MALLM} enables multiple agents to collaboratively generate and refine solutions, thereby enhancing the overall problem-solving process. My experiments reveal that consensus protocols, which require agents to collaboratively agree on a solution, excel in knowledge-intensive domains. In contrast, voting protocols, where agents cast votes on proposed answers, and judge protocols, where a designated agent evaluates and selects the best solution, are more effective for logic-based tasks. The number of agents participating in the discussion can increase performance, but longer discussions hurt task performance. Furthermore, increasing response diversity through independent solution generation improves task performance, while variations in information provision during the voting phase have minimal impact on overall performance. Smaller language models achieve significantly greater performance gains than larger models when utilizing the \ac{MALLM} framework, highlighting its ability to enhance models trained with limited resources. However, this improvement requires an increase in computational power during inference. When challenging the final solution of the decision protocol with a single agent, it can be seen that these agents can detect incorrect answers and, if provided with enough reasoning details from the discussion, can understand the solution. They cannot directly improve the solution as they lack the multi-agent collaboration.

\newpage
\section{Related Work}
The area of decision protocols for \ac{MAS} spans a wide array of research topics, ranging from artificial intelligence \cite{dubey_llama_2024}, social choice theory \cite{list_social_2022}, to multi-agent frameworks \cite{zhuge_language_2024, wu_autogen_2023, hong_metagpt_2024}. Decision-making within \ac{MAS} combines the language agents' different objectives, strategies, and knowledge bases to improve the final solution. Advancements in reasoning capabilities and computation cost allow the usage of \acp{LLM} to simulate multi-agent discussions and use them to decide on the final solution \cite{yang_llm_2024}. In this section, I analyze related work on decision-making, \acp{LLM}, and agents, while also reviewing existing frameworks for simulating multi-agent discussions. This analysis identifies gaps in functionality and the need for a novel framework for the experiments proposed in \Cref{sec:experiments}.

\subsection{Decision-Making}
A study conducted by \textcite{jones_comparison_1994} in 1994 compared the relative strengths and weaknesses of the voting-based and consensus-based decision-making procedures. It is focused on a case that involved a citizen task force that dealt with a controversial public housing proposal. In the first year, they used a consensus-based approach, which came to a final result, but afterward, some group members reported that they were not satisfied with the outcome. In general, achieving consensus was a lengthy process that could not satisfy everyone. In the second year, they used a voting-based approach, which had to reach a two-thirds majority. The group members reported that they were more comfortable with the approach as they felt that they had more impact on the overall result. Ultimately, this approach failed as some members disturbed the voting process with strategic absences. Similarly \textcite{list_social_2022} showed that voting systems are vulnerable to strategic manipulation, but some methods exist to mitigate this. For example, reducing the voting protocol's complexity increases the resistance to manipulation. This can be done by limiting the number of votes for each participant. Some approaches created complex rules to protect voting integrity with computational complexity, but these are not useful in a small scope \cite{bartholdi_computational_1989}.

More recent work has already used hybrid consensus and voting-based approaches combined with \ac{MAS} as seen in ReConcile from \textcite{chen_reconcile_2024}. Here, each agent generates an answer to a given task. The final answer is determined by a weighted voting system for each round. Each agent has to generate a confidence score for their solution. The confidence scores are averaged for the same solutions and the solution with the highest combined confidence score gets selected. After that, the next round starts, and the cycle is repeated. The process ends when all agents give the same solution. This approach significantly outperforms the baseline but also uses different \acp{LLM} for each agent. This expands the knowledge base, which can lead to better task performance. Multi-agent discussions with only one model, such as the approach of \textcite{yang_llm_2024}, showed smaller improvements. They use a voting-based approach that introduces a variety of voting methods. One is, approval voting, where each agent can vote for any number of answers, and the most voted answer wins. For ranked voting, agents have to sort the answers according to their preferences, and the answer with the highest combined ranking wins. Another method is cumulative voting. Here, the agent can assign a certain number of points to each answer, and the answer with the most points wins. The study showed that \acp{LLM} are susceptible to the order in which the possible answers are presented. Overall, \acp{LLM} are not consistent with their votes, but this problem can be reduced by using more powerful models or introducing personas to the agents \cite{yang_llm_2024}.

In my research, I build on these findings by considering both voting-based and consensus-based protocols. Additionally, I introduce a third class, the authoritarian judge decision protocol. I focus on adapting them to multi-agent language models rather than human-based discussions as seen in \textcite{jones_comparison_1994}. Unlike previous studies that used several different models or introduced complex voting mechanisms, I deliberately keep the underlying model structure consistent and emphasize simplicity in the decision protocols to ensure comparability.

\subsection{LLMs and Agents}
Multi-agent discussions are expensive as they require a lot of messages exchanged between all participants to reach a final answer. Therefore, it is important to use an efficient \ac{LLM}, which has a sufficient level of communication and role-playing skills even with a smaller model size. The recently introduced Llama 3 model family provides many different model sizes \cite{dubey_llama_2024}. The original Llama 3 model is available with 8 billion and 70 billion model parameters. After conducting the experiments, they also released Llama 3.1 models with 8 billion, 70 billion, and 405 billion parameters. The Llama 3.1 models have a longer context length and come with multilingual pre-training, which makes them a good choice for future experiments. All models also exist as an instruction-fine-tuned version, which is important for using the \ac{LLM} as a chatbot or discussion participant. The two smaller models (i.e., Llama 3 8B and Llama 3 70B) are very interesting as they are small enough to fit on accessible hardware, such as described in \Cref{sec:setup}. 

There are many possibilities to further improve the performance of these models. Most of them rely on increasing computational cost during inference to get better results \cite{snell_scaling_2024}. \textcite{wei_chain--thought_2023} introduced a very simple method called \acf{CoT}, which requires only a small change to the prompt. The model is asked to provide reasoning steps before reaching its final conclusion. This can lead to substantial improvements in task performance, but this also depends on the dataset used. As this method only requires a small change in the prompt, it can be used in almost every situation. Another more sophisticated approach, called self-refinement, was shown by \textcite{madaan_self-refine_2023}, which uses an \ac{LLM} to iteratively improve its own answer. First, the model creates an initial solution. This is followed by a feedback step, in which the model critiques its own answer. The feedback is then used to create a new and improved answer. This process can be repeated indefinitely, but subsequent rounds have diminishing returns. The study showed that this works better for more powerful models, as smaller models lack the capability to provide adequate feedback. In contrast to the previous approach, \textcite{wang_self-consistency_2022} introduced a new method called self-consistency. Here, one model creates many diverse answers to the same problem, also utilizing \ac{CoT}. With the help of a decision protocol, all these possible answers are combined into one final answer by selecting the most occurring answer. This approach showed a good improvement over the baseline. 

To increase the variability of the answers from the agents, they can be prompted to role-play a specific persona. \textcite{jiang_personallm_2024} has shown that larger \acp{LLM} such as GPT-3.5 \cite{brown_language_2020} and GPT-4 \cite{openai_gpt-4_2024} are capable of accurately displaying specific character traits based on personality tests. Later, \textcite{samuel_personagym_2024} developed an automatic evaluation framework to test how well a \ac{LLM} adheres to its assigned persona in various environments, allowing a large-scale and efficient persona agent evaluation. This showed that also smaller models, especially Llama 3 8B, are able to achieve a high persona score, even outperforming much larger models like GPT-3.5. \textcite{kim_persona_2024} evaluated the performance of persona-prompted agents in multiple datasets and discovered that this can have positive and negative impacts. It often comes with an increase in performance, but the persona can introduce unwanted biases or get distracted from the task, and as \textcite{shi_large_2023} showed, irrelevant context can hinder task performance. \textcite{kim_persona_2024} introduced a method for automatically generating personas for a given problem statement, removing the need to create agent personas manually and making it possible to have specific personas for a wide range of tasks.

In my work, I integrate \ac{CoT} prompting and persona-based role-playing into multi-agent discussions, drawing on the strengths of self-refinement and self-consistency. This combination aims to maximize performance gains without resorting to overly large models, keeping computational overhead at a manageable level with accessible hardware. More details for the hardware used can be found in \Cref{sec:setup}.

\subsection{Multi-Agent Frameworks}
In the concept of a \ac{SOM} \cite{minsky_society_1987, zhuge_mindstorms_2023}, higher-level intelligence is understood not to arise from a singular entity but from the interaction and collaboration of simpler modular cognitive components. Each of these components, or agents, operates with its own limited function and purpose, yet when combined, they contribute to a dynamic, adaptive, and emergent form of intelligence that mirrors the complexities of human cognition \cite{minsky_society_1987}.

This paradigm has inspired recent advances in multi-agent frameworks within artificial intelligence, where agents work together to solve problems beyond the capability of any single agent \cite{chen_reconcile_2024, du_improving_2023}. The frameworks proposed by \textcite{wu_autogen_2023} and \textcite{zhuge_language_2024} are similar and both enable researchers to set up multi-agent discussions. AutoGen is able to simulate the interactions of agents with different skills. Some have specific personas, others can access tools to better solve complex tasks, and others can access other knowledge bases to improve factual correctness. Agents can collaborate over multiple turns to adjust to feedback and find a collective decision \cite{wu_autogen_2023}. SwarmGPT works similarly to AutoGen but defines the agent interactions as graphs, where nodes represent operations (such as LLM queries or tool use), and edges indicate information flow. This allows the framework to dynamically optimize the graph to improve the efficiency of the framework \cite{zhuge_language_2024}.

MetaGPT allows the researcher to create \acp{SOP}. These can be applied to different tasks and define how a task can be divided between multiple agents. For example, a management agent defines requirements that need to be fulfilled by the programmer agent, who is responsible for implementing all of these requirements. This allows to solve some complex tasks very accurately as long as a \ac{SOP} exists \cite{hong_metagpt_2024}.

I build upon the ideas of these frameworks by designing a simpler \ac{MAS} that emphasizes direct collaboration through iterative discussion rather than extensive tool use. I focus on conversational tasks, keeping agent interactions manageable while still allowing each agent to bring a unique persona to the table. This approach preserves the core idea of combining specialised agents, but streamlines the process to better fit the conversational paradigm and maintain efficiency.

\subsection{Overview}

This study extends the work form \textcite{jones_comparison_1994} that showed differences between the voting-based and consensus-based decision protocols. It adds a new class where a single authoritarian language agent can decide on the final answer, similar to how a judge decides the final verdict. Voting-based decision protocols are inspired by \textcite{yang_llm_2024} as they already showed good results for \ac{MAS}. The Llama 3 family of models is used for all experiments, as they offer a good variety of models with high task performance for relatively low resource usage. The existing multi-agent frameworks are not designed for conversational task solving, which would require substantial modifications to adapt them. In addition, they often lack built-in mechanisms for efficient parallelization, making them less suitable for large datasets. Therefore, for this work, a new framework has been developed from the ground up with parallelization in mind, enabling the seamless integration of any HuggingFace dataset without altering the underlying code. Details can be found in \Cref{sec:mallm}.

\newpage
\section{Methodology}
\label{sec:methodology}
In this section, I introduce a new multi-agent framework for conversational task solving, specifically designed to address the research questions outlined in \Cref{sec:intro}. This framework enables collaborative interactions between multiple agents to enhance problem-solving skills. Additionally, I provide detailed information on the datasets used in this study.

\subsection{Multi-Agent Discussion Framework}
\label{sec:mallm}
I present \acf{MALLM}\footnote{\href{https://github.com/Multi-Agent-LLMs/mallm}{github.com/Multi-Agent-LLMs/mallm}}, a newly developed framework for multi-agent conversational task solving with language agents. Multi-agent conversations are inherently expensive as they require many requests per round and agent. Therefore, this framework is specially focused on efficiency and modularity as I want to conduct many experiments with different parameters and datasets.  On top of that, these discussions can have a dynamic number of rounds, which can make it even more expensive. To combat this, I designed \ac{MALLM} to use the available resources efficiently by performing many discussions asynchronously. The framework is aimed at researchers. Therefore, modularity is especially important because it allows researchers to use the existing infrastructure provided by \ac{MALLM} and change certain functionalities, such as agent persona generation, agent answer generation, discussion protocol, and decision protocol.

\subsubsection{Architecture Overview}
\label{sec:architecture}
\begin{figure}
    \centering
    \includegraphics[width=0.95\linewidth]{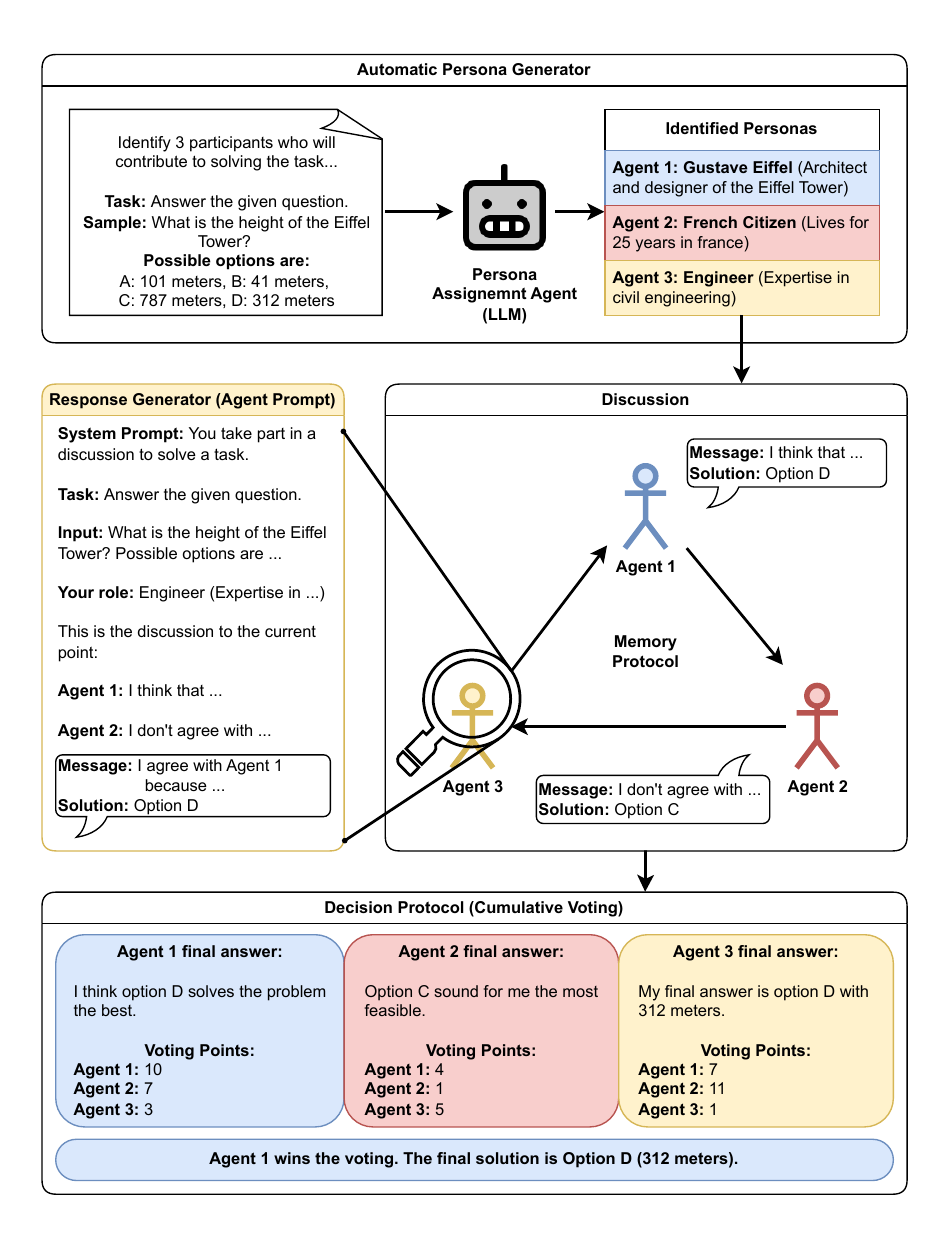}
    \caption{Example multi-agent discussion conducted in the \ac{MALLM} framework. It showcases the functionality of the four modules and how they work together to get an improved final solution.}
    \label{fig:mallm_overview}
\end{figure}

To better understand the different modules, I take a closer look at each component and what role it plays in creating multi-agent discussions. An overview can be found in \Cref{fig:mallm_overview} as it provides an example workflow for the framework and how a discussion is created. The discussion starts with generating personas relevant to the given task and assigning them to the participating agents. The personas are generated using the same \ac{LLM} which is later used for the agents. After that the agents start to generate solutions and improve the suggestions from the other agents. The turn order of the agents is defined by the \textit{discussion paradigm}. This also defines which answers are visible to other agents and who can talk to whom. The \textit{response generator} defines how an agent receives the other answers and also the way it responds. After a certain number of rounds or when enough agents agree, a \textit{decision protocol} is used to select the best answer either via voting, using a judge agent, or just by looking for a certain consensus threshold. If the decision protocol fails, for example, due to a tied vote, the discussion continues for another round. In the framework a parameter can be defined to terminate discussions after a certain number of rounds to make sure they do not communicate forever.

\begin{figure}[H]
    \centering
    \includegraphics[width=0.9\textwidth]{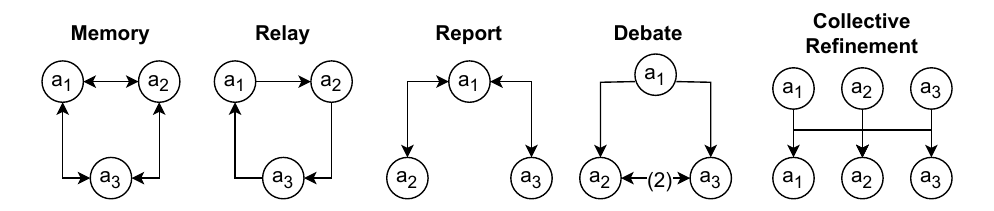}
    \caption{Illustration of Discussion Paradigms available for use in \ac{MALLM}}
    \label{fig:discussion_paradigms}
\end{figure}

\paragraph{Agent Personalities.}
The first step of the discussion is the generation of agent personas. Each of the agents participating in the discussion has a certain persona assigned to them. This can unlock more knowledge for the \ac{LLM} on a specific topic \cite{kim_persona_2024}. To get the best results, I want as diverse personas as possible while still maintaining them to be relevant to the task. The default setting for the framework is to prompt a \ac{LLM} and ask for a persona relevant to the given task \cite{wang_unleashing_2024}. After each generation it also provides the generated personas to avoid duplication. This way of generating personas provides a good starting point, but as this is built as a modular component, it can be swapped out with another function, which, for example, generates half of the agents with this method and initializes the other ones as neutral agents without a persona.
\paragraph{Response Generators.}
Another important part of multi-agent discussions is how each agent responds to the previous responses. Do I use \ac{CoT} to improve performance, or does this result in too long answers? By changing the way an agent is prompted, a lot of performance can be gained or lost. Therefore, it is key to make this as customizable as possible. The researcher has the possibility to change the default behavior (neutral answers), for example, by prompting the agent to be more critical or changing the way the discussion history is presented. The system prompt for the agent's persona can also be adjusted. \ac{MALLM} already has many different built-in response generators. The ones relevant for this work are the following.
\begin{itemize}
    \item \textbf{Free Text} is the most basic form of the agent prompt. Each agent gets a predefined number of discussion history rounds as memory. The prompt language is neutral, and the task is presented each round to mitigate the potential drift from the topic of the discussion \cite{becker_multi-agent_2024}. In addition, the agent is always asked to agree or disagree with the answer of the previous agent.
    \item \textbf{Simple} behaves very similar to the Free Text response generator, but the prompt is a bit simpler to make it easier to understand for the \ac{LLM} and reduce the context length.
    \item \textbf{Critical} forces the agent to respond very critically to the previous answer and try to find new solutions. Some studies have shown that \acp{LLM} can show some form of sycophancy, which is not helpful for a constructive discussion \cite{sharma_towards_2023}. Encouraging them to be more critical may reduce this.
    \item \textbf{Reasoning} doesn't allow the agents to communicate their final solution with the other agents. They can only share reasoning that can be used to find a final solution. In the end, each agent has to come up with its own solution without being directly influenced by other agents.
\end{itemize}
\paragraph{Discussion Paradigms.}
These paradigms define the discussion format for the entire task. They can control the order in which the agents communicate with each other, and which answers are visible only to certain agents. Currently, all the built-in discussion paradigms are static, meaning that the turn order is predefined and cannot be changed based on specific events during the discussion. However, due to the modular nature of \ac{MALLM}, a new discussion paradigm can be added, for example using an \ac{LLM} as a moderator to dynamically decide which agent should respond next. Current research by \textcite{yin_exchange--thought_2023} and \textcite{becker_multi-agent_2024} shows that discussion protocols have little impact on downstream task performance. \ac{MALLM} includes the following discussion paradigms, which are illustrated in \Cref{fig:discussion_paradigms}. The first four paragdims are inspired by the work of \textcite{yin_exchange--thought_2023}, while the fifth was developed as part of this work.

\bigskip

\begin{itemize}
    \item \textbf{Memory} is the most basic discussion paradigm. The agents respond to the solution of the previous agents with feedback or an improved solution. All answers are visible to the other agents. 
    \item \textbf{Relay} behaves similarly to the memory paradigm. The turn order is the same, but each agent can only see the answer from the previous agent.
    \item \textbf{Report} introduces one agent as a moderator that can communicate with other agents. The other agents can only communicate with the moderator and have access to these messages only. Only the moderator can see all messages.
    \item \textbf{Debate} is similar to the report paradigm, as it also needs a moderator. Here, the other agents can communicate for a predefined number of rounds before they forward their reasoning to the moderator agent, which starts the next round of debate.
    \item \textbf{Collective Refinement} In this protocol, each agent first generates an answer independently. In each subsequent round, every agent receives the responses from all other agents at the same time. Using this shared information, each agent refines their own answer. This process continues throughout the rounds, helping agents gradually reach a shared and improved solution. There is no turn order, and all agents have the same level of knowledge in each round.
\end{itemize}

\begin{figure}[H]
    \centering
    \includegraphics[width=0.9\textwidth]{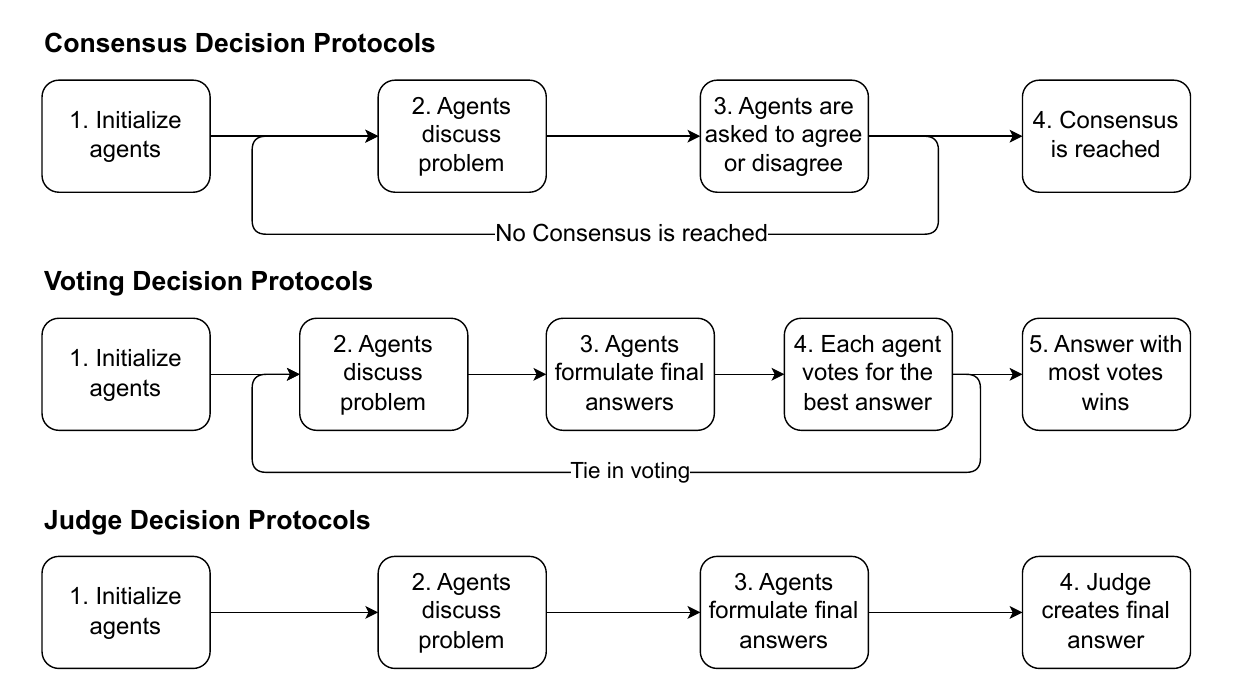}
    \caption{Illustration of the workflow for the different decision protocol families included in \ac{MALLM}.}
    \label{fig:decision_protocols}
\end{figure}

\paragraph{Decision Protocols.}
These are crucial for the framework as they decide which answer gets presented as the final answer to the problem. Multi-agent discussions produce multiple results for the same problem because each agent has its own reasoning and ideas on how to solve the problem. Therefore, some process is needed to decide which answer looks the most promising. I divide these decision protocols into three subtypes that I want to analyze. An overview of how each of these decision protocols works theoretically can be found in \Cref{fig:decision_protocols} and all prompts used for them can be found in \Cref{sec:app_prompts}.
\vspace{-1em}
\subparagraph{Consensus Based Decision Protocols.} 
These are the simplest kinds of decision protocols. After each answer, the next agent has to agree or disagree with the previous statement. Depending on the response generator, this happens in the same message, and the agreement is extracted with a regular expression, or this is split into multiple answers. If enough agents agree in order, there is a consensus. The final answer is extracted by instructing the last agent to solve the given task with the information available in the latest messages. The prompt used for this can be found in \Cref{sec:app_final_answer_prompt}. There are several types of consensus decision protocols available in \ac{MALLM}. \textbf{Majority consensus} requires 50\% of the agents to agree. \textbf{Supermajority consensus} requires 66\% of the agents to agree, and \textbf{unanimity consensus} requires all agents to agree.

\subparagraph{Voting Based Decision Protocols.}
For voting based decision protocols, the process differs slightly compared to consensus-based decision protocols. The agents are forced to discuss for a predefined number of turns and afterward create a final solution. In the default setting, they have to discuss for three rounds, as current research such as \textcite{du_improving_2023} shows that this allows for reasonable strong improvements considering computing resources. If there happens to be a tie in the voting, the agents have to discuss it for another round, and after that, they are asked to vote again. If they do not reach a final decision before exceeding the maximum number of rounds (defined in the discussion configuration), the solution of the first agent is used. To analyze the impact of the voting procedure, different processes similar to the work of \textcite{yang_llm_2024} are implemented.
\begin{itemize}
    \item \textbf{Simple Voting} Each of the agents has only one vote. They can vote for any other agent or for themselves. The agent with the most votes wins.
    \item \textbf{Ranked Voting} The agents have to rank all final answers. The best solution is chosen by adding the ranking indices for a given agent and then selecting the answer with the best cumulative rank.
    \item \textbf{Cumulative Voting} Each agent has to distribute up to 25 points to all possible answers. They can also give fewer points and freely divide the points between all agents (even themselves). The winner is selected by adding all the points for a given agent and selecting the final answer with the most points.
    \item \textbf{Approval Voting} The agent has to provide a list of solutions that it approves. After that, the approvals from all agents are counted, and the answer with the most approvals wins the vote.
\end{itemize}

\subparagraph{Judge Decision Protocol.}
The judge decision protocol starts similarly to the voting-based decision protocols. The agents first have to discuss for three rounds, and then each agent has to create a final solution on their own. Then a new neutral agent is created who is tasked with judging all the answers and creating a new final answer based on all the reasoning presented in the agents' final answers. This approach is also used in other research, as seen in \textcite{zheng_judging_2023}, where it is used to select the best answers from different \acp{LLM}. In this work, it is not used to judge the output of different models but rather the same model with different personas.

\subsubsection{Setup for Experiments}
\label{sec:setup}
For all experiments in this study, I use the \ac{MALLM} framework with Llama 3 8B or 70B model\footnote{The two models used for this study can be found at \href{https://huggingface.co/meta-llama/Meta-Llama-3-8B-Instruct}{meta-llama/Meta-Llama-3-8B-Instruct} and \href{https://huggingface.co/meta-llama/Meta-Llama-3-70B-Instruct}{meta-llama/Meta-Llama-3-70B-Instruct}}. The hardware for the smaller model are two NVIDIA RTX5000 with 16 GB of VRAM each, and the larger model uses eight NVIDIA A100 with 40GB VRAM. A list of default parameters and experiment-specific parameters can be found in \Cref{sec:mallm_setup_app}.

\subsection{Datasets}
\label{sec:datasets}
The dataset selection is very important for this work. It needs to be tested whether decision protocols perform well in multiple domains and whether there are some protocols that perform better with specific tasks than others. Therefore, I selected many datasets from different domains and divided them into two groups. There are the \textbf{Knowledge-based Datasets} like MMLU, MMLU-Pro and GPQA. These datasets still require a lot of reasoning, but also a lot of very specific domain knowledge. On the other hand, there are the \textbf{Logic-based Datasets} like StrategyQA, MuSR, Math-lvl-5, and SQuAD 2.0. An overview of all these datasets can be found in \Cref{tab:datasets} with a description and the number of samples used for evaluation.

\bigskip
\begin{table}[h]
\centering
\begin{tabular}{@{}l| p{7cm} |r| r@{}}
\textbf{Dataset} & \textbf{Description} & \textbf{Samples} & \textbf{Eval-Samples} \\
\midrule
\multicolumn{4}{l}{\textbf{Knowledge-based}} \\
\midrule
\textbf{MMLU} & Massive Multitask Language Understanding benchmark covering 57 subjects & 14,042 & 375 (x3) \\
\textbf{MMLU Pro} & Professional-level extension of MMLU with advanced questions & 12,032 & 374 (x3) \\
\textbf{GPQA} & Challenging dataset of multiple-choice questions written by domain experts in biology, physics, and chemistry & 546 & 250 (x3)\\
\midrule
\multicolumn{4}{l}{\textbf{Logic-based}} \\
\midrule
\textbf{StrategyQA} & Dataset of questions requiring implicit multi-hop reasoning & 2,289 & 330 (x3)\\
\textbf{MuSR} & Logic reasoning for solving murder mystery stories & 250 & 152 (x3)\\
\textbf{Math-lvl-5} & Tests high-level mathematical reasoning and problem-solving abilities & 721 & 252 (x3)\\
\textbf{SQuAD 2.0} & Stanford QA Dataset with answerable and unanswerable questions & 11,873 & 373 (x3)\\
\end{tabular}
\caption{All datasets used for evaluation with a short description, number of samples in the test set, and number of samples used in this study. The datasets are divided into knowledge-based tasks and logic-based tasks.}
\label{tab:datasets}
\end{table}

MMLU was chosen because it covers a wide range of topics from algebra to religion with relatively simple questions \cite{hendrycks_measuring_2021}. To test these decision protocols on more difficult questions that also require a bit more reasoning, GPQA and MMLU-Pro were selected. These two consist of difficult questions on specific domains and, especially with GPQA, they are difficult to answer with web search, which also means that the model is probably not pre-trained on these questions \cite{wang_mmlu-pro_2024, rein_gpqa_2023}. StrategyQA tests the reasoning performance of \ac{MALLM} and the decision protocols, as it requires multiple steps to reach the correct decision. Many of these questions are structured as two different tasks that need to be solved, and the answers have to be combined with some reasoning to get the final answer. This can be difficult because each mistake can influence the overall result \cite{geva_did_2021}. MuSR also requires some multistep reasoning similar to StrategyQA but also comes with a long story that needs to be understood to solve the murder mystery. This tests the performance of the framework with long context length, which can lead to drift from the real task as agents are distracted by all the additional information \cite{sprague_musr_2024, becker_multi-agent_2024}. For the Math-lvl-5 dataset, agents need to prove that they can use really sophisticated logic and reasoning to find the correct solution, as this dataset does not come with multiple choice options, but the agents must come up with the solution themselves \cite{hendrycks_measuring_2021-1}. The final dataset is SQuAD 2.0, where agents have to extract information from a given text to answer a question. The challenge is that sometimes the needed information is not in the text. For these cases, the agents must agree that the information is not given in the text \cite{rajpurkar_know_2018}.

As previously mentioned, it is very expensive to run these multi-agent discussions. Therefore, for this work, I won't run all samples from each dataset, but rather only a small subset, which still gives a good representation of the whole dataset. Similar methods were used in \textcite{yin_exchange--thought_2023, chen_reconcile_2024, becker_multi-agent_2024}. To estimate performance across multiple data sets of varying sizes without processing each sample individually, a sampling strategy is applied that ensures a 95\% confidence level with a 5\% margin of error. For each dataset, an initial sample size \( n_0 \) is calculated using the formula

\[
n_0 = \frac{Z^2 \cdot p \cdot (1 - p)}{d^2}\textup{,}
\]

with \( Z = 1.96 \) (corresponding to 95\% confidence), \( p = 0.5 \) (assuming maximum variability), and \( d = 0.05 \) (representing a 5\% margin of error) \cite{thompson_sampling_2012}. For finite datasets, the parameter \( n_0 \) is adjusted with a finite population correction via

\[
n = \frac{n_0}{1 + \frac{n_0 - 1}{N}}\textup{,}
\]

where \( N \) is the total number of samples in each dataset, to avoid oversampling \cite{cochran_sampling_1953}. The specific sample sizes for each dataset, reflecting this calculation, are provided in \Cref{tab:datasets}.  To ensure repeatability, each data set was tested three times \cite{reimers_reporting_2017, chen_reconcile_2024, becker_multi-agent_2024}, to obtain a standard deviation value.

\newpage
\section{Experiments}
\label{sec:experiments}
I propose four different experiments to answer the research questions. The first experiment tests the performance of various decision protocols across different tasks and compares the results to multiple baselines to determine any significant improvements. To assess whether certain decision protocols are more effective in specific domains, a diverse range of datasets is used. The second experiment evaluates how different prompting methods affect the diversity of solutions generated by agents. The third experiment examines the stability of voting and judge decision protocols by either withholding or providing additional information to see if these changes impact the results and enhance their stability. The final experiment challenges the agents' final answers to determine their reliability and whether their conclusions can be easily altered after extended discussions. The following sections provide detailed descriptions of each experiment.

\paragraph{Task Performance of Decision Protocols.} 

The first experiment is designed to test the effectiveness of the multi-agent system in combination with decision protocols in solving specific datasets. I use three different types of decision protocols to test if there is any advantage in using a specific architecture over another. The first two types are voting and consensus-based decision protocols. \textcite{jones_comparison_1994} already showed some advantages, such as greater comfort for the participants if a decision is achieved by consensus. However, it can also be more challenging to find a solution via consensus and get a certain percentage of participants to agree. Here, voting can help to find a solution faster. The last type of decision protocol I analyze is the judge decision protocol. In this case, a specific agent has the power to create a final solution based on the other answers. All results are compared with a solution from the same base model that is used for the agents and another solution created using the model with \ac{CoT} prompting \cite{wei_chain--thought_2023}. In addition, each experiment is conducted with different sized \acp{LLM} to test whether multi-agent discussions have a greater advantage for smaller or larger models. For all results, the task performance and standard deviation in three runs are recorded. I also take a look at the additional compute time needed for these multi-agent conversations.

This experiment is expanded by incorporating additional datasets to compare task performance in different domains to check if there is a noticeable difference. In addition, it can further analyze the usefulness of certain more strict decision protocols, such as unanimity consensus, as it takes more time to reach this kind of consensus. Some studies from the European Union are already investigating the usefulness of these decision-making protocols compared to the greater flexibility if a less strict consensus protocol is used \cite{cecilia_navarra_qualified_2023, gozi_working_2021}.

\paragraph{Response Diversity.}

In this experiment, both the response generators \textit{Critical} and \textit{Reasoning} and the discussion paradigm \textit{Collective Refinement} are used to encourage more diverse responses from each agent. By using these different response generators and discussion paradigms, the experiment aims to increase the variety of solutions produced, thereby increasing the effectiveness of the decision protocols. It is expected that the decision protocol will have a greater impact if agents generate unique responses, as this diversity increases the probability of including the correct solution. However, increased diversity may also present challenges in accurately selecting the optimal response, requiring more robust decision mechanisms.

\paragraph{Decision Protocol Alterations.} 
In this experiment, only the voting and judge decision protocols are analyzed. It is designed to track the impact of adding or removing specific information during the voting process. As \textcite{shi_large_2023} showed, the addition of irrelevant information can have a big impact on the downstream task performance of \acp{LLM}. To test the effect of this in the voting step, I add or remove certain information from the voting prompt and calculate the downstream task performance. These changes include access to the complete discussion history, providing additional information relevant to the task, calculating a confidence score of the agents in their final answer, and using public or anonymous voting. Additional information is provided with the help of ContextPlus\footnote{A library developed by Florian Wunderlich. Available under: \href{https://github.com/Multi-Agent-LLMs/context-plus}{github.com/Multi-Agent-LLMs/context-plus}}. This package scans relevant Wikipedia pages and combines the most important paragraphs from the page into one coherent final piece of additional information. The confidence values for each final answer are obtained in different ways to test which approach yields the best results. \textcite{chen_reconcile_2024} proposes a method for directly prompting the agent to estimate its own confidence in the answer. This is very simple, but as \textcite{yang_can_2024} showed, \acp{LLM} tend to be overconfident with their answer. Therefore, I propose two more methods. The first one depends on the log-probabilities of the \ac{LLM} output tokens. They are combined using the equation

\[
c_{\text{logprob}}(a) = \frac{\sum_{t \in \text{tokens}(a)} \exp{(\text{logprob}(t))}}{|a|}\textup{,}
\]

Here, \( c_{\text{logprob}}(a) \) represents the confidence of an agent in its final response \( a \). The function iterates over all tokens in \( a \), calculating the mean probability of each token. Since the exponential function is the inverse of the logarithm, it transforms the log probabilities back to probabilities, yielding values between \( 0 \) and \( 1 \) for each token. The last method uses the similarity of the answer in the following rounds as a proxy for the confidence of an agent in its answer. If a model stays with its answer for multiple rounds and cannot be deterred by other agents, the agent seems to have a higher confidence in its answer. This is estimated by calculating the cosine similarity between the answer embeddings produced by an SBERT model \cite{reimers_sentence-bert_2019}.

\paragraph{Challenge of Results.}
The last experiment focuses on testing the consistency of the agents. Previous studies show that people's opinions can be changed even without discussion or persuasive arguments \cite{levitan_conformity_2016}. Especially, social pressure has a large impact on the opinion of people and can also be used to change it \cite{mallinson_effects_2018, dohmen_social_2005}. To answer whether this behavior also exists in language agents, the final answer of each agent is challenged. The agent is then allowed to correct or confirm its final answer. The change in correctness is analyzed for the cases when the agent changes its answer.

\subsection{Performance of Decision Protocols}
\label{sec:experiment1}
This experiment analyzes the difference in task performance for all decision protocols explained in \Cref{sec:architecture} for all datasets introduced in \Cref{sec:datasets}. The setup for this experiment can be found in \Cref{sec:setup} and additional details in \Cref{sec:mallm_setup_app}. It aims to answer the following research question:

\textbf{How do decision-making mechanisms influence LLM performance in conversational tasks?}
\vspace{-0.5em}
\begin{enumerate}
    \itemsep-0.5em 
    \item Is there a decision protocol that consistently performs best?
    \item How effectively can these mechanisms adapt to different tasks?
    \item Does the round in which voting starts affect task performance?
    \item Does the number of agents proposing solutions matter?
\end{enumerate}

\bigskip
\begin{table}[h!]
\centering
\resizebox{\textwidth}{!}{  
\begin{tabular}{p{3.5cm}|ccccccc}
\textbf{Option} & \textbf{MMLU} & \textbf{MMLU-Pro} & \textbf{GPQA} & \textbf{SQuAD} & \textbf{StrategyQA} & \textbf{MuSR} & \textbf{Math-lvl-5} \\ \midrule
Baseline             & 44.8 \tiny{$\pm$ 1.9} & 28.5 \tiny{$\pm$ 1.3} & 28.9 \tiny{$\pm$ 1.2} & 52.3 \tiny{$\pm$ 2.2} & 51.7 \tiny{$\pm$ 2.5} & 25.8 \tiny{$\pm$ 1.6} & 8.8 \tiny{$\pm$ 1.3} \\
Baseline with \ac{CoT}         & 53.1 \tiny{$\pm$ 4.6} & 32.2 \tiny{$\pm$ 4.7} & 29.9 \tiny{$\pm$ 0.6} & 53.8 \tiny{$\pm$ 1.2} & 55.5 \tiny{$\pm$ 1.3} & 29.5 \tiny{$\pm$ 2.6} & 8.7 \tiny{$\pm$ 0.6} \\
\midrule
Simple Voting               & 53.3 \tiny{$\pm$ 1.8} & 32.0 \tiny{$\pm$ 2.7} & 30.5 \tiny{$\pm$ 0.9} & 56.2 \tiny{$\pm$ 0.5} & 58.5 \tiny{$\pm$ 0.9} & 55.2 \tiny{$\pm$ 1.5} & 9.5 \tiny{$\pm$ 1.7} \\
Ranked Voting               & 49.2 \tiny{$\pm$ 1.5} & 33.1 \tiny{$\pm$ 4.6} & 27.3 \tiny{$\pm$ 3.9} & \textbf{58.0 \tiny{$\pm$ 0.8}} & 56.2 \tiny{$\pm$ 3.4} & 52.5 \tiny{$\pm$ 0.0} & 6.8 \tiny{$\pm$ 1.3} \\
Cumulative Voting           & 52.6 \tiny{$\pm$ 4.0} & 28.3 \tiny{$\pm$ 3.1} & 31.3 \tiny{$\pm$ 2.8} & 55.8 \tiny{$\pm$ 3.4} & \textbf{61.2 \tiny{$\pm$ 1.6}} & 56.8 \tiny{$\pm$ 4.2} & 9.0 \tiny{$\pm$ 1.5} \\
Approval Voting             & 43.0 \tiny{$\pm$ 2.1} & 29.2 \tiny{$\pm$ 5.2} & 31.3 \tiny{$\pm$ 2.6} & 46.5 \tiny{$\pm$ 1.4} & 58.7 \tiny{$\pm$ 0.4} & 50.9 \tiny{$\pm$ 4.0} & 7.8 \tiny{$\pm$ 2.6} \\
Average\tablefootnote{Approval Voting is left out as it consistently fails to reach a voting decision as seen in \Cref{tab:termination_percentages}.} & 51.7 \tiny{$\pm$ 2.4} & 31.1 \tiny{$\pm$ 3.5} & 29.7 \tiny{$\pm$ 2.5} & 56.7 \tiny{$\pm$ 1.6} & 58.6 \tiny{$\pm$ 2.0} & 54.8 \tiny{$\pm$ 1.9} &  8.4 \tiny{$\pm$ 1.5} \\
\midrule
Judge              & 53.7 \tiny{$\pm$ 4.7} & 33.5 \tiny{$\pm$ 0.8} & 27.6 \tiny{$\pm$ 2.3} & 57.2 \tiny{$\pm$ 0.6} & 53.7 \tiny{$\pm$ 2.0} & \textbf{59.3 \tiny{$\pm$ 1.0}} & 9.2 \tiny{$\pm$ 3.0} \\
\midrule
Majority Consensus           & 53.2 \tiny{$\pm$ 2.5} & \textbf{36.4 \tiny{$\pm$ 2.1}} & \textbf{32.3 \tiny{$\pm$ 2.9}} & 43.1 \tiny{$\pm$ 2.1} & 59.9 \tiny{$\pm$ 0.1} & 27.8 \tiny{$\pm$ 2.5} & 9.2 \tiny{$\pm$ 3.0} \\
Supermajority Con.      & \textbf{54.6 \tiny{$\pm$ 3.6}} & 35.2 \tiny{$\pm$ 3.0} & 30.7 \tiny{$\pm$ 2.1} & 44.4 \tiny{$\pm$ 0.4} & 56.4 \tiny{$\pm$ 2.1} & 29.3 \tiny{$\pm$ 2.6} & 9.2 \tiny{$\pm$ 0.8} \\
Unanimity Con.           & 54.2 \tiny{$\pm$ 1.0} & 36.3 \tiny{$\pm$ 0.4} & 30.0 \tiny{$\pm$ 2.3} & 43.4 \tiny{$\pm$ 2.0} & 58.8 \tiny{$\pm$ 2.6} & 28.2 \tiny{$\pm$ 2.8} & \textbf{10.8 \tiny{$\pm$ 1.6}} \\
Average              & 54.0 \tiny{$\pm$ 2.7} & 36.0 \tiny{$\pm$ 1.8} & 31.0 \tiny{$\pm$ 2.4} & 43.6 \tiny{$\pm$ 1.5} &  58.4 \tiny{$\pm$ 1.6} & 28.4 \tiny{$\pm$ 2.6} & 9.7 \tiny{$\pm$ 1.8} \\
\midrule
\textbf{Metric} & \textbf{Accuracy} & \textbf{Accuracy} & \textbf{Accuracy} & \textbf{F1 Score} & \textbf{Accuracy} & \textbf{Accuracy} & \textbf{Accuracy} \\
\end{tabular}}
\caption{Task performance of \ac{MALLM} using memory discussion paradigm with different decision protocols. Agents use Llama 8B model as the backbone model. Bold performance values performed the best on a given dataset. Standard deviation is calculated on three independent runs.}
\label{tab:results_8b}
\end{table}

The first question can be answered by looking at the results presented in \Cref{tab:results_8b}. There is no clear trend that one decision protocol performs best on all datasets. In all tasks, the results produced by \ac{MALLM} with some decision protocol outperform the simple baseline and also outperform the baseline that uses \ac{CoT}. For some datasets like MMLU, GPQA, and Math-lvl-5, this is only by a small margin. An overall trend can be observed that on the first three knowledge-based tasks, consensus decision protocols achieve a better score than the other decision protocols. This suggests that these tasks benefit from the sharing of knowledge and reaching of a collective agreement. The later logic-based tasks are better solved using voting and judge decision protocols, likely because they require more individual reasoning or an authoritative decision. The math task stands out as an outlier in this trend. The best improvements are seen in SQuAD 2.0 with $4.2\%$, StrategyQA with $5.7\%$, and MuSR with $19.8\%$ improvement compared to the \ac{CoT} baseline. It is necessary to consider the results from the MuSR dataset with a degree of caution since the baseline and consensus-based methods encountered difficulties in adhering to the prompt with the smaller Llama 8B model.

\bigskip 
\begin{table}[h!]
\centering
\resizebox{\textwidth}{!}{  
\begin{tabular}{p{3.5cm}|ccccccc}
\textbf{Option} & \textbf{MMLU} & \textbf{MMLU-Pro} & \textbf{GPQA} & \textbf{SQuAD} & \textbf{StrategyQA} & \textbf{MuSR} & \textbf{Math-lvl-5} \\ \midrule
Baseline             & 72.7 \tiny{$\pm$ 0.8} & 57.5 \tiny{$\pm$ 0.7} & 45.2 \tiny{$\pm$ 2.0} & 69.5 \tiny{$\pm$ 0.8} & 76.6 \tiny{$\pm$ 2.3} & 63.2 \tiny{$\pm$ 1.2} & 26.5 \tiny{$\pm$ 0.9} \\
Baseline with \ac{CoT}         & \textbf{75.5 \tiny{$\pm$ 0.8}} & \textbf{60.8 \tiny{$\pm$ 1.0}} & 45.9 \tiny{$\pm$ 1.5} & 68.0 \tiny{$\pm$ 1.4} & 78.2 \tiny{$\pm$ 1.0} & \textbf{66.8 \tiny{$\pm$ 1.6}} & \textbf{26.7 \tiny{$\pm$ 0.6}} \\
\midrule
Simple Voting               & 75.5 \tiny{$\pm$ 1.3} & 56.5 \tiny{$\pm$ 5.4} & 45.7 \tiny{$\pm$ 0.3} & 69.5 \tiny{$\pm$ 0.5} & 81.2 \tiny{$\pm$ 1.4} & 59.3 \tiny{$\pm$ 3.0} & 12.0 \tiny{$\pm$ 0.9} \\
Ranked Voting               & 73.3 \tiny{$\pm$ 1.9} & 54.3 \tiny{$\pm$ 0.9} & 44.2 \tiny{$\pm$ 2.1} & \textbf{70.6 \tiny{$\pm$ 1.1}} & 80.3 \tiny{$\pm$ 1.0} & 59.5 \tiny{$\pm$ 0.5} & 12.0 \tiny{$\pm$ 1.3} \\
Cumulative Voting           & 72.5 \tiny{$\pm$ 1.9} & 53.0 \tiny{$\pm$ 0.4} & 43.9 \tiny{$\pm$ 3.1} & 69.7 \tiny{$\pm$ 1.1} & 80.0 \tiny{$\pm$ 0.8} & 60.0 \tiny{$\pm$ 1.7} & 11.9 \tiny{$\pm$ 2.4} \\
Approval Voting             & 50.2 \tiny{$\pm$ 2.1} & 36.3 \tiny{$\pm$ 3.7} & 33.0 \tiny{$\pm$ 3.4} & 24.3 \tiny{$\pm$ 5.0} & 46.4 \tiny{$\pm$ 13.9} & 49.1 \tiny{$\pm$ 12.4} & 7.9 \tiny{$\pm$ 2.3} \\
Average\tablefootnote{Approval Voting is left out as it consistently fails to reach a voting decision as seen in \Cref{tab:termination_percentages}.} & 73.8 \tiny{$\pm$ 1.7} & 54.6 \tiny{$\pm$ 2.2} & 44.6 \tiny{$\pm$ 1.8} & 69.9 \tiny{$\pm$ 0.9} & 80.5 \tiny{$\pm$ 0.7} & 59.6 \tiny{$\pm$ 1.7} & 12.0 \tiny{$\pm$ 1.5}\\
\midrule
Judge              & 72.2 \tiny{$\pm$ 1.4} & 58.0 \tiny{$\pm$ 3.9} & 42.9 \tiny{$\pm$ 5.1} & 65.7 \tiny{$\pm$ 1.1} & \textbf{82.9 \tiny{$\pm$ 1.3}} & 64.2 \tiny{$\pm$ 1.0} & 19.2 \tiny{$\pm$ 0.6} \\
\midrule
Majority Consensus   & 74.0 \tiny{$\pm$ 1.4} & 57.3 \tiny{$\pm$ 3.0} & 43.7 \tiny{$\pm$ 1.0} & 58.2 \tiny{$\pm$ 1.0} & 80.1 \tiny{$\pm$ 0.3} & 61.3 \tiny{$\pm$ 3.3} & 23.2 \tiny{$\pm$ 3.3} \\
Supermajority Con.   & 71.9 \tiny{$\pm$ 1.2} & 57.0 \tiny{$\pm$ 1.7} & \textbf{46.6 \tiny{$\pm$ 0.8}} & 54.3 \tiny{$\pm$ 1.9} & 80.3 \tiny{$\pm$ 1.3} & 60.2 \tiny{$\pm$ 0.3} & 25.7 \tiny{$\pm$ 2.4} \\
Unanimity Con.       & 72.2 \tiny{$\pm$ 2.4} & 57.3 \tiny{$\pm$ 1.5} & 45.3 \tiny{$\pm$ 2.5} & 56.7 \tiny{$\pm$ 2.2} & 78.1 \tiny{$\pm$ 2.3} & 61.3 \tiny{$\pm$ 0.8} & 20.7 \tiny{$\pm$ 1.5} \\
Average & 72.7 \tiny{$\pm$ 1.7} & 57.2 \tiny{$\pm$ 2.1} & 45.2 \tiny{$\pm$ 1.4} & 56.4 \tiny{$\pm$ 1.7} & 79.5 \tiny{$\pm$ 1.3} & 60.9 \tiny{$\pm$ 1.5} & 23.2 \tiny{$\pm$ 2.4}\\
\midrule
\textbf{Metric} & \textbf{Accuracy} & \textbf{Accuracy} & \textbf{Accuracy} & \textbf{F1 Score} & \textbf{Accuracy} & \textbf{Accuracy} & \textbf{Accuracy} \\
\end{tabular}}
\caption{Task performance of \ac{MALLM} using memory discussion paradigm with different decision protocols. Agents use the Llama 70B model as the backbone model. Bold performance values performed the best on a given dataset. The standard deviation is calculated on three independent runs.}
\label{tab:results_70b}
\end{table}

Compared to these results, the larger Llama 3 70B model performs much better overall, as seen in \Cref{tab:results_70b}. Most of the results are a bit better than the baseline, but the multi-agent discussions are only in a few cases able to outperform the \ac{CoT} baseline. This model does not fail to follow the prompt for the MuSR baseline and consensus-based decision protocols. Therefore, the big performance gain from the smaller model cannot be observed here. SQuAD 2.0 and StrategyQA had the largest performance gains, even outperforming the \ac{CoT} baseline, similar to the results from the smaller model. This difference in task performance can have many reasons. As \textcite{li_dawn_2024} showed, smaller models are more likely to hallucinate, which reduces task performance. This can be mitigated by using multiple agents because it is less likely that two agents hallucinate the same things. Larger models tend to hallucinate less, reducing this effect for the Llama 3 70B model \cite{li_dawn_2024}. In general, Llama 3 70B has a much higher baseline for task performance, making it more difficult to improve baseline results. Many of the improvements by the Llama 3 8B model are quite small, except for the ones where the Llama 3 70B model also outperforms the \ac{CoT} baseline. This can be taken as evidence that these multi-agent discussions require specific problem structures, or else the agents are just talking about the same results for multiple rounds and agreeing with each other. If these discussions continue too long, they can drift away from the original task, which reduces task performance. This has also been observed by \textcite{becker_multi-agent_2024} and an example can be seen in \Cref{example:failed_discussion}. A positive example of how discussion can help task performance can be seen in \Cref{example:good_discussion}.

\begin{figure}[H]
    \centering
    \includegraphics[width=\linewidth]{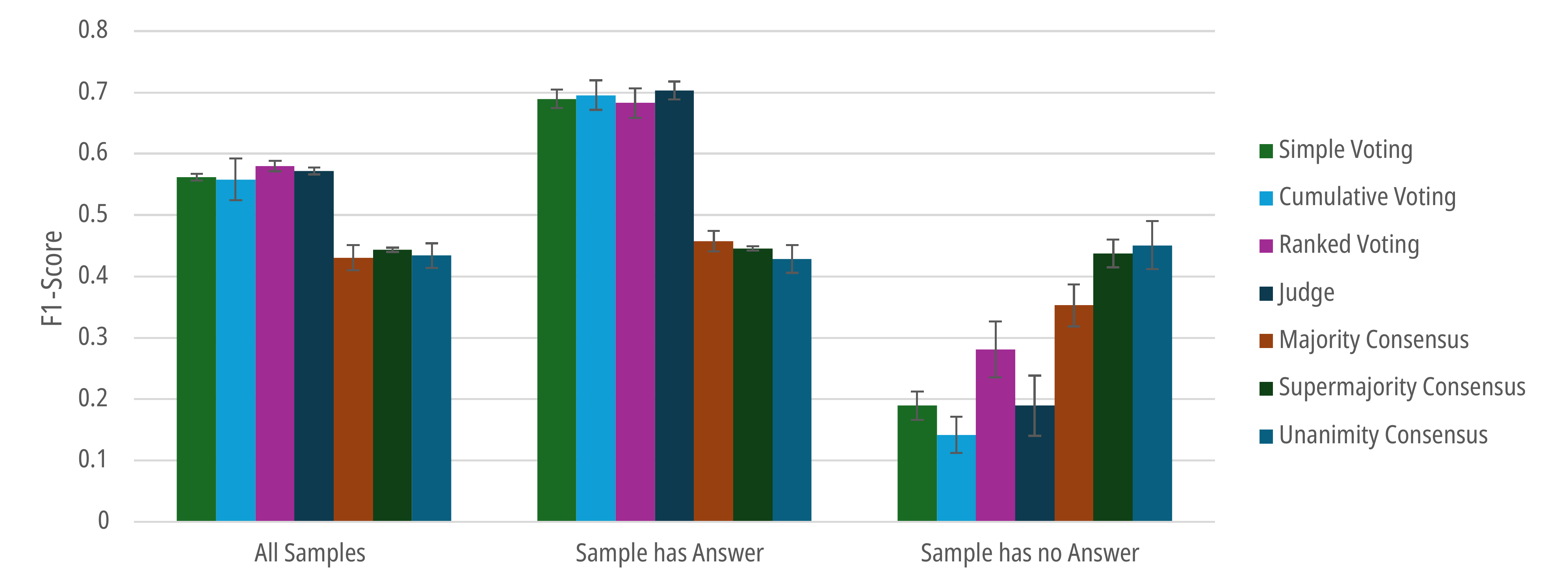}
    \caption{Task performance of \ac{MALLM} on SQuAD 2.0 divided into three groups. All samples, samples that provide the answer in the context, and samples that don't provide the answer. Approval voting is left out as it consistently fails to reach a voting decision as seen in \Cref{tab:termination_percentages}. The standard deviation is calculated on three independent runs.}
    \label{fig:squad_divided}
\end{figure}

To further analyze the strengths and weaknesses of different decision protocols, it is helpful to take a closer look at more specific results from the SQuAD 2.0 dataset. The goal of this task is to extract some specific information from a provided context. If this information is not included within the context, the agents have to mark the answer as unsolvable. In \Cref{fig:squad_divided}, the task performance for all decision protocols is divided into three different classes \cite{rajpurkar_know_2018}. The first class includes all answers. The second class only contains samples that have the answer provided in the given context. The last class contains all samples that are unanswerable with the help of the given context. 
\bigskip

There is a big performance difference between voting and consensus-based decision protocols for SQuAD 2.0. Overall voting has a higher F1-Score, and by only looking at the samples that are answerable, this difference is even more pronounced. This changes with unanswerable samples, as consensus-based decision protocols consistently outperform the other decision protocols. This shows that these different protocols have specific tasks in which they perform better. For voting decision protocols it is more difficult to agree that this sample has no answer as this requires them to agree on the final verdict that it is unanswerable. These samples are often constructed to fool the reader, as seen in \Cref{example:trick_discussion}. This makes it difficult for voting as some agents are tricked, and then propose this as a solution which also tricks other agents into voting for them. In consensus decision protocols, this does not happen that easily, as multiple agents have to agree in a row. It can also be observed that the consensus protocols with a more strict consensus threshold tend to perform better here. Although all types of decision protocols have their specific strengths and weaknesses, voting decision protocols outperform consensus decision protocols here because there are more samples with answers than without answers. This results in a higher overall score for the voting and judge decision protocols.

The data in \Cref{tab:termination_percentages} shows the number of turns that are needed for each decision protocol to reach a final decision for the MMLU dataset. Most of the voting decision protocols are able to vote for a final answer already in the first round in which they are allowed to vote. Simple voting has the highest agreement rate, but also cumulative and ranked voting only need in a few cases another round. In contrast, the approval decision protocol only achieves this in $\sim27\%$ of the cases. About $14\%$ need another round and the rest is canceled after the fifth round. This happens because these models like to agree with each other, and therefore they tend to vote for many of the answers, which often leads to a tie. Therefore, more restrictive voting decision protocols can reach a decision more easily, as a tie is less likely. The judge decision protocol always finishes after the third round, as it just generates a final answer based on the other solutions, and no tie can occur. The consensus decision protocols require only one to two rounds to reach consensus and still achieve a higher task performance because these results are based on the MMLU dataset. In \Cref{sec:appendix_turns_until_consensus} the results for all the datasets and models can be found. The decision protocols tend to behave similarly in terms of rounds needed to create a final answer, independent of task and model.

\bigskip
\begin{table}[H]
\centering
\begin{tabular}{l|ccccc|c}
\toprule
\textbf{Group} & \textbf{Turn 1} & \textbf{Turn 2} & \textbf{Turn 3} & \textbf{Turn 4} & \textbf{Turn 5} & \textbf{Task Performance Score} \\
\midrule
Voting         & 0.00\%      & 0.00\%      & 99.33\%  & 0.50\%   & 0.17\%   & 53.3 \tiny{$\pm$ 1.8} \\
Cumulative     & 0.00\%      & 0.00\%      & 94.00\%  & 5.50\%   & 0.50\%   & 52.6 \tiny{$\pm$ 4.0} \\
Ranked         & 0.00\%      & 0.00\%      & 91.17\%  & 7.83\%   & 1.00\%   & 49.2 \tiny{$\pm$ 1.5} \\
Approval       & 0.00\%      & 0.00\%      & 26.67\%  & 14.33\%  & 59.00\%  & 43.0 \tiny{$\pm$ 2.1} \\
\midrule
Judge        & 0.00\%      & 0.00\%      & 100.00\% & 0.00\%      & 0.00\%      & 53.7 \tiny{$\pm$ 4.7} \\
\midrule
Majority       & 80.00\%  & 13.67\%  & 4.83\%   & 1.00\%   & 0.50\%   & 53.2 \tiny{$\pm$ 2.5} \\
Supermaj.  & 79.33\%  & 14.33\%  & 4.83\%   & 1.00\%   & 0.50\%   & 54.6 \tiny{$\pm$ 3.6} \\
Unanimity      & 59.50\%  & 21.67\%  & 12.67\%  & 3.50\%   & 2.67\%   & 54.2 \tiny{$\pm$ 1.0} \\
\bottomrule
\end{tabular}
\caption{Number of rounds needed for each decision protocol to reach a final decision for the MMLU dataset.}
\label{tab:termination_percentages}
\end{table}

The next analysis focuses on the third and fourth points of the first research question. Here I track performance gains and losses for agents that must communicate for more rounds before they are allowed to vote. In the current setup, agents communicate for three rounds, as this showed promising results in \textcite{du_improving_2023}. The previous analysis based on \Cref{tab:termination_percentages} shows that consensus decision protocols achieve better task performance even with fewer turns. If this also works for the voting decision protocols, less compute power is needed, which improves the efficiency of the multi-agent discussions.

\bigskip
\begin{figure}[H]
\begin{subfigure}{.5\textwidth}
  \centering
  \includegraphics[width=0.9\linewidth]{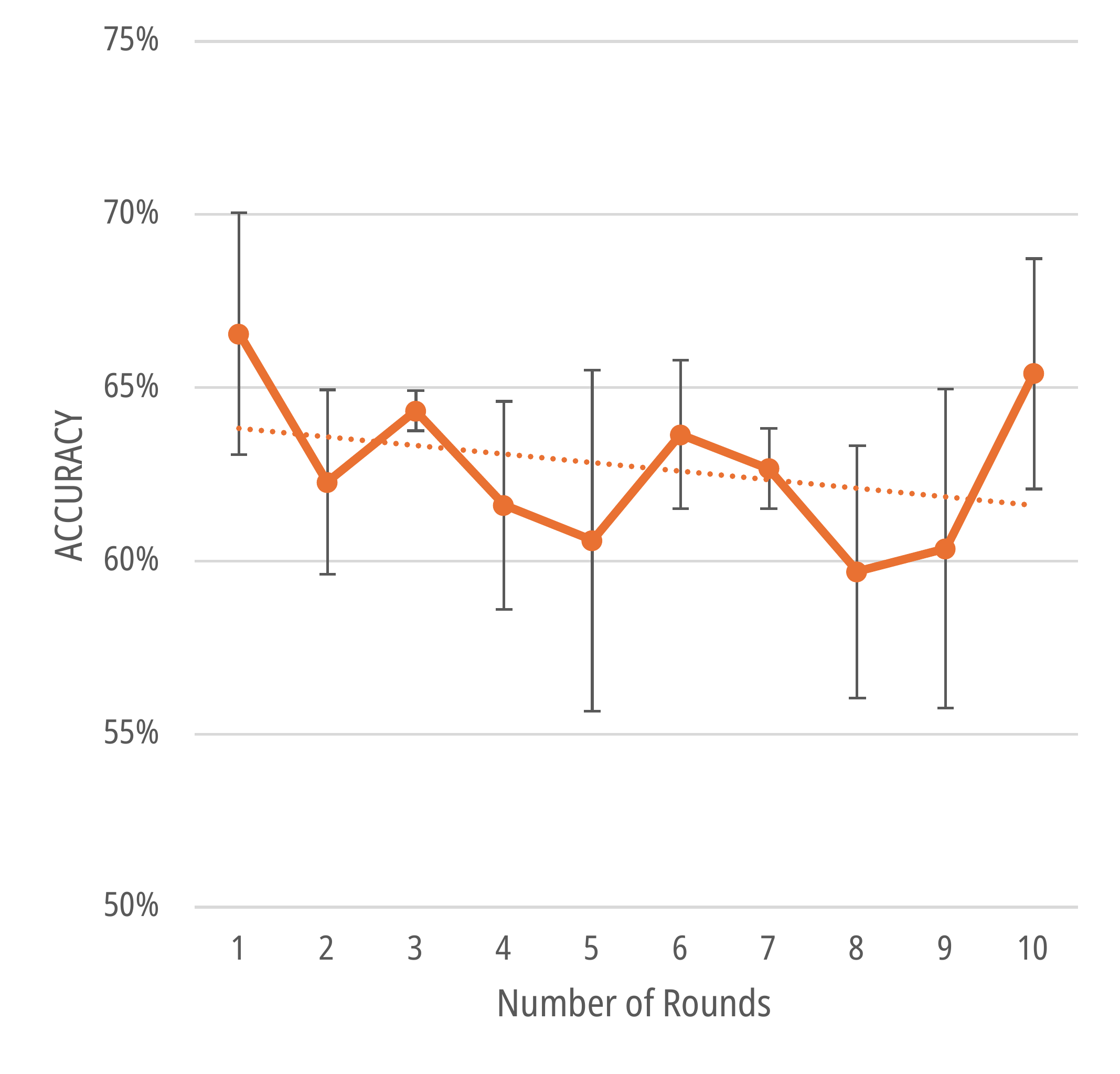}
  \caption{}
  \label{fig:num_rounds}
\end{subfigure}%
\begin{subfigure}{.5\textwidth}
  \centering
  \includegraphics[width=0.9\linewidth]{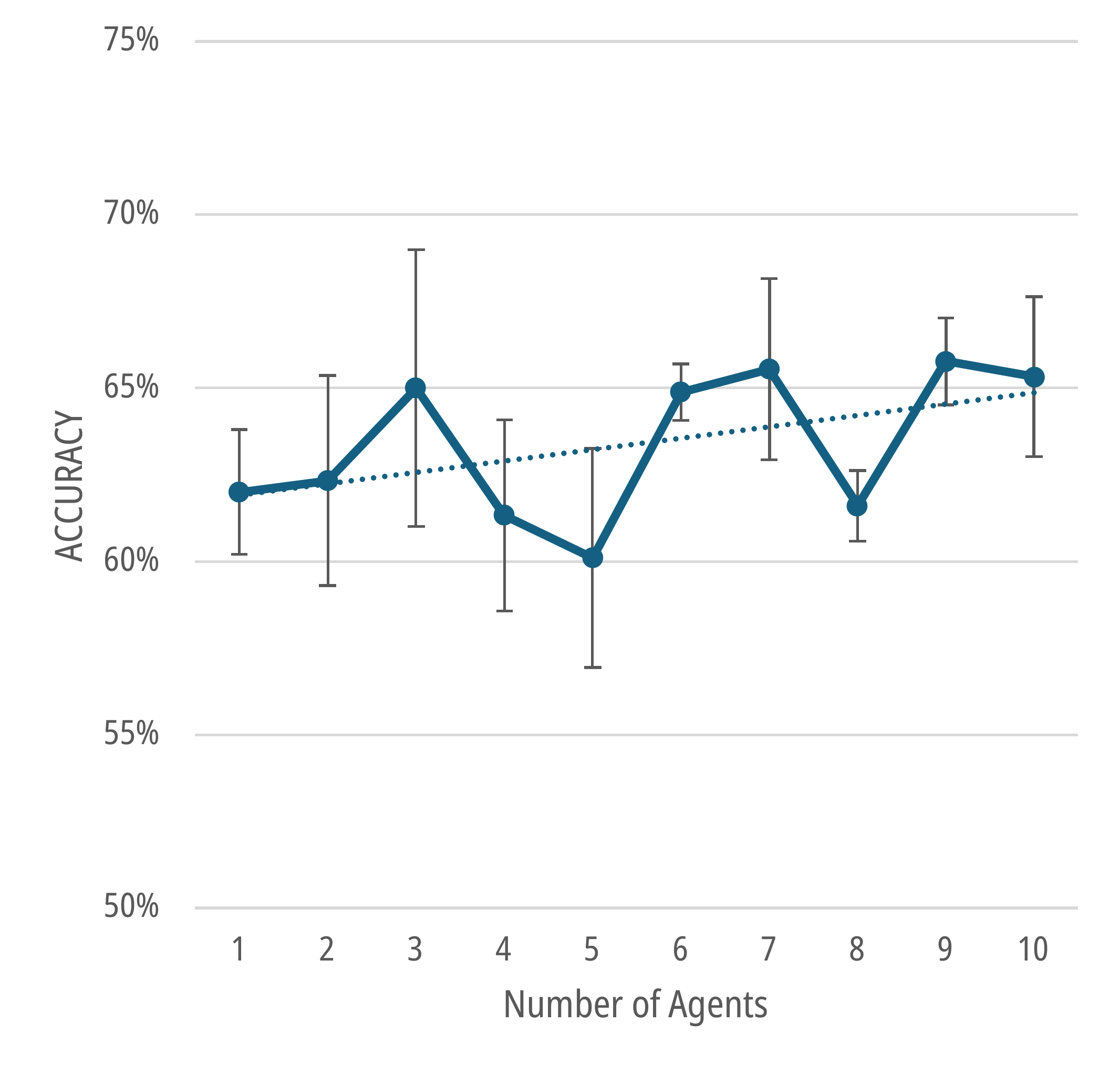}
  \caption{}
  \label{fig:num_agents}
\end{subfigure}
\caption{\textbf{Graph (a)} shows the task performance on StrategyQA when the agents have to talk for a given number of rounds before they are allowed to vote using simple voting decision protocol. \textbf{Graph (b)} shows the task performance on StrategyQA with a different number of agents participating in the discussion. The final answer is also created using the simple voting decision protocol. The standard deviation is calculated on three independent runs.}
\label{fig:num_rounds_agents}
\end{figure}

To analyze the impact of the first round that allows for voting, \Cref{fig:num_rounds} shows the task performance, compared to the number of rounds until the first vote. The data is generated using the simple voting decision protocol in the StrategyQA dataset. Although the results are quite noisy, a downward trend can be observed, visualized by the dotted line. This shows that more discussion rounds can hinder task performance, further supporting the hypothesis of \textcite{becker_multi-agent_2024} that these agents tend to drift away from the real problem, resulting in lower accuracy of the final answer. \Cref{fig:num_agents} analyzes the same decision protocol and dataset but scales the number of agents who collaborate on the problem. In this case, the dotted line shows an upward trend in task performance with an increasing number of agents. Recent work, such as \textcite{wang_rethinking_2024}, shows that an increase in the number of agents does not automatically lead to better task performance. \textcite{chen_reconcile_2024} uses several different \ac{LLM} base models for multi-agent discussion. They showed that the more different models used, the better the performance improvement. Therefore, the most likely reason why an upward trend can be seen in this case is due to the personas of the agents. This allows the model to access different information and makes the discussion more diverse. Similarly, different base models also increase the diversity and knowledge base.

These results are similar to the observed improvement in self-consistency, where an increase in the number of answers increases task performance \cite{wang_self-consistency_2022}. Here, having more agents generating solutions also tends to produce better scores. However, increasing the number of rounds reduces performance, which should not occur if self-refinement is reliable. Recent research questions the results of \textcite{madaan_self-refine_2023}, with \textcite{huang_large_2024} showing that these results may be unreliable, which is also supported by these findings.

\subsubsection{Takeaways}
\begin{enumerate}
    \item \textbf{Task-specific Protocol Performance:} Voting protocols excel in knowledge-based tasks, while consensus protocols perform better in logic-based and unanswerable tasks.
    \item \textbf{Model Size Effects:} Smaller models benefit more from multi-agent discussions, while larger models show limited improvement due to higher baseline performance and lower hallucination.
    \item \textbf{Impact of Discussion Rounds:} Limiting discussion rounds before voting enhances performance by minimizing possible task drift.
    \item \textbf{Answer Diversity:} More agents improve task performance through enriched discussions and greater access to information.
\end{enumerate}

\subsection{Answer Diversity}
In multi-agent systems, response diversity is a critical factor in improving decision protocols; therefore, this research question focuses on understanding how response diversity can be enhanced and used effectively.

\textbf{How does the diversity of responses affect the performance of decision protocols?}
\vspace{-0.5em}
\begin{enumerate}
    \itemsep-0.5em 
    \item Is it more efficient to start with one solution and iterate on it, or should each agent propose its own initial solution?
    \item Can different discussion protocols or prompting help increase the answer diversity?
\end{enumerate}

During all experiments, a pattern could be observed in which these agents tend to agree with each other, resulting in similar answers. To achieve better results with a decision protocol, it is important to have multiple possible answers, as it tends to be easier to select the correct answer from a pool of possible answers than to come up with a solution on our own. This is also a basis for how self-consistency achieves better results \cite{wang_self-consistency_2022} and why multiple-choice tests tend to be easier. Currently, the first agent comes up with a solution and all other agents can either generate a new solution or improve on the first one. As these models like to agree, they often try to improve the first solution instead of providing a new solution based on their expertise. This leads to a smaller knowledge base for the whole discussion and possibly hinders task performance. This hypothesis is tested by not allowing the agents to improve any other answer but forcing them to generate their initial answer on their own. In the following rounds, they can iterate on their own or on any other solution. This creates at least one answer based on the expertise of each agent.

\begin{figure}[H]
    \centering
    \includegraphics[width=0.95\linewidth]{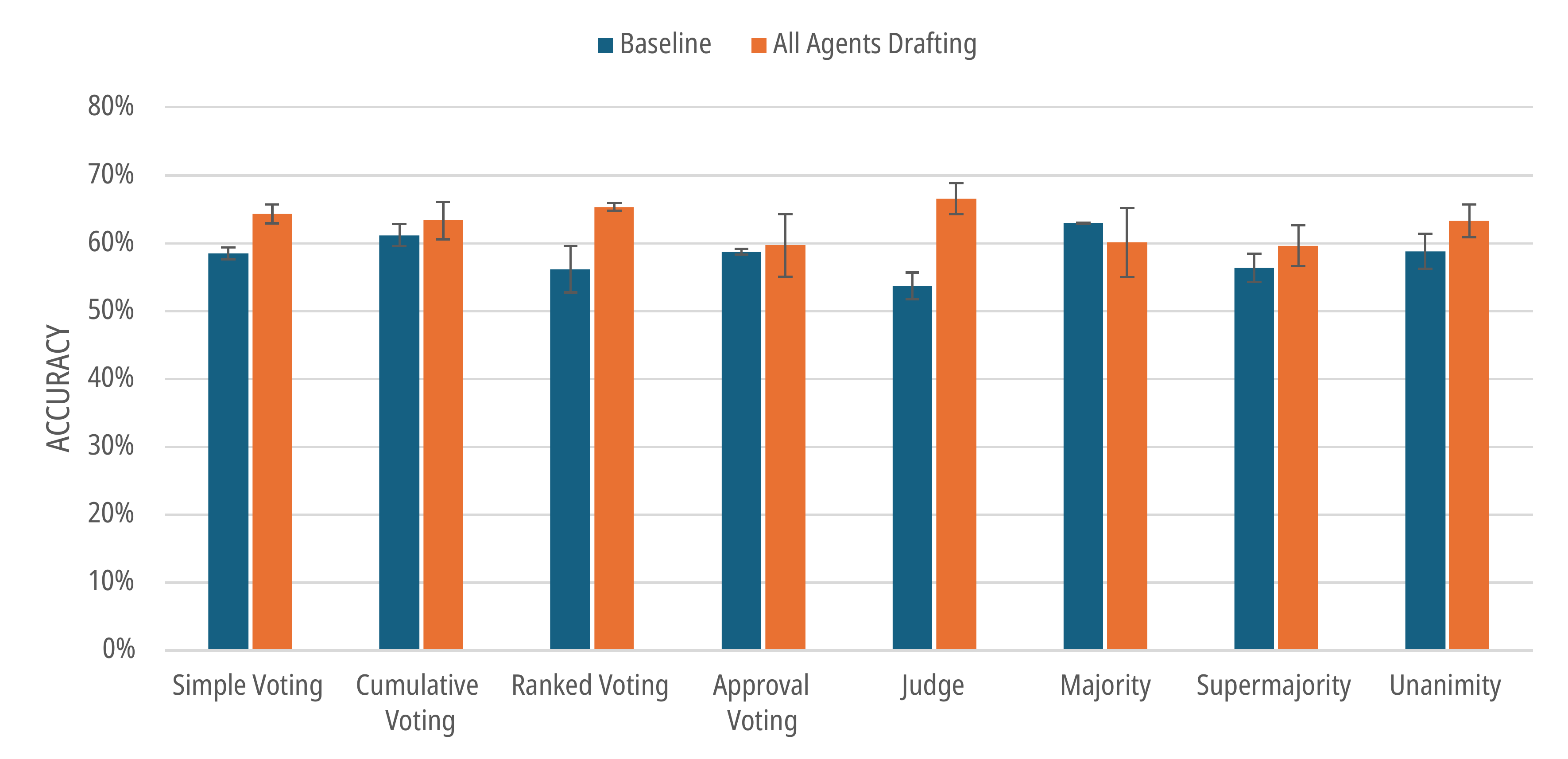}
    \caption{Comparison of agents iterating on one initial draft (baseline) vs. each agent generating one initial draft on the StrategyQA dataset. The standard deviation is calculated on three independent runs.}
    \label{fig:all_agents_draft}
\end{figure}

The results of this approach can be seen in \Cref{fig:all_agents_draft}. All decision protocols are tested, with only the first agent drafting a solution and all agents drafting an initial solution. For almost all decision protocols, the approach with all agents generating an initial solution gives a higher average score. Only for the majority consensus protocol this is different, but it is still within the standard deviation. This shows that increased answer diversity seems to help with task performance.

\begin{figure}[H]
    \centering
    \includegraphics[width=0.75\linewidth]{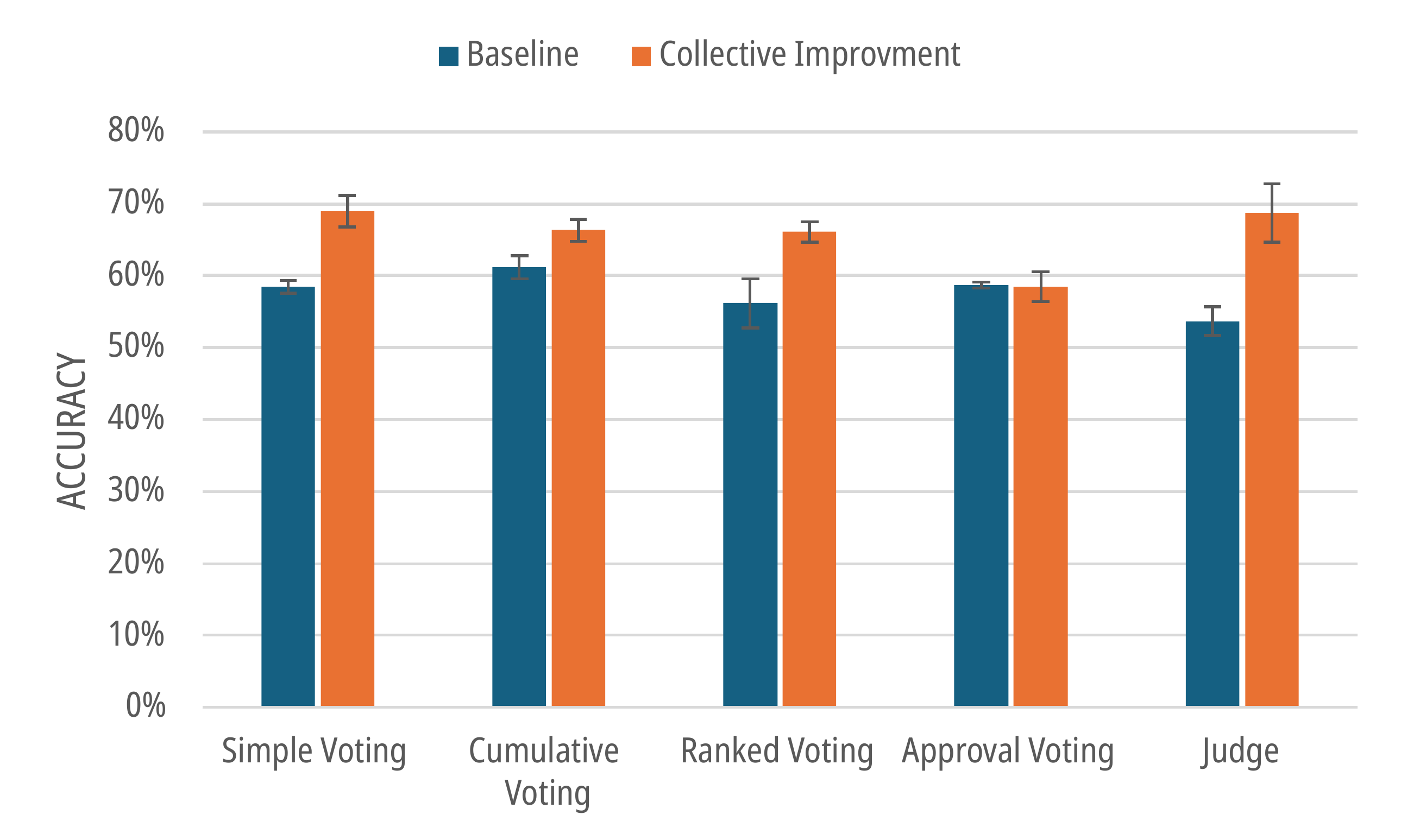}
    \caption{Comparison of agents discussing using the memory discussion protocol (baseline) compared to the collective refinement discussion protocol on the StrategyQA dataset. The standard deviation is calculated on three independent runs.}
    \label{fig:collective}
\end{figure}

The next experiment analyzes the impact of the collective refinement discussion protocol. This builds on the idea of the previous experiment. In fact, this discussion protocol always forces all agents to come up with a solution and does not allow any inter-round communication. Only the solutions from the previous round are presented to all agents, which results in no active turn order. Therefore, less communication takes place and the ideas can evolve more independently, which should help with answer diversity. Having no turn order of the agents makes this discussion protocol unsuitable for consensus-based decision protocols, as they require some specific order, which allows them to agree or disagree with the previous answers. Therefore, this experiment was performed only using the voting and judge decision protocols. The results of this approach can be seen in \Cref{fig:collective}. Compared to \Cref{fig:all_agents_draft}, the improvements are slightly stronger, but this also benefits from the fact that this discussion protocol forces each agent to come up with an initial solution on their own. Nevertheless, limiting active communication between agents seems to also help with task performance.

\bigskip

\begin{figure}[H]
    \centering
    \includegraphics[width=\linewidth]{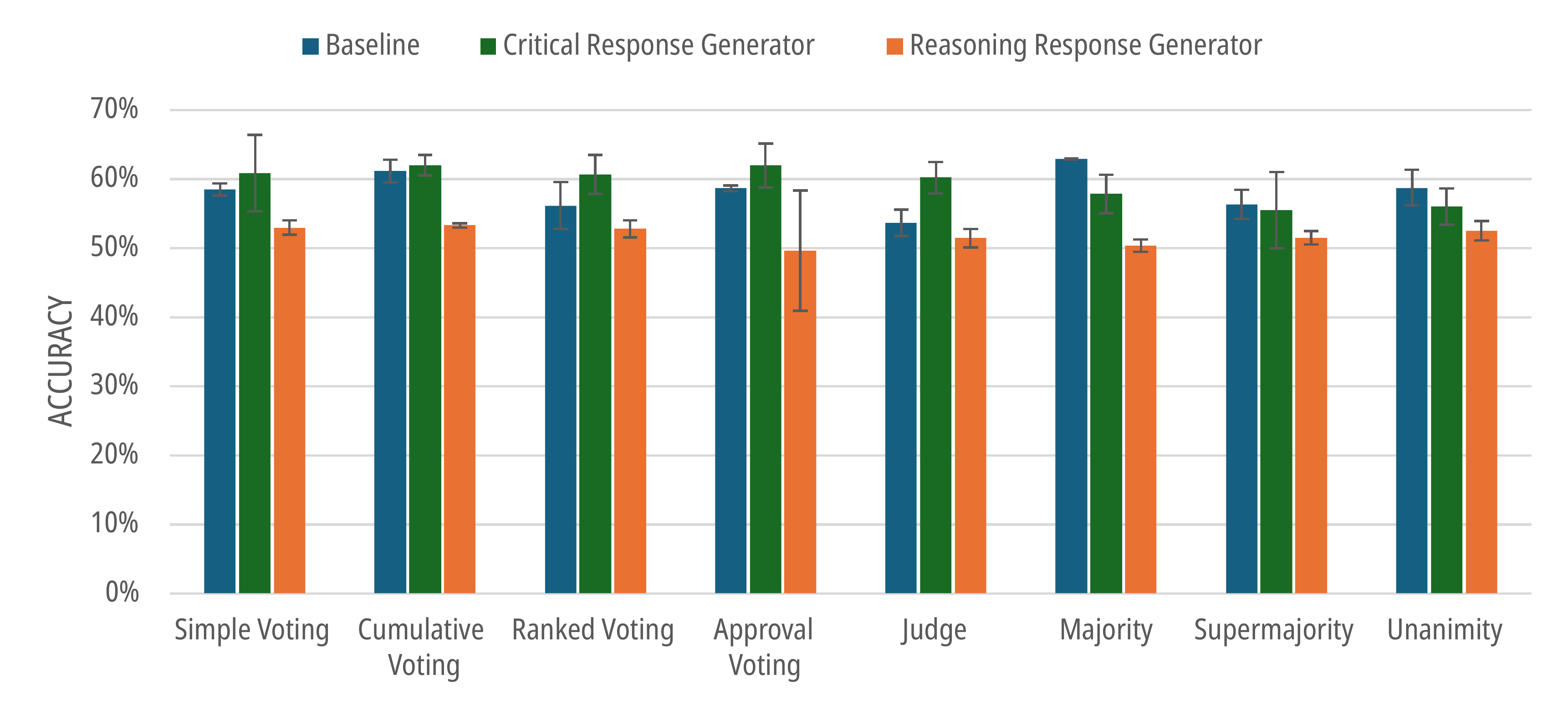}
    \caption{Comparison of agents using the default response generator compared to the reasoning and critical response generator on the StrategyQA dataset. The standard deviation is calculated on three independent runs.}
    \label{fig:response_generator}
\end{figure}

In contrast to changing the discussion protocol, the next experiment focuses on the response generator. An explanation of its role in \ac{MALLM} can be found in \Cref{fig:mallm_overview}. There are two response generators implemented that are focused on improving the answer diversity within multi-agent conversations. The \textit{critical response generator} changes the agent prompt to analyze the answers of the other agents more critically and point out all errors. This should reduce the agreeability of the agents and overall result in more diverse solutions. The \textit{reasoning response generator} does not allow the agents to suggest a final solution to the other agents but only share the reasoning that is important to get to a final decision. With this approach, each agent should be able to come up with a solution on its own and be less influenced by other agents. The results in \Cref{fig:response_generator} show that the critical response generator achieves better scores with voting-based decision protocols. The reasoning response generator seems to perform always worse than the baseline.

\bigskip

To further analyze the results, I created a similarity score that computes the mean cosine similarity between the embeddings of all the agents' final answers on a given task. The embeddings are created using the sentence transformer SBERT\footnote{\href{https://sbert.net/}{sbert.net}}. This is done for all methods that aim to improve response diversity. The results can be found in \Cref{tab:embed-strategies-transposed}. Overall, it can be seen that all of these similarities are pretty high, but this is expected for multiple-choice datasets. All strategies that achieve a better mean task performance in the StrategyQA dataset also show a decreased similarity of answers. Only the reasoning response generator shows an increased answer similarity score, which also results in a large drop in task performance. The critical response generator is an outlier, as it has a strongly decreased similarity score but only a small improvement in task performance. This can probably be attributed to the decrease in discussion quality as the agents are forced to disagree more often with each other.

\bigskip
\begin{table}[ht]
\centering
\begin{tabular}{l|lllll}
\toprule
 \textbf{Strategy}& \textbf{Baseline} & \textbf{All draft} & \textbf{Collective} & \textbf{Critical} & \textbf{Reasoning} \\ \midrule
\textbf{Similarity Score}    & 0.8880      & 0.8704     & 0.8446       & 0.8431      & 0.9162        \\ 
\textbf{Mean Accuracy}    & 58.3\%      & 62.8\%  & 65.7\%   & 59.4\% & 51.9\%               \\ 
\textbf{Delta Baseline}    & 0.0\%    & 4.5\%& 7.4\%   & 1.1\%     & -6.4\%              \\ \bottomrule
\end{tabular}
\caption{Final answer similarity based on average cosine similarity between SBERT embeddings of final answers compared to task performance on StrategyQA dataset.}
\label{tab:embed-strategies-transposed}
\end{table}

\subsubsection{Takeaways}
\begin{enumerate}
    \item \textbf{Diverse Ideas:} Encouraging agents to independently generate solutions leads to a wider range of ideas, which improves decision-making and overall performance.
    \item \textbf{Effective Collaboration:} Using strategies that promote independent thinking and limit group influence improves performance.
\end{enumerate}

\subsection{Alterations in Decision Protocols}
\label{sec:alterations}

This section focuses on the voting and judge decision protocols, as both have a final decision step that can be influenced, in contrast to consensus decision protocols, where a final solution is reached during discussion. The research question primarily analyses task performance when some information is withheld or additional information is added during this final step. 

\textbf{How do small variations in information provided during the decision phase impact decision protocols in multi-agent LLM systems?}
\vspace{-0.5em}
\begin{enumerate}
    \itemsep-0.5em 
    \item How do small changes in the amount of information provided during the decision phase affect task performance and decision stability? 
    \item How does providing agents with new information (e.g., confidence scores or additional facts) after the discussion phase influence their voting decisions and final results?
\end{enumerate}

\begin{figure}[H]
    \centering
    \includegraphics[width=\linewidth]{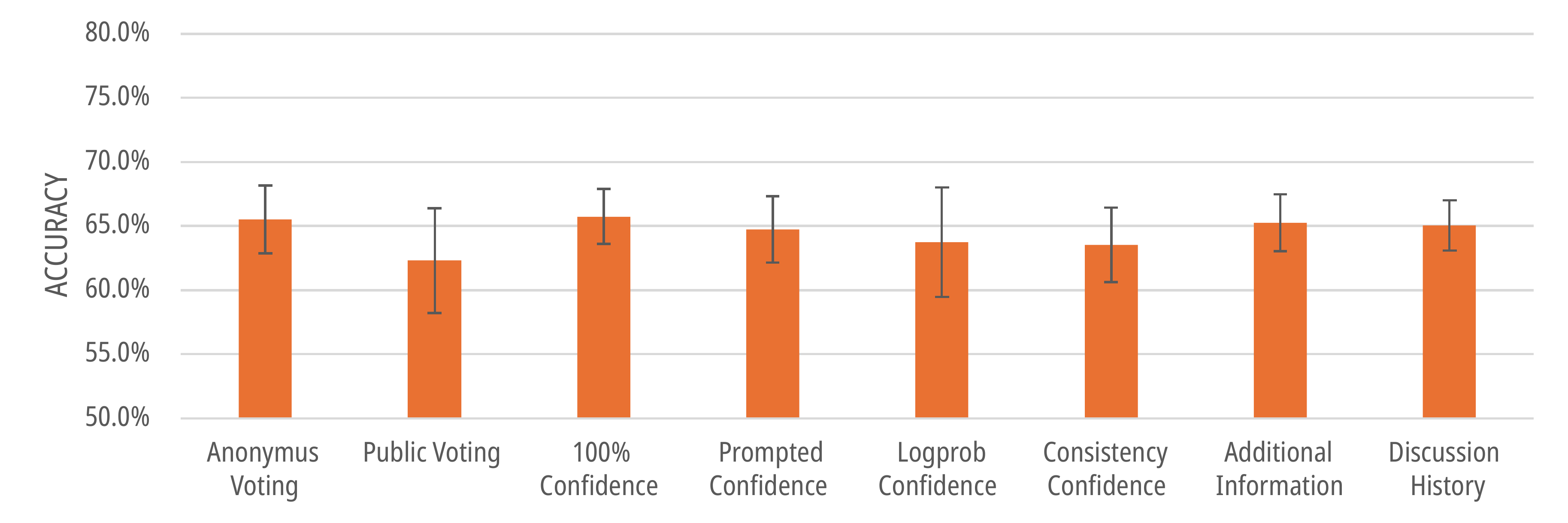}
    \caption{Mean task performance on the StrategyQA dataset for all voting and judge decision protocols with all variations. A description of each alteration can be found in \Cref{sec:alterations}. Standard deviation is calculated on three independent runs.}
    \label{fig:alt-strat-mean}
\end{figure}

This experiment focuses only on the voting and judge decision protocol as explained earlier, but additionally, the approval voting decision protocol is also left out. This is done because this decision protocol cannot consistently arrive at a final solution, as seen in \Cref{tab:termination_percentages}. This would make it less comparable, since the final answer is simply the last solution provided by the last agent and not created with a voting step. Only four datasets are used in this experiment, as some alterations are computationally expensive. These datasets are MMLU, MMLU-Pro, StrategyQA, and SQuAD 2.0, as they showed good results and provide an even mix between knowledge and logic-based datasets. \Cref{fig:alt-strat-mean} shows the mean task performance of all decision protocols in the StrategyQA dataset. The following alterations were applied to the decision protocols:

\begin{itemize} 
    \item \textbf{Anonymous Voting.} This is the default behavior for all decision protocols. The final answers are presented without any additional information to anonymize decisions and prevent bias. 
    \item \textbf{Public Voting.} In this protocol, solutions are linked to their proposers, but individual votes remain anonymous. This provides transparency on the origin of ideas while preserving the anonymity of voters. 
    \item \textbf{100\% Confidence.} Assumes agents are completely confident in their final answers, treating each response as definitive without any explicit confidence calculation. This alteration is added as a baseline for the following dynamic confidence alterations.
    \item \textbf{Prompted Confidence.} Agents are directly asked to estimate their confidence in their final response, as proposed by \textcite{chen_reconcile_2024}. 
    \item \textbf{Logprob Confidence.} Confidence is calculated based on the log probabilities of the tokens in the output of the model, as described in the equation provided in \Cref{sec:experiments}. 
    \item \textbf{Consistency Confidence.} Measures confidence based on the consistency of the answers in multiple rounds, using cosine similarity between embeddings generated by SBERT \cite{reimers_sentence-bert_2019}. 
    \item \textbf{Additional Information.} Relevant external knowledge, extracted using ContextPlus, is provided to agents to improve decision quality. 
    \item \textbf{Discussion History.} Access to the complete dialogue history is provided, allowing agents to build on past exchanges and improve their decisions. 
\end{itemize}

\bigskip

\begin{figure}[H]
    \centering
    \includegraphics[width=\linewidth]{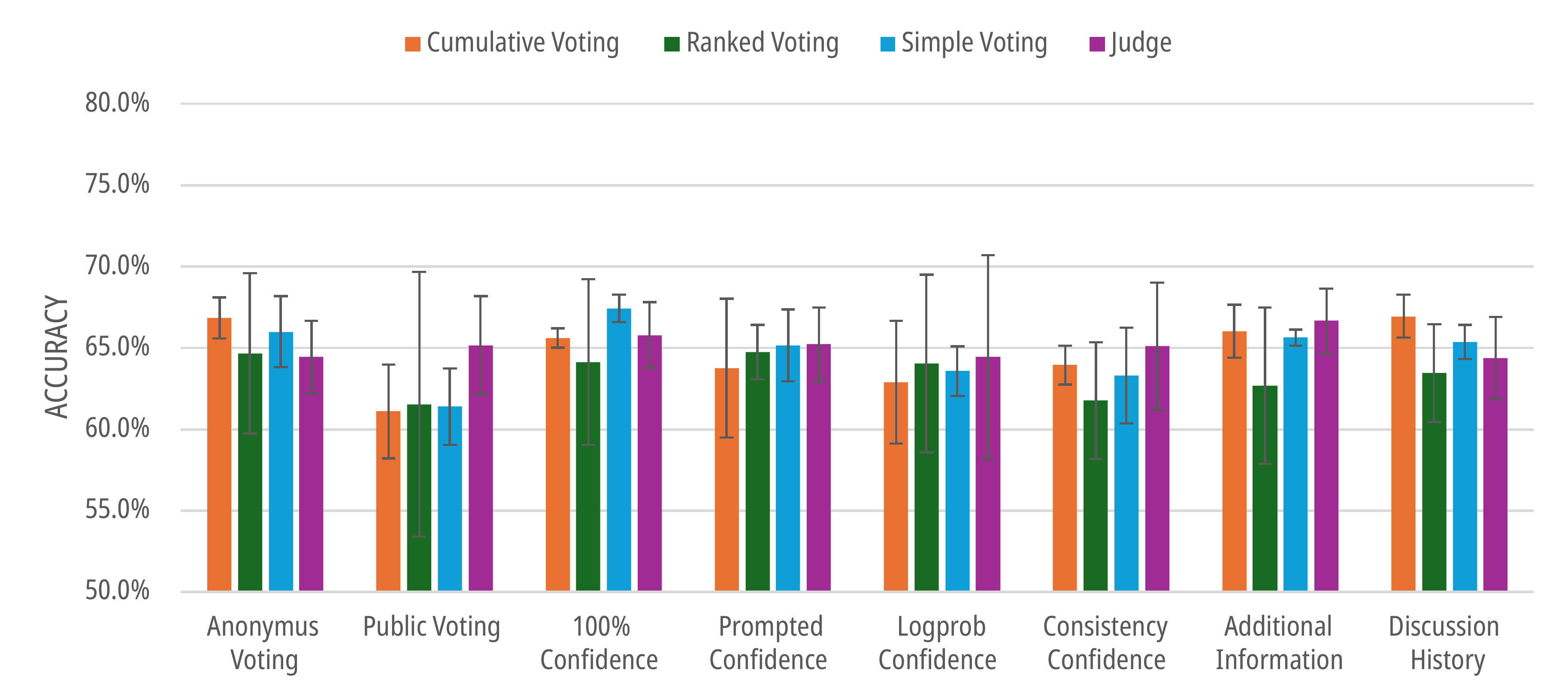}
    \caption{Task performance on the StrategyQA dataset for all voting and judge decision protocols with all variations. A description of each alteration can be found in \Cref{sec:alterations}. The standard deviation is calculated on three independent runs.}
    \label{fig:alt-strat-all}
\end{figure}

Overall, the alterations perform very similarly, without a clear improvement over the baseline. The results for the other datasets can be found in \Cref{sec:appendix_decision_alterations} and there is also no visible improvement or trend that some alterations perform differently. Sometimes, anonymous solutions perform better than public solutions, and for other datasets, vice versa. Adding confidence scores does not seem to have a big impact either, as all scores are very close to each other. Providing additional information and the full discussion history also does not provide a measurable improvement. All task performances are within one standard deviation of each other. In \Cref{fig:alt-strat-all} the same results are shown but divided for each decision protocol. All results for the other datasets divided for each decision protocol are available in \Cref{sec:appendix_decision_alterations}.

In conclusion, the alterations applied to the voting and judge decision protocols, ranging from variations in anonymity and confidence estimation to the provision of additional information, did not lead to significant differences in task performance in the datasets tested. As illustrated in \Cref{fig:alt-strat-all}, all modifications yielded performance results within one standard deviation of each other. This consistency suggests that these decision protocols are robust to small changes in the information provided during the final decision-making phase. The lack of measurable impact implies that the agents do not fully utilize the additional information or that the existing protocols already capture the essential elements required for effective decision-making.

\subsubsection{Takeaways} 
\begin{enumerate} 
    \item \textbf{Minimal Impact of Information Variations:} Changes in the amount or type of information provided during the voting phase do not significantly affect task performance. 
    \item \textbf{Limited Influence of Confidence Scores and Additional Facts:} Providing agents with confidence estimates or additional factual information after the discussion phase does not substantially alter their voting decisions or final outcomes. 
    \item \textbf{Robustness of Decision Protocols:} The stability (standard deviation) of voting and judge decision protocols remain largely unaffected by alterations in the information presented during the final decision step. 
\end{enumerate}

\subsection{Challenge of Final Solution}

The final experiment investigates the confidence of agents in the solutions they reach collaboratively. After reaching a decision through discussion and decision protocols, each agent is individually challenged to reconsider their agreement with the group's answer. They must decide whether to maintain the decision or propose a new solution. This experiment aims to answer the following questions:

\textbf{How do language agents react when challenged in a multi-agent setting?}
\vspace{-0.5em}
\begin{enumerate}
    \itemsep-0.5em 
    \item How often do language agents change their decisions when challenged?
    \item Can challenging an answer improve task performance?
\end{enumerate}

The challenge of the answer takes place with five different levels of access to information. All agents are asked if they agree or disagree with the proposed solution. If they disagree, they have to generate a new solution. In the first setting, only the generated solution is presented to the agents. This is done to test whether the second setting, which provides the discussion history, has an effect on the agreement with the solution. This would show that the reasoning within the discussion can help convince the agent of the correctness of the answer. The third setting tests whether providing additional information using the ContextPlus library makes it more or less likely that the agents are convinced by the solution. The fourth and fifth approaches do not present the discussed solution, but an incorrect and irrelevant solution generated using the QWEN 2 7B model \cite{yang_qwen2_2024}. This allows to test if the model is able to find and correct obviously wrong solutions as a baseline. The answers are generated using a different model than Llama 3, to remove any bias that might occur if the same model used to detect the wrong answer also generated it \cite{wahle_are_2021}. Similarly to the last experiment, all these experiments are run only on the StrategyQA, MMLU, MMLU-Pro, and SQuAD 2.0 datasets due to limited computing resources.

\begin{figure}[H]
    \centering
    \includegraphics[width=0.9\linewidth]{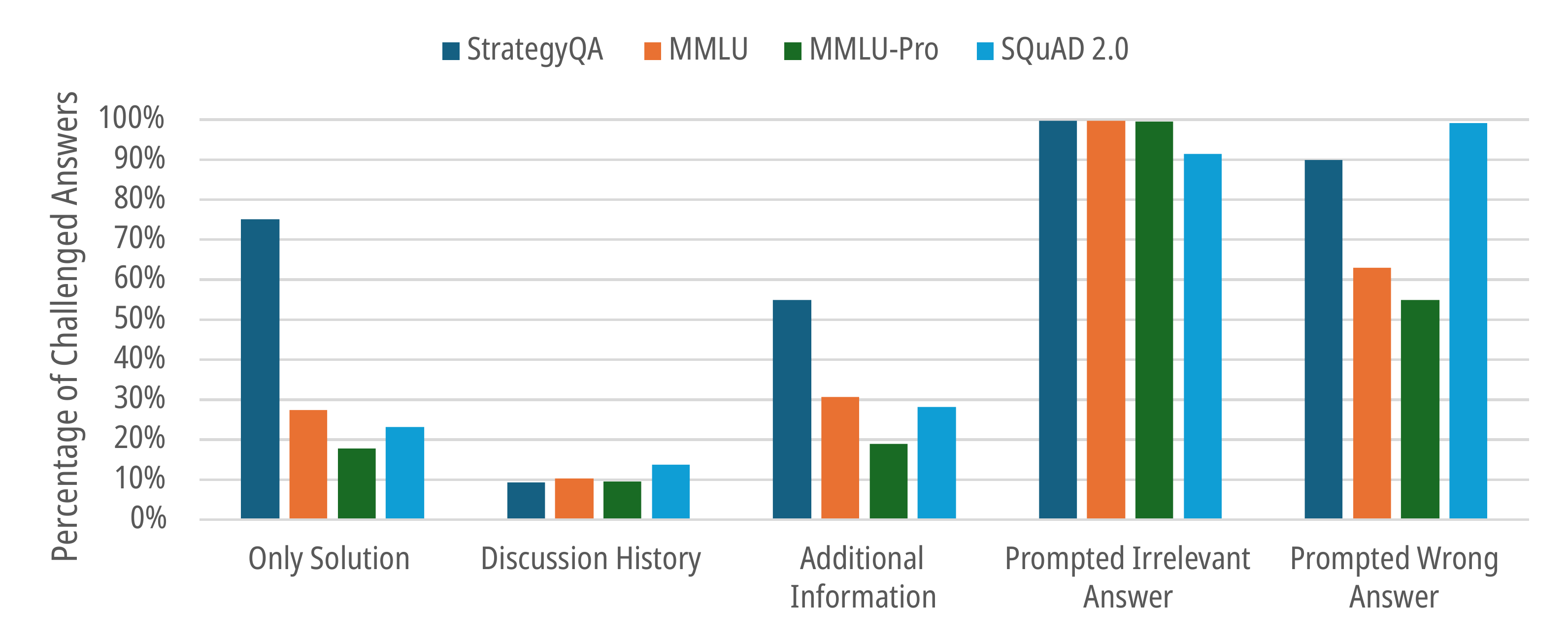}
    \caption{Percentage of agents challenging the final results after being presented with different levels of information. The experiment was performed on the StrategyQA, MMLU, MMLU-Pro and SQuAD 2.0 dataset.}
    \label{fig:challenge}
\end{figure}

\Cref{fig:challenge} shows the percentages of solutions challenged using all four data sets. If only the solution is provided, around $20\%$ of the agents challenge it. The only exception is the StrategyQA dataset, with around $70\%$ of solutions challenged. In the setting where the discussion history is provided, it can be seen that agents are more likely to agree with only around $10\%$ of the answers challenged. This shows that providing the reasoning from the discussion helps to understand the solution and makes it more likely that the agent is also agreeing with the solution. The StrategyQA dataset had the greatest change as the challenge rate decreased by $\sim60\%$. In this dataset, multi-agent discussions had the best task performance improvements as seen in the first experiment (\Cref{sec:experiment1}). Therefore, a single agent will not agree with the solution as it lacks reasoning capabilities. By providing the discussion history, the agent can understand the solution and is much more likely to agree. The same trend can be seen for all other datasets, but not as pronounced as for the StrategyQA dataset.

If an irrelevant solution generated by the QWEN 2 model is proposed as the final solution, it can be seen that the majority of agents decide to challenge the solution. If the model is deliberately prompted to generate a wrong solution, the agents only challenge around $60\%$ of the solutions because sometimes the agents are tricked into believing the wrong solution. Only the SQuAD 2.0 dataset also achieves a challenge rate of $99.0\%$, since this is a free text task, making it much easier to generate an incorrect solution. The StrategyQA dataset also has a higher challenge rate of around $90\%$, which is consistent with the previous results.

\begin{figure}[H]
    \centering
    \includegraphics[width=0.9\linewidth]{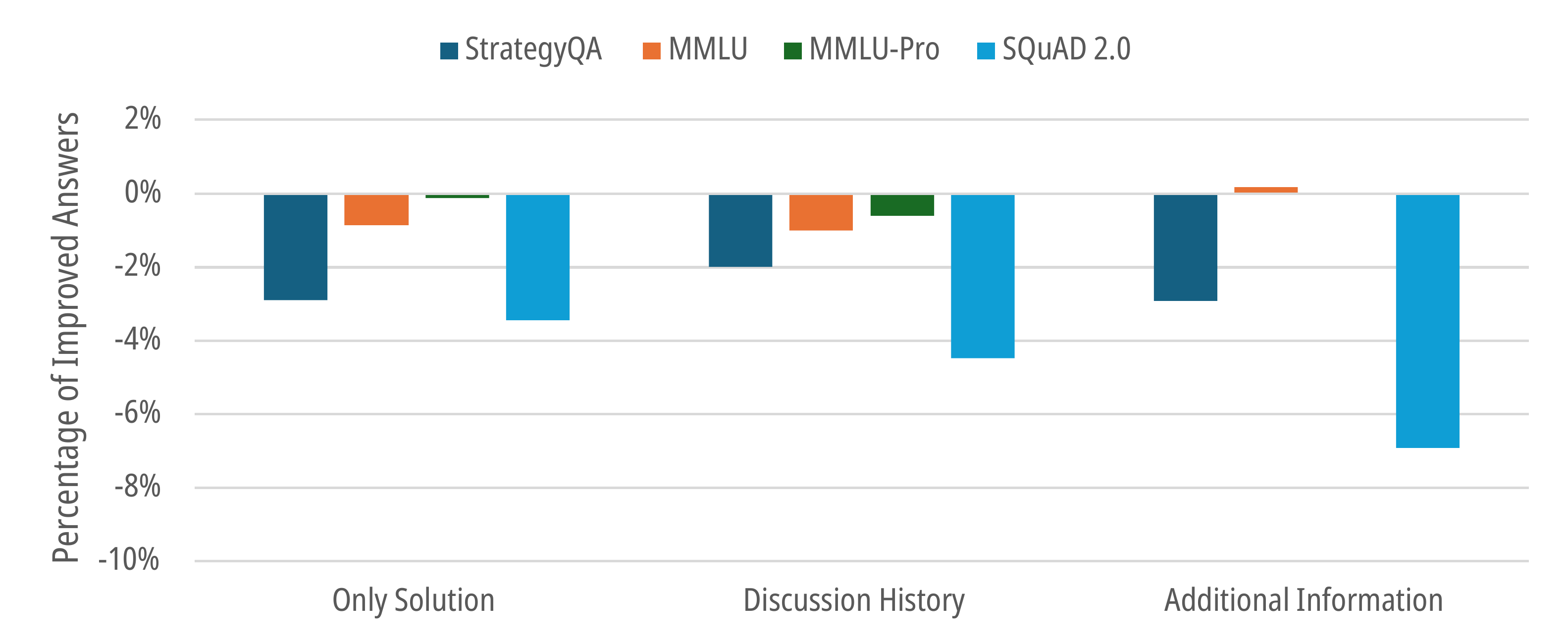}
    \caption{Percentage of answers that improve or worsen after being challenged using different levels of information. The experiment was performed on the StrategyQA, MMLU, MMLU-Pro, and SQuAD 2.0 datasets.}
    \label{fig:challenge_improve}
\end{figure}

The percentages of improved answers can be seen in \Cref{fig:challenge_improve}. The task performance changes are only presented for the first three cases in which only the answer, the discussion history, or additional information was provided since the other samples are only used as a baseline to validate the challenge and show that the agents are capable of detecting wrong solutions. On average, there are more samples that were changed to an incorrect solution than the other way around. For StrategyQA and SQuAD 2.0 the performance dropped in around $3-6\%$ of the samples. MMLU and MMLU-Pro did not have significant changes compared to the discussion result.

\subsubsection{Takeaways}
\begin{enumerate}
    \item \textbf{Impact of discussion history:} Providing agents with the discussion history significantly reduces the likelihood that they challenge the solution, as it helps them understand the reasoning behind it.
    \item \textbf{Challenging the final solution:} Agents tend to challenge their decisions more often when presented with irrelevant or incorrect solutions, especially when they lack discussion history.
    \item \textbf{Effectiveness of extra information:} Adding additional context or information does not significantly improve the accuracy of agent decisions compared to just providing the discussion history.
    \item \textbf{Performance changes after challenges:} Challenging these final solutions with a single agent leads in the majority of cases to a decrease in task performance, as it lacks the multi-agent reasoning.
\end{enumerate}
\subsection{Compute Time}

It is important to put all of these results in context by analyzing the needed compute time compared to other methods. Only when compute times are reasonable compared to the expected performance gain, multi-agent systems are feasible to exist in real-world applications. Multi-agent discussions are inherently expensive as they require many messages from a \ac{LLM}. They consist of multiple rounds with multiple agents. Although the number of agents is fixed at the beginning of the discussion, the number of rounds can scale dynamically depending on how fast the agents agree. This can result in a significant increase in the required computing power compared to the baselines, as they only require one response. All messages exchanged during the discussions also use \ac{CoT} which results in much longer responses. For this evaluation, the compute time from start to finish was recorded. This is not a very accurate measure as the multi-agent responses could be stuck in a queue as many discussions are processed simultaneously. This cannot happen for the baseline as they only require one response. Therefore, it is necessary to exercise caution when interpreting these values.

\bigskip
\begin{figure}[H]
    \centering
    \includegraphics[width=\linewidth]{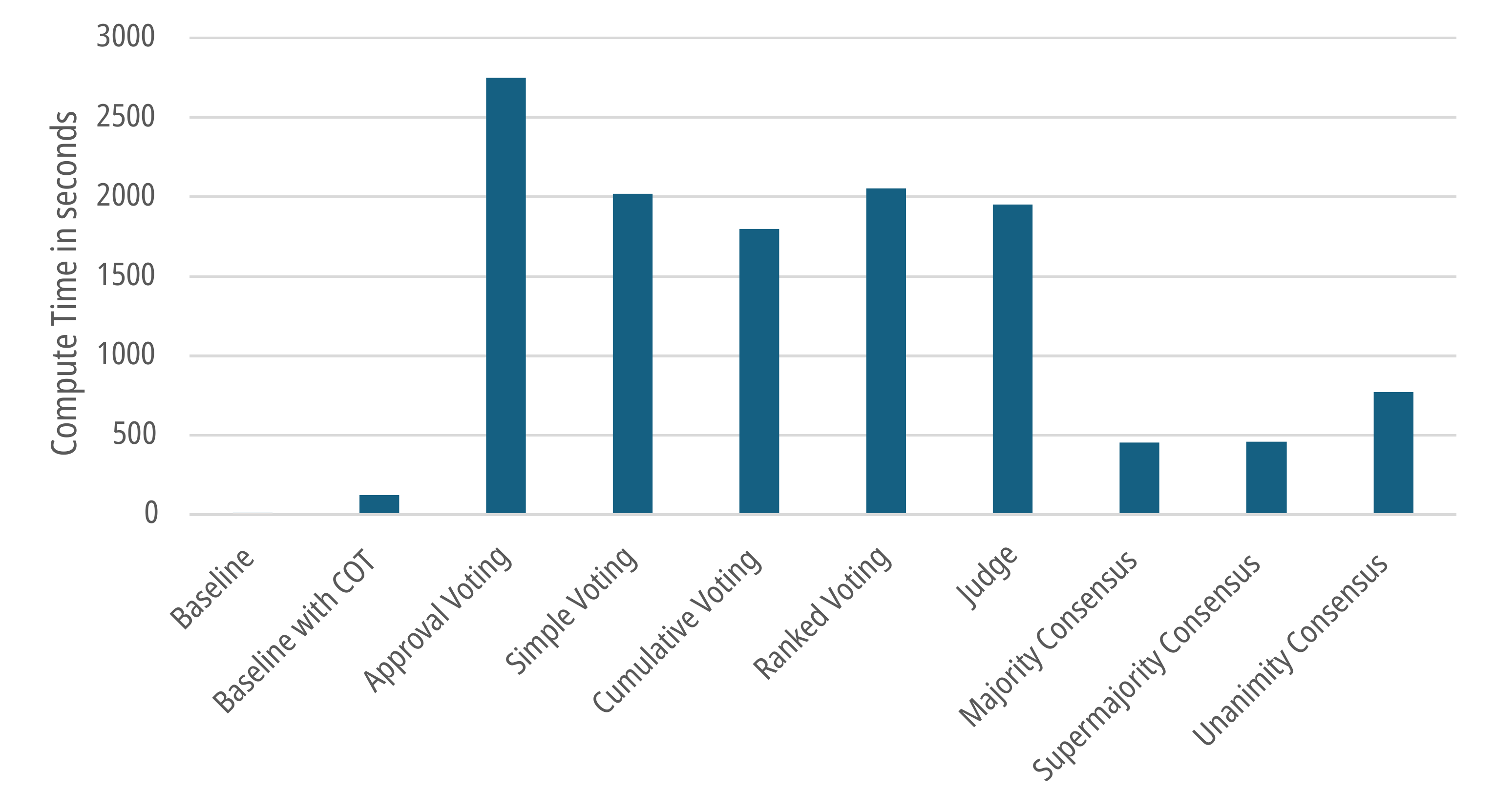}
    \caption{Compute time for all decision protocols averaged over all datasets for the Llama 3 8B model. Hardware is described in \Cref{sec:setup}.}
    \label{fig:compute_time}
\end{figure}

\Cref{fig:compute_time} shows the computation time for each method averaged on all datasets for the Llama 3 8B model. There is already approximately ten times the computation cost for the baseline with \ac{CoT} compared to the single-agent baseline. The voting and judge decision protocols are the most expensive, as they run on average for more discussion rounds and require an extra decision step. The approval voting protocol is even more expensive, as it needs even more rounds to come to a final decision. These protocols are approximately 15 times more expensive than the baseline with \ac{CoT}. The consensus-based decision protocols are much cheaper as they only require one to two rounds to reach a decision. This corresponds to a five-fold increase in computation time.

\subsubsection{Takeaways}
\begin{enumerate}
\item \textbf{High Compute Cost of Multi-Agent Discussions:} Multi-agent discussions require significantly more computation due to multiple rounds and long responses.
\item \textbf{Costly Voting Protocols:} Voting-based decision protocols are much more computationally expensive compared to consensus-based ones, requiring more rounds and additional decision steps.
\end{enumerate}

\newpage
\section{Epilogue}
\subsection{Conclusion}
In this work, I propose a collection of decision-making protocols for \ac{MAS}. The efficiency of these decision-making protocols is analyzed on seven different datasets from different domains, with a smaller Llama 3 8B model and a larger Llama 3 70B model. All of these experiments are performed using the framework \ac{MALLM}, which was developed as part of this work and is fully open-source and available to other researchers. It is capable of simulating multi-agent discussions for conversational task-solving using a wide array of discussion protocols, response generators, and decision protocols. To conclude the results of this study, in this section, I revisit the research questions that were initially proposed.

\paragraph{How do decision-making mechanisms influence LLM performance in conversational
tasks?}
Overall, it can be seen that the decision protocols help to outperform the baselines. The consensus decision protocols perform better on knowledge-based tasks, while the judge and voting decision protocols are stronger on logic-based tasks. The size of the model has a strong impact on the improvement gained from the multi-agent discussion. For the smaller Llama 3 8B each baseline for a dataset is exceeded by a multi-agent discussion with a decision protocol. This changes for the larger model, as it was not able to beat the baseline in some cases. It is also observed that some decision protocols struggle to reach a final decision because the voting step has a high chance of being tied. Therefore, more restrictive voting protocols are superior, as they help to reach a decision.

Additional experiments show that limiting the number of rounds the agents have to discuss before voting for a final decision can bring some improvements, as this decreases the chance that the agents drift away from the main topic. By increasing the number of agents that collaborate on the task, it could be seen that the task performance also increased. This is probably due to a wider range of expertise.

\paragraph{How does the diversity of responses affect the performance of decision protocols?}

This experiment tests multiple methods to improve the diversity of responses. By forcing each agent to start with its own solution and only iterate after that, an increase in task performance is observed. An even greater improvement can be achieved by using the collective refinement discussion protocol, which limits group influences and promotes independent thinking. These two methods show a decrease in the similarity of the final responses. Forcing agents to share only reasoning steps or answer hypercritically by changing the response generator did not have a positive impact on task performance.

\paragraph{How do small variations in information provided during the voting phase impact
decision protocols in multi-agent LLM systems?}

The alterations to the voting step have minimal impact on task performance and often introduce a lot of computational overhead. There is no clear trend as to which alteration performs best. The task performances for all decision protocols are within one standard deviation of each other. Therefore, these small variations did not yield any significant results.

\paragraph{How do language agents react when challenged in a multi-agent setting?}

The challenge experiment demonstrates that the confidence of a language agent in their collaborative solution is significantly influenced by the information available during the decision-making process. When agents only receive the final solution, they are more likely to challenge it. In contrast, when agents have access to the discussion history, they are less likely to challenge the consensus, as the reasoning behind the decision helps build confidence in the solution. However, when presented with obviously incorrect or irrelevant solutions, agents consistently challenged the proposed answer, with challenge rates close to $100\%$. This highlights the agent's ability to recognize and correct poor solutions. Despite this, challenging the final decision in normal cases, where the solution was not deliberately wrong, resulted in small performance decreases.

\subsection{Future Work}
\label{sec:future_work}
An interesting future direction for this project is opening up with the emergence of newer and more powerful models, especially in the area of reasoning skills. Models such as the OpenAI o1 preview\footnote{\href{https://openai.com/index/introducing-openai-o1-preview/}{openai.com/index/introducing-openai-o1-preview}} or the QwQ-32B preview model\cite{team_qwq_2024}, whose weights are even publicly available, allow the model to use better reasoning because they are pre-trained on many successful \ac{CoT} examples. This could improve the voting steps, as they require more reasoning to determine which answers are right and which are wrong. The judge decision protocol could benefit from these capabilities for the same reasons. Overall, it would be interesting to compare the multi-agent discussion of these newer reasoning pre-trained models with older generations of models with a similar number of parameters, as this may provide insight into whether these newer models achieve a higher quality of discussion and, therefore, a higher difference to their baseline task performance.

Applying the lessons learned from this study to improve the performance of decision protocols is another interesting direction. Especially with the voting and judge-based decision protocols, there is room for improvement. This could be done by allowing the agents to start voting earlier or create a pre-vote to decide whether the majority of agents believe that they are ready for a decision. This would still allow longer conversations for more complex tasks and reduce computing costs for the other cases. The compute cost could be further decreased by also limiting the answer length of the agents. Overall, it can be observed that these agents tend to give long answers, which can pollute the context of the other agents. By prompting the agents to repeat less information and answer more concisely, the efficiency could be increased. It needs to be tested whether this leads to an improvement or decrease in task performance.

\subsection{Limitations}
Due to the complex nature of multi-agent discussions and the high compute cost, it is not possible to test all possible combinations of parameters. The default parameters and configurations for the experiments can be found in \Cref{sec:mallm_setup_app}. By evaluating the decision protocols across a variety of datasets from different domains, it was possible to determine whether these findings were universally applicable or limited to specific tasks. This wide range of datasets further increased the parameter space. Therefore, dataset sampling and calculating the standard deviation on three different runs was used to estimate task performance as described in \Cref{sec:datasets}. 

The calculation of compute time is not very accurate, as it also includes the wait time for the API. The current results can only be seen as an approximation of the actual values. But as these are calculated over many samples with all decision protocols using the same API and similar wait times, it's still comparable. Additionally, there are still some possible improvements that can be applied to the voting and judge-based decision protocols as mentioned in \Cref{sec:future_work}. This could close the gap in compute cost difference between these decision protocols and the consensus-based decision protocols.

\newpage
\printbibliography
\newpage
\pagenumbering{roman}

\newpage
\appendix
\addtocontents{toc}{\protect\setcounter{tocdepth}{1}}%
\section{Results}
Additional results for all experiments to provide further information.
\subsection{Alterations in Decision Protocols}
\label{sec:appendix_decision_alterations}

\begin{figure}[H]
    \centering
    \includegraphics[width=\linewidth]{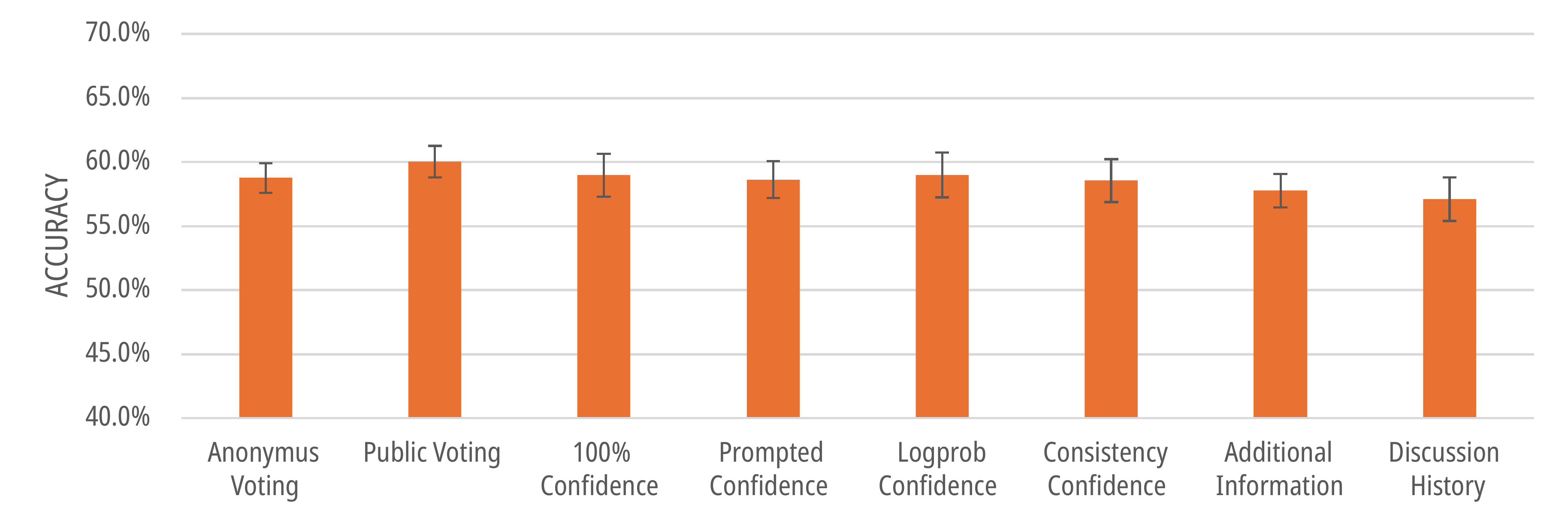}
    \caption{Mean task performance on the SQuAD 2.0 dataset for all voting and judge decision protocols with all variations.}
    \label{fig:alt-squad-mean}
\end{figure}

\begin{figure}[H]
    \centering
    \includegraphics[width=\linewidth]{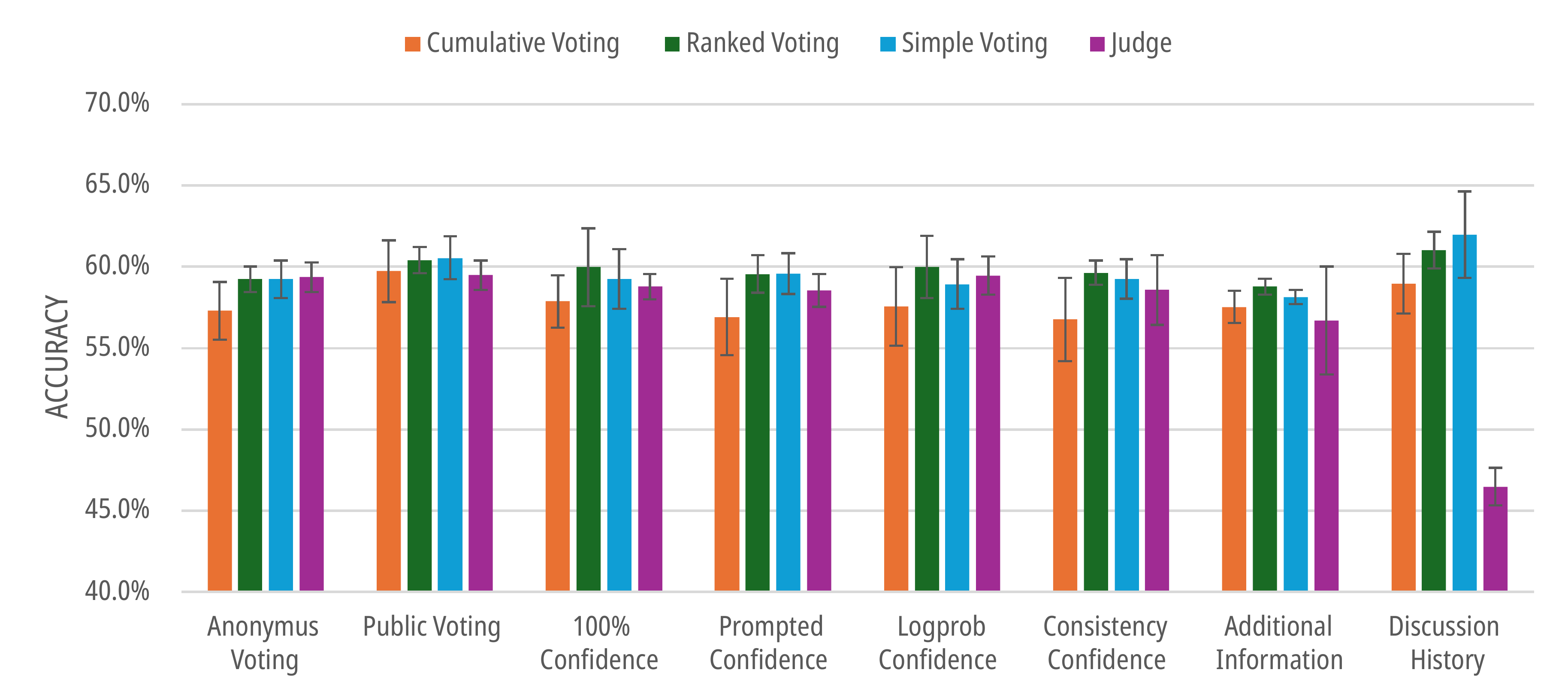}
    \caption{Task performance on the SQuAD 2.0 dataset for all voting and judge decision protocols with all variations.}
    \label{fig:alt-squad-all}
\end{figure}

\begin{figure}[H]
    \centering
    \includegraphics[width=\linewidth]{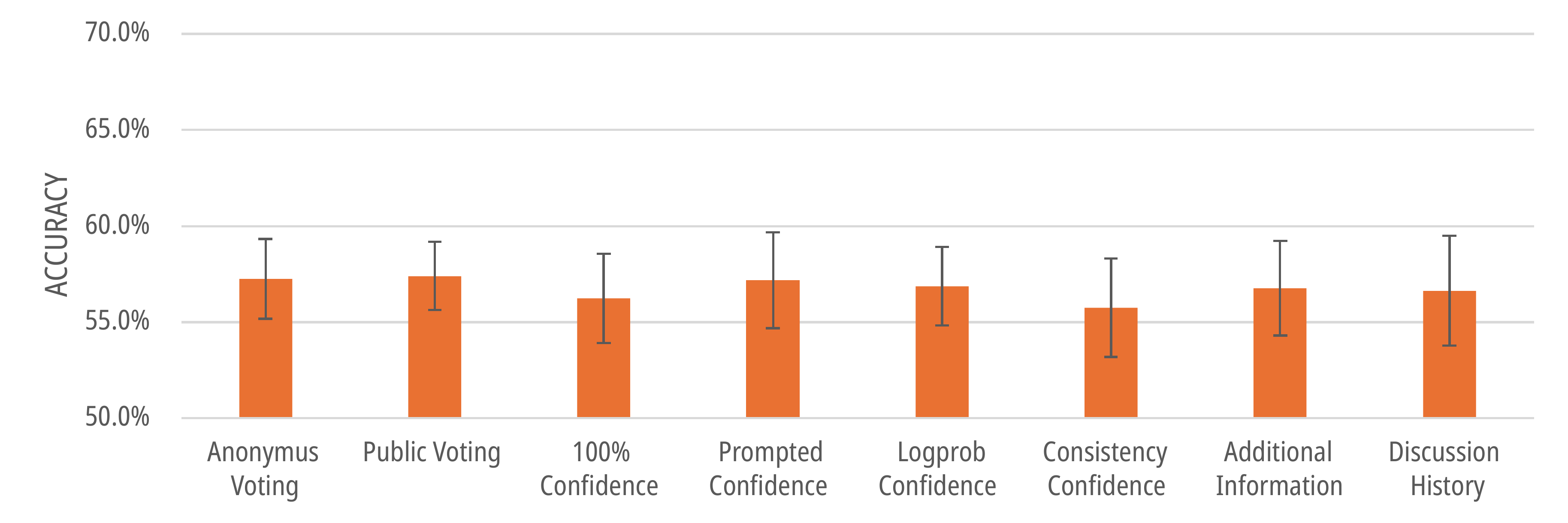}
    \caption{Mean task performance on the MMLU dataset for all voting and judge decision protocols with all variations.}
    \label{fig:alt-mmlu-mean}
\end{figure}

\begin{figure}[H]
    \centering
    \includegraphics[width=\linewidth]{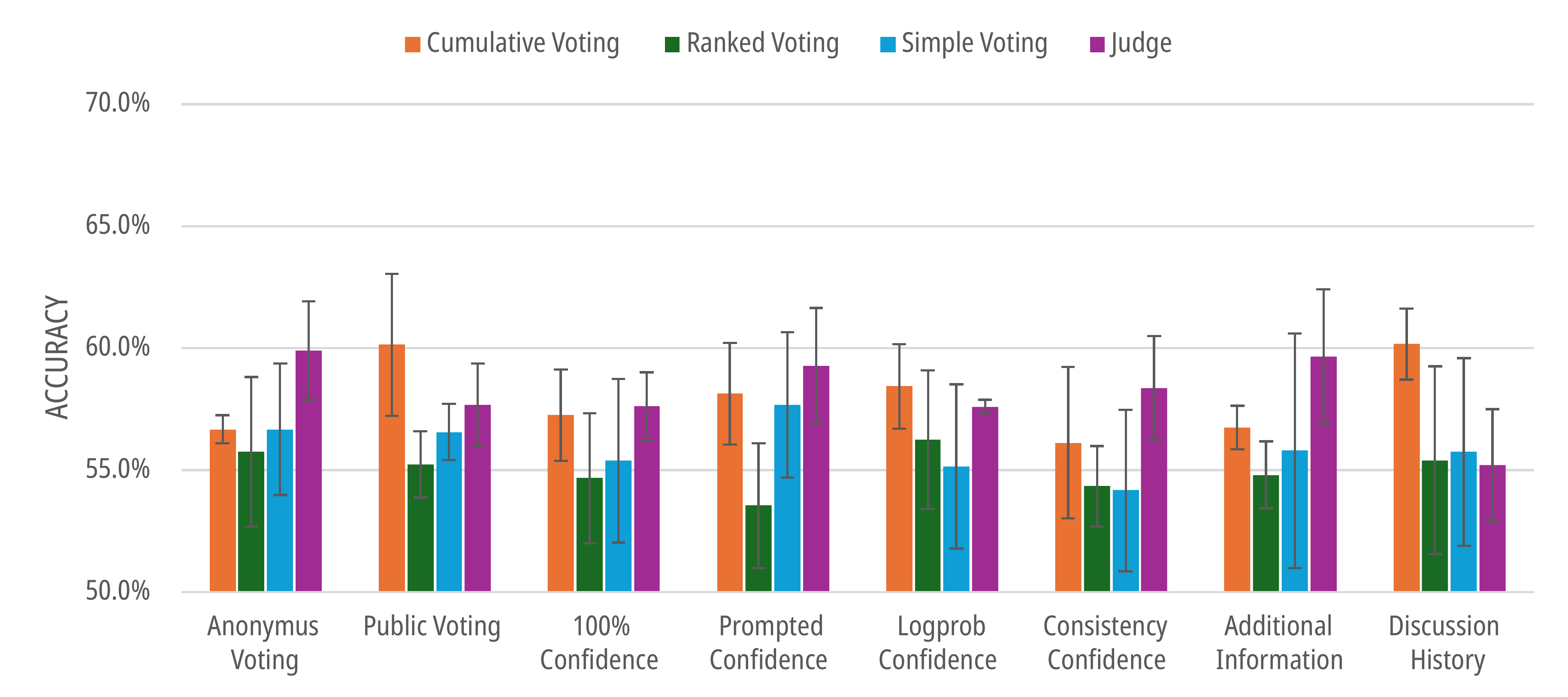}
    \caption{Task performance on the MMLU dataset for all voting and judge decision protocols with all variations.}
    \label{fig:alt-mmlu-all}
\end{figure}

\begin{figure}[H]
    \centering
    \includegraphics[width=\linewidth]{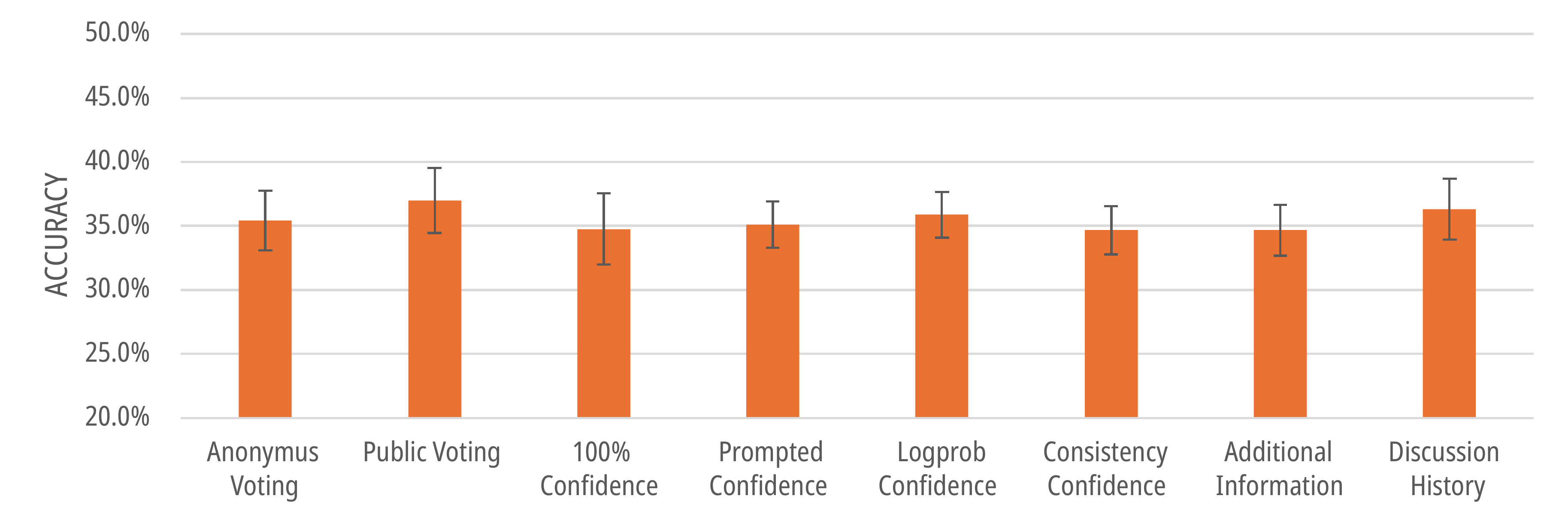}
    \caption{Mean task performance on the MMLU-Pro dataset for all voting and judge decision protocols with all variations.}
    \label{fig:alt-mmlu-pro-mean}
\end{figure}

\begin{figure}[H]
    \centering
    \includegraphics[width=\linewidth]{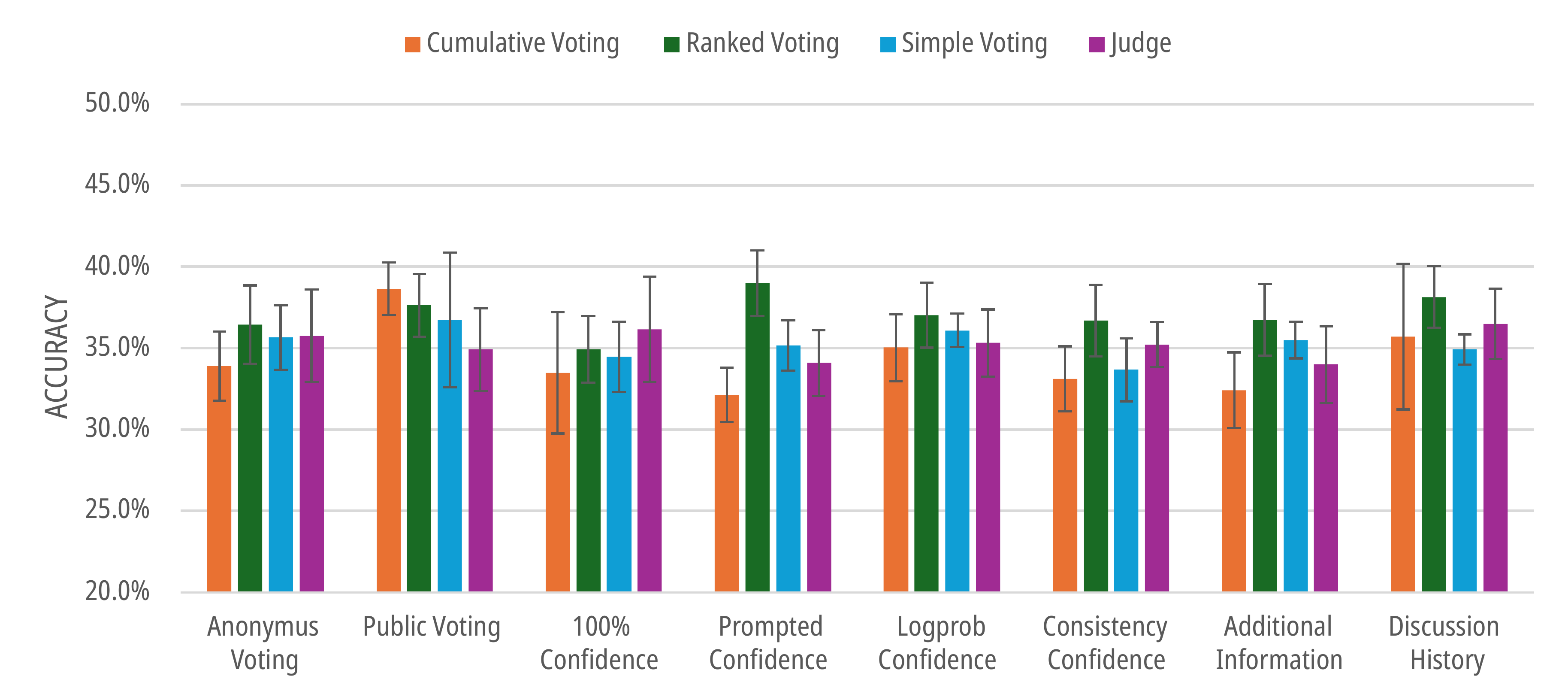}
    \caption{Task performance on the MMLU-Pro dataset for all voting and judge decision protocols with all variations.}
    \label{fig:alt-mmlu-pro-all}
\end{figure}
\newpage
\subsection{Turns until Decision}
\label{sec:appendix_turns_until_consensus}
All tables for how many discussion rounds are needed to reach a decision for each dataset with a Llama 3 8B and 70B model. The voting and judge decision protocols are forced to discuss for three rounds, as \textcite{du_improving_2023} showed that this yields good results.  
\subsubsection{8B Model}
\bigskip
\begin{table}[H]
\centering
\begin{tabular}{l|ccccc|c}
\toprule
\textbf{Group} & \textbf{Turn 1} & \textbf{Turn 2} & \textbf{Turn 3} & \textbf{Turn 4} & \textbf{Turn 5} & \textbf{Task Performance Score} \\
\midrule
Voting & 0.00\% & 0.00\% & 98.17\% & 1.83\% & 0.00\% & 30.5 \tiny{$\pm$ 0.9} \\
Cumulative & 0.00\% & 0.00\% & 94.33\% & 5.50\% & 0.17\% & 31.3 \tiny{$\pm$ 2.8} \\
Ranked & 0.00\% & 0.00\% & 88.00\% & 10.67\% & 1.33\% & 27.3 \tiny{$\pm$ 3.9} \\
Approval & 0.00\% & 0.00\% & 36.67\% & 20.50\% & 42.83\% & 31.3 \tiny{$\pm$ 2.6} \\
\midrule
Judge & 0.00\% & 0.00\% & 100.00\% & 0.00\% & 0.00\% & 27.6 \tiny{$\pm$ 2.3} \\
\midrule
Majority & 78.00\% & 16.00\% & 4.50\% & 0.50\% & 1.00\% & 32.3 \tiny{$\pm$ 2.9} \\
Supermaj. & 77.83\% & 17.17\% & 3.83\% & 1.00\% & 0.17\% & 30.7 \tiny{$\pm$ 2.1} \\
Unanimity & 61.50\% & 20.50\% & 11.50\% & 3.00\% & 3.50\% & 30.0 \tiny{$\pm$ 2.3} \\
\bottomrule
\end{tabular}
\caption{Termination Percentage by Turn for Each Decision Protocol GPQA}
\label{tab:termination_percentages_gpqa}
\end{table}

\bigskip

\begin{table}[H]
\centering
\begin{tabular}{l|ccccc|c}
\toprule
\textbf{Group} & \textbf{Turn 1} & \textbf{Turn 2} & \textbf{Turn 3} & \textbf{Turn 4} & \textbf{Turn 5} & \textbf{Task Performance Score} \\
\midrule
Voting & 0.00\% & 0.00\% & 99.33\% & 0.50\% & 0.17\% & 53.3 \tiny{$\pm$ 1.8} \\
Cumulative & 0.00\% & 0.00\% & 94.00\% & 5.50\% & 0.50\% & 52.6 \tiny{$\pm$ 4.0} \\
Ranked & 0.00\% & 0.00\% & 91.17\% & 7.83\% & 1.00\% & 49.2 \tiny{$\pm$ 1.5} \\
Approval & 0.00\% & 0.00\% & 26.67\% & 14.33\% & 59.00\% & 43.0 \tiny{$\pm$ 2.1} \\
\midrule
Judge & 0.00\% & 0.00\% & 100.00\% & 0.00\% & 0.00\% & 53.7 \tiny{$\pm$ 4.7} \\
\midrule
Majority & 80.00\% & 13.67\% & 4.83\% & 1.00\% & 0.50\% & 53.2 \tiny{$\pm$ 2.5} \\
Supermaj. & 79.33\% & 14.33\% & 4.83\% & 1.00\% & 0.50\% & 54.6 \tiny{$\pm$ 3.6} \\
Unanimity & 59.50\% & 21.67\% & 12.67\% & 3.50\% & 2.67\% & 54.2 \tiny{$\pm$ 1.0} \\
\bottomrule
\end{tabular}
\caption{Termination Percentage by Turn for Each Decision Protocol MMLU}
\label{tab:termination_percentages_mmlu}
\end{table}

\bigskip

\begin{table}[H]
\centering
\begin{tabular}{l|ccccc|c}
\toprule
\textbf{Group} & \textbf{Turn 1} & \textbf{Turn 2} & \textbf{Turn 3} & \textbf{Turn 4} & \textbf{Turn 5} & \textbf{Task Performance Score} \\
\midrule
Voting & 0.00\% & 0.00\% & 97.33\% & 2.67\% & 0.00\% & 32.0 \tiny{$\pm$ 2.7} \\
Cumulative & 0.00\% & 0.00\% & 95.17\% & 4.67\% & 0.17\% & 28.3 \tiny{$\pm$ 3.1} \\
Ranked & 0.00\% & 0.00\% & 86.67\% & 10.83\% & 2.50\% & 33.1 \tiny{$\pm$ 4.6} \\
Approval & 0.00\% & 0.00\% & 32.33\% & 19.00\% & 48.67\% & 29.2 \tiny{$\pm$ 5.2} \\
\midrule
Judge & 0.00\% & 0.00\% & 100.00\% & 0.00\% & 0.00\% & 33.5 \tiny{$\pm$ 0.8} \\
\midrule
Majority & 79.83\% & 15.83\% & 3.00\% & 0.50\% & 0.83\% & 36.4 \tiny{$\pm$ 2.1} \\
Supermaj. & 76.33\% & 17.33\% & 4.83\% & 1.00\% & 0.50\% & 35.2 \tiny{$\pm$ 3.0} \\
Unanimity & 59.67\% & 23.00\% & 11.17\% & 3.33\% & 2.83\% & 36.3 \tiny{$\pm$ 0.4} \\
\bottomrule
\end{tabular}
\caption{Termination Percentage by Turn for Each Decision Protocol MMLU-Pro}
\label{tab:termination_percentages_mmlu-pro}
\end{table}

\bigskip

\begin{table}[H]
\centering
\begin{tabular}{l|ccccc|c}
\toprule
\textbf{Group} & \textbf{Turn 1} & \textbf{Turn 2} & \textbf{Turn 3} & \textbf{Turn 4} & \textbf{Turn 5} & \textbf{Task Performance Score} \\
\midrule
Voting & 0.00\% & 0.00\% & 94.33\% & 5.50\% & 0.17\% & 58.5 \tiny{$\pm$ 0.9} \\
Cumulative & 0.00\% & 0.00\% & 94.00\% & 6.00\% & 0.00\% & 61.2 \tiny{$\pm$ 1.6} \\
Ranked & 0.00\% & 0.00\% & 92.00\% & 7.50\% & 0.50\% & 56.2 \tiny{$\pm$ 3.4} \\
Approval & 0.00\% & 0.00\% & 22.50\% & 11.00\% & 66.50\% & 58.7 \tiny{$\pm$ 0.4} \\
\midrule
Judge & 0.00\% & 0.00\% & 100.00\% & 0.00\% & 0.00\% & 53.7 \tiny{$\pm$ 2.0} \\
\midrule
Majority & 61.17\% & 30.00\% & 6.50\% & 1.67\% & 0.67\% & 59.9 \tiny{$\pm$ 0.1} \\
Supermaj. & 57.00\% & 33.50\% & 7.17\% & 1.50\% & 0.83\% & 56.4 \tiny{$\pm$ 2.1} \\
Unanimity & 30.33\% & 42.83\% & 18.67\% & 6.00\% & 2.17\% & 58.8 \tiny{$\pm$ 2.6} \\
\bottomrule
\end{tabular}
\caption{Termination Percentage by Turn for Each Decision Protocol StrategyQA}
\label{tab:termination_percentages_strategy}
\end{table}

\bigskip

\begin{table}[H]
\centering
\begin{tabular}{l|ccccc|c}
\toprule
\textbf{Group} & \textbf{Turn 1} & \textbf{Turn 2} & \textbf{Turn 3} & \textbf{Turn 4} & \textbf{Turn 5} & \textbf{Task Performance Score} \\
\midrule
Voting & 0.00\% & 0.00\% & 99.67\% & 0.33\% & 0.00\% & 55.2 \tiny{$\pm$ 1.5} \\
Cumulative & 0.00\% & 0.00\% & 94.88\% & 5.12\% & 0.00\% & 56.8 \tiny{$\pm$ 4.2} \\
Ranked & 0.00\% & 0.00\% & 89.33\% & 9.00\% & 1.67\% & 52.5 \tiny{$\pm$ 0.0} \\
Approval & 0.00\% & 0.00\% & 28.50\% & 18.00\% & 53.50\% & 50.9 \tiny{$\pm$ 4.0} \\
\midrule
Judge & 0.00\% & 0.00\% & 100.00\% & 0.00\% & 0.00\% & 59.3 \tiny{$\pm$ 1.0} \\
\midrule
Majority & 71.33\% & 20.33\% & 5.67\% & 1.83\% & 0.83\% & 27.8 \tiny{$\pm$ 2.5} \\
Supermaj. & 68.33\% & 22.67\% & 6.50\% & 1.33\% & 1.17\% & 29.3 \tiny{$\pm$ 2.6} \\
Unanimity & 38.83\% & 31.83\% & 19.17\% & 6.00\% & 4.17\% & 28.2 \tiny{$\pm$ 2.8} \\
\bottomrule
\end{tabular}
\caption{Termination Percentage by Turn for Each Decision Protocol MuSR}
\label{tab:termination_percentages_musr}
\end{table}

\bigskip

\begin{table}[H]
\centering
\begin{tabular}{l|ccccc|c}
\toprule
\textbf{Group} & \textbf{Turn 1} & \textbf{Turn 2} & \textbf{Turn 3} & \textbf{Turn 4} & \textbf{Turn 5} & \textbf{Task Performance Score} \\
\midrule
Voting & 0.00\% & 0.00\% & 99.17\% & 0.83\% & 0.00\% & 56.2 \tiny{$\pm$ 0.5} \\
Cumulative & 0.00\% & 0.00\% & 96.33\% & 3.50\% & 0.17\% & 55.8 \tiny{$\pm$ 3.4} \\
Ranked & 0.00\% & 0.00\% & 93.00\% & 6.17\% & 0.83\% & 58.0 \tiny{$\pm$ 0.8} \\
Approval & 0.00\% & 0.00\% & 32.50\% & 13.67\% & 53.83\% & 46.5 \tiny{$\pm$ 1.4} \\
\midrule
Judge & 0.00\% & 0.00\% & 100.00\% & 0.00\% & 0.00\% & 57.2 \tiny{$\pm$ 0.6} \\
\midrule
Majority & 81.83\% & 11.33\% & 3.67\% & 1.67\% & 1.50\% & 43.1 \tiny{$\pm$ 2.1} \\
Supermaj. & 82.00\% & 10.33\% & 4.67\% & 1.33\% & 1.67\% & 44.4 \tiny{$\pm$ 0.4} \\
Unanimity & 66.83\% & 16.00\% & 8.83\% & 4.17\% & 4.17\% & 43.4 \tiny{$\pm$ 2.0} \\
\bottomrule
\end{tabular}
\caption{Termination Percentage by Turn for Each Decision Protocol SQuAD 2.0}
\label{tab:termination_percentages_squad}
\end{table}

\bigskip

\begin{table}[H]
\centering
\begin{tabular}{l|ccccc|c}
\toprule
\textbf{Group} & \textbf{Turn 1} & \textbf{Turn 2} & \textbf{Turn 3} & \textbf{Turn 4} & \textbf{Turn 5} & \textbf{Task Performance Score} \\
\midrule
Voting & 0.00\% & 0.00\% & 96.50\% & 3.50\% & 0.00\% & 9.5 \tiny{$\pm$ 1.7} \\
Cumulative & 0.00\% & 0.00\% & 95.17\% & 4.50\% & 0.33\% & 9.0 \tiny{$\pm$ 1.5} \\
Ranked & 0.00\% & 0.00\% & 89.50\% & 9.33\% & 1.17\% & 6.8 \tiny{$\pm$ 1.3} \\
Approval & 0.00\% & 0.00\% & 59.83\% & 23.83\% & 16.33\% & 7.8 \tiny{$\pm$ 2.6} \\
\midrule
Judge & 0.00\% & 0.00\% & 100.00\% & 0.00\% & 0.00\% & 9.2 \tiny{$\pm$ 3.0} \\
\midrule
Majority & 62.83\% & 29.33\% & 4.67\% & 1.67\% & 1.50\% & 9.2 \tiny{$\pm$ 3.0} \\
Supermaj. & 64.67\% & 27.67\% & 5.50\% & 0.83\% & 1.33\% & 9.2 \tiny{$\pm$ 0.8} \\
Unanimity & 42.33\% & 34.83\% & 15.33\% & 3.83\% & 3.67\% & 10.8 \tiny{$\pm$ 1.6} \\
\bottomrule
\end{tabular}
\caption{Termination Percentage by Turn for Each Decision Protocol Math lvl 5}
\label{tab:termination_percentages_math}
\end{table}

\subsubsection{70B Model}
\bigskip

\begin{table}[H]
\centering
\begin{tabular}{l|ccccc|c}
\toprule
\textbf{Group} & \textbf{Turn 1} & \textbf{Turn 2} & \textbf{Turn 3} & \textbf{Turn 4} & \textbf{Turn 5} & \textbf{Task Performance Score} \\
\midrule
Voting & 0.00\% & 0.00\% & 99.21\% & 0.73\% & 0.06\% & 45.7 \tiny{$\pm$ 0.3} \\
Cumulative & 0.00\% & 0.00\% & 94.33\% & 5.33\% & 0.33\% & 43.9 \tiny{$\pm$ 3.1} \\
Ranked & 0\% & 0\% & 89.71\% & 8.11\% & 2.18\% &  44.2 \tiny{$\pm$ 2.1}\\
Approval & 0.00\% & 0.00\% & 7.35\% & 3.71\% & 88.94\% & 33.0 \tiny{$\pm$ 3.4} \\
\midrule
Judge & 0.00\% & 0.00\% & 100.00\% & 0.00\% & 0.00\% & 42.9 \tiny{$\pm$ 5.1} \\
\midrule
Majority & 62.33\% & 31.67\% & 5.67\% & 0.33\% & 0.00\% & 43.7 \tiny{$\pm$ 1.0} \\
Supermaj. & 66.17\% & 28.50\% & 4.67\% & 0.67\% & 0.00\% & 46.6 \tiny{$\pm$ 0.8} \\
Unanimity & 34.33\% & 45.83\% & 15.67\% & 3.33\% & 0.83\% & 45.3 \tiny{$\pm$ 2.5} \\
\bottomrule
\end{tabular}
\caption{Termination Percentage by Turn for Each Decision Protocol GPQA}
\label{tab:termination_percentages_gpqa_70}
\end{table}

\bigskip

\begin{table}[H]
\centering
\begin{tabular}{l|ccccc|c}
\toprule
\textbf{Group} & \textbf{Turn 1} & \textbf{Turn 2} & \textbf{Turn 3} & \textbf{Turn 4} & \textbf{Turn 5} & \textbf{Task Performance Score} \\
\midrule
Voting & 0.00\% & 0.00\% & 98.50\% & 1.50\% & 0.00\% & 75.5 \tiny{$\pm$ 1.3} \\
Cumulative & 0.00\% & 0.00\% & 95.50\% & 4.17\% & 0.33\% & 72.5 \tiny{$\pm$ 1.9} \\
Ranked & 0.00\% & 0.00\% & 98.00\% & 1.50\% & 0.50\% & 73.3 \tiny{$\pm$ 1.9} \\
Approval & 0.00\% & 0.00\% & 7.83\% & 3.00\% & 89.17\% & 50.2 \tiny{$\pm$ 2.1} \\
\midrule
Judge & 0.00\% & 0.00\% & 100.00\% & 0.00\% & 0.00\% & 72.2 \tiny{$\pm$ 1.4} \\
\midrule
Majority & 87.67\% & 10.33\% & 1.67\% & 0.00\% & 0.33\% & 74.0 \tiny{$\pm$ 1.4} \\
Supermaj. & 87.17\% & 11.17\% & 1.67\% & 0.00\% & 0.00\% & 71.9 \tiny{$\pm$ 1.2} \\
Unanimity & 67.50\% & 23.17\% & 7.00\% & 1.33\% & 1.00\% & 72.2 \tiny{$\pm$ 2.4} \\
\bottomrule
\end{tabular}
\caption{Termination Percentage by Turn for Each Decision Protocol MMLU}
\label{tab:termination_percentages_mmlu_70}
\end{table}

\bigskip

\begin{table}[H]
\centering
\begin{tabular}{l|ccccc|c}
\toprule
\textbf{Group} & \textbf{Turn 1} & \textbf{Turn 2} & \textbf{Turn 3} & \textbf{Turn 4} & \textbf{Turn 5} & \textbf{Task Performance Score} \\
\midrule
Voting & 0.00\% & 0.00\% & 98.83\% & 1.00\% & 0.17\% & 56.5 \tiny{$\pm$ 5.4} \\
Cumulative & 0.00\% & 0.00\% & 92.67\% & 6.17\% & 1.17\% & 53.0 \tiny{$\pm$ 0.4} \\
Ranked & 0.00\% & 0.00\% & 97.00\% & 2.83\% & 0.17\% & 54.3 \tiny{$\pm$ 0.9} \\
Approval & 0.00\% & 0.00\% & 12.00\% & 5.00\% & 83.00\% & 36.3 \tiny{$\pm$ 3.7} \\
\midrule
Judge & 0.00\% & 0.00\% & 100.00\% & 0.00\% & 0.00\% & 58.0 \tiny{$\pm$ 3.9} \\
\midrule
Majority & 87.83\% & 10.83\% & 0.67\% & 0.33\% & 0.33\% & 57.3 \tiny{$\pm$ 3.0} \\
Supermaj. & 85.83\% & 12.00\% & 1.50\% & 0.33\% & 0.33\% & 57.0 \tiny{$\pm$ 1.7} \\
Unanimity & 62.50\% & 28.67\% & 6.83\% & 1.00\% & 1.00\% & 57.3 \tiny{$\pm$ 1.5} \\
\bottomrule
\end{tabular}
\caption{Termination Percentage by Turn for Each Decision Protocol MMLU-Pro}
\label{tab:termination_percentages_mmlu_pro_70}
\end{table}

\bigskip

\begin{table}[H]
\centering
\begin{tabular}{l|ccccc|c}
\toprule
\textbf{Group} & \textbf{Turn 1} & \textbf{Turn 2} & \textbf{Turn 3} & \textbf{Turn 4} & \textbf{Turn 5} & \textbf{Task Performance Score} \\
\midrule
Voting & 0.00\% & 0.00\% & 99.17\% & 0.83\% & 0.00\% & 81.2 \tiny{$\pm$ 1.4} \\
Cumulative & 0.00\% & 0.00\% & 92.17\% & 3.33\% & 4.50\% & 80.0 \tiny{$\pm$ 0.8} \\
Ranked & 0.00\% & 0.00\% & 98.50\% & 1.17\% & 0.33\% & 80.3 \tiny{$\pm$ 1.0} \\
Approval & 0.00\% & 0.00\% & 12.33\% & 1.83\% & 85.83\% & 46.4 \tiny{$\pm$ 13.9} \\
\midrule
Judge & 0.00\% & 0.00\% & 100.00\% & 0.00\% & 0.00\% & 82.9 \tiny{$\pm$ 1.3} \\
\midrule
Majority & 89.17\% & 9.17\% & 1.50\% & 0.17\% & 0.00\% & 80.1 \tiny{$\pm$ 0.3} \\
Supermaj. & 89.83\% & 8.17\% & 1.67\% & 0.17\% & 0.17\% & 80.3 \tiny{$\pm$ 1.3} \\
Unanimity & 73.33\% & 18.17\% & 6.00\% & 1.50\% & 1.00\% & 78.1 \tiny{$\pm$ 2.3} \\
\bottomrule
\end{tabular}
\caption{Termination Percentage by Turn for Each Decision Protocol StrategyQA}
\label{tab:termination_percentages_strategy_70}
\end{table}

\bigskip

\begin{table}[H]
\centering
\begin{tabular}{l|ccccc|c}
\toprule
\textbf{Group} & \textbf{Turn 1} & \textbf{Turn 2} & \textbf{Turn 3} & \textbf{Turn 4} & \textbf{Turn 5} & \textbf{Task Performance Score} \\
\midrule
Voting & 0.00\% & 0.00\% & 97.00\% & 2.83\% & 0.17\% & 59.3 \tiny{$\pm$ 3.0} \\
Cumulative & 0.00\% & 0.00\% & 97.17\% & 2.67\% & 0.17\% & 60.0 \tiny{$\pm$ 1.7} \\
Ranked & 0.00\% & 0.00\% & 98.33\% & 1.67\% & 0.00\% & 59.5 \tiny{$\pm$ 0.5} \\
Approval & 0.00\% & 0.00\% & 4.17\% & 2.50\% & 93.33\% & 49.1 \tiny{$\pm$ 12.4} \\
\midrule
Judge & 0.00\% & 0.00\% & 100.00\% & 0.00\% & 0.00\% & 64.2 \tiny{$\pm$ 1.0} \\
\midrule
Majority & 86.33\% & 8.67\% & 2.00\% & 1.17\% & 1.83\% & 61.3 \tiny{$\pm$ 3.3} \\
Supermaj. & 85.83\% & 9.50\% & 1.83\% & 1.17\% & 1.67\% & 60.2 \tiny{$\pm$ 0.3} \\
Unanimity & 74.00\% & 13.17\% & 7.17\% & 2.50\% & 3.17\% & 61.3 \tiny{$\pm$ 0.8} \\
\bottomrule
\end{tabular}
\caption{Termination Percentage by Turn for Each Decision Protocol MuSR}
\label{tab:termination_percentages_musr_70}
\end{table}

\bigskip

\begin{table}[H]
\centering
\begin{tabular}{l|ccccc|c}
\toprule
\textbf{Group} & \textbf{Turn 1} & \textbf{Turn 2} & \textbf{Turn 3} & \textbf{Turn 4} & \textbf{Turn 5} & \textbf{Task Performance Score} \\
\midrule
Voting & 0.00\% & 0.00\% & 99.50\% & 0.50\% & 0.00\% & 69.5 \tiny{$\pm$ 0.5} \\
Cumulative & 0.00\% & 0.00\% & 94.33\% & 4.83\% & 0.83\% & 69.7 \tiny{$\pm$ 1.1} \\
Ranked & 0.00\% & 0.00\% & 99.83\% & 0.17\% & 0.00\% & 70.6 \tiny{$\pm$ 1.1} \\
Approval & 0.00\% & 0.00\% & 6.17\% & 2.83\% & 91.00\% & 24.3 \tiny{$\pm$ 5.0} \\
\midrule
Judge & 0.00\% & 0.00\% & 100.00\% & 0.00\% & 0.00\% & 65.7 \tiny{$\pm$ 1.1} \\
\midrule
Majority & 98.33\% & 1.67\% & 0.00\% & 0.00\% & 0.00\% & 58.2 \tiny{$\pm$ 1.0} \\
Supermaj. & 97.00\% & 2.50\% & 0.00\% & 0.17\% & 0.33\% & 54.3 \tiny{$\pm$ 1.9} \\
Unanimity & 86.83\% & 11.33\% & 1.67\% & 0.00\% & 0.17\% & 56.7 \tiny{$\pm$ 2.2} \\
\bottomrule
\end{tabular}
\caption{Termination Percentage by Turn for Each Decision Protocol SQuAD 2.0}
\label{tab:termination_percentages_squad_70}
\end{table}

\bigskip

\begin{table}[H]
\centering
\begin{tabular}{l|ccccc|c}
\toprule
\textbf{Group} & \textbf{Turn 1} & \textbf{Turn 2} & \textbf{Turn 3} & \textbf{Turn 4} & \textbf{Turn 5} & \textbf{Task Performance Score} \\
\midrule
Voting & 0.00\% & 0.00\% & 98.83\% & 1.17\% & 0.00\% & 12.0 \tiny{$\pm$ 0.9} \\
Cumulative & 0.00\% & 0.00\% & 71.00\% & 17.67\% & 11.33\% & 11.9 \tiny{$\pm$ 2.4} \\
Ranked & 0.00\% & 0.00\% & 98.67\% & 1.17\% & 0.17\% & 12.0 \tiny{$\pm$ 1.3} \\
Approval & 0.00\% & 0.00\% & 32.33\% & 16.83\% & 50.83\% & 7.9 \tiny{$\pm$ 2.3} \\
\midrule
Judge & 0.00\% & 0.00\% & 100.00\% & 0.00\% & 0.00\% & 19.2 \tiny{$\pm$ 0.6} \\
\midrule
Majority & 92.50\% & 7.17\% & 0.00\% & 0.00\% & 0.33\% & 23.2 \tiny{$\pm$ 3.3} \\
Supermaj. & 90.67\% & 8.67\% & 0.33\% & 0.00\% & 0.33\% & 25.7 \tiny{$\pm$ 2.4} \\
Unanimity & 66.50\% & 30.50\% & 2.33\% & 0.17\% & 0.50\% & 20.7 \tiny{$\pm$ 1.5} \\
\bottomrule
\end{tabular}
\caption{Termination Percentage by Turn for Each Decision Protocol Math lvl 5}
\label{tab:termination_percentages_math_70}
\end{table}

\subsection{Challenge Results}
\label{sec:appendix-challenge}

\begin{figure}[H]
    \centering
    \includegraphics[width=\linewidth]{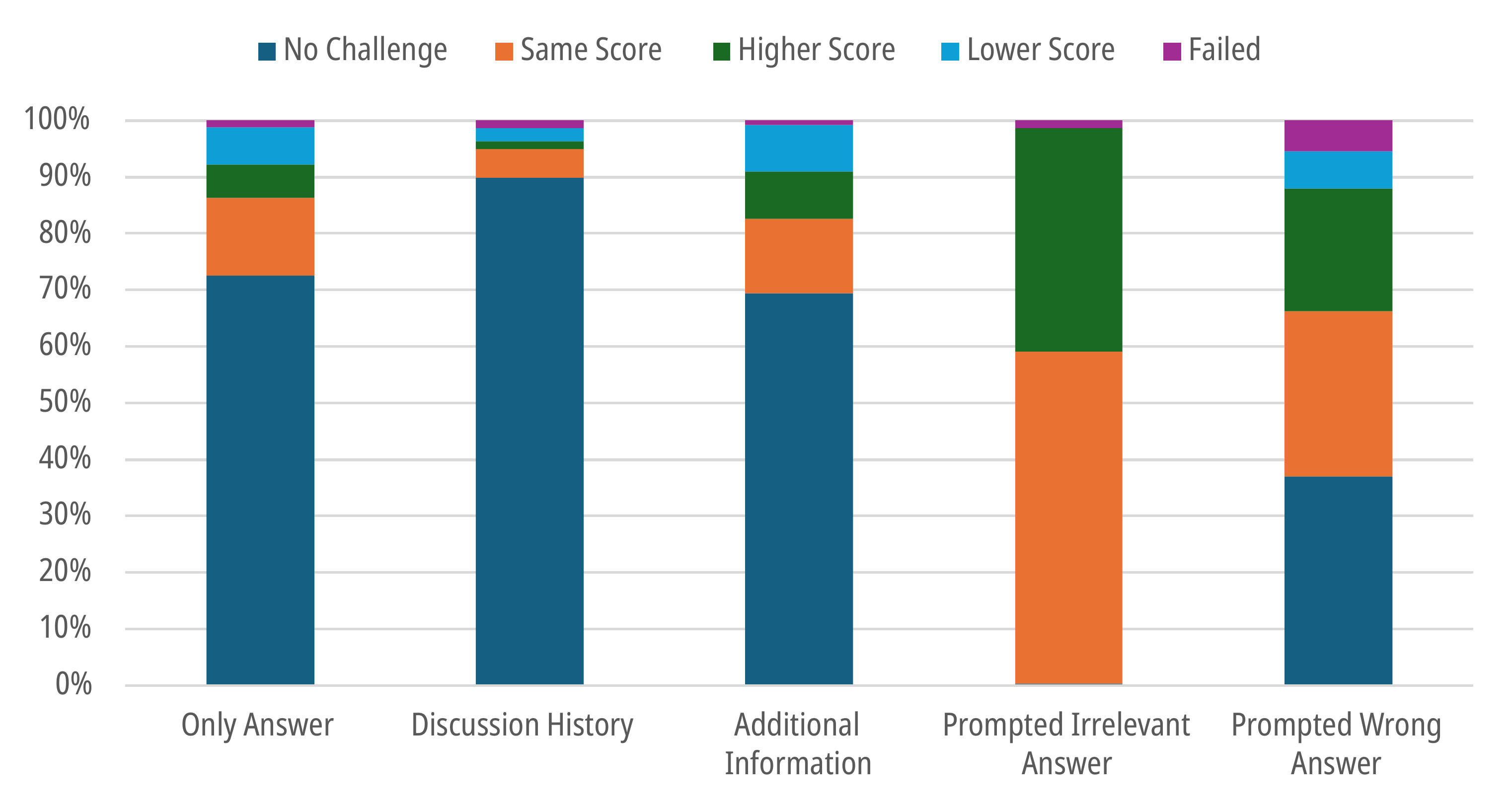}
    \caption{Percentage of agents challenging the answer for the MMLU dataset. If the answer is challenged the result is compared to the previous answer which results in a higher, lower or same score. If the agent generates an unusable answer, the challenge failed.}
    \label{fig:challenge-mmlu}
\end{figure}

\begin{figure}[H]
    \centering
    \includegraphics[width=\linewidth]{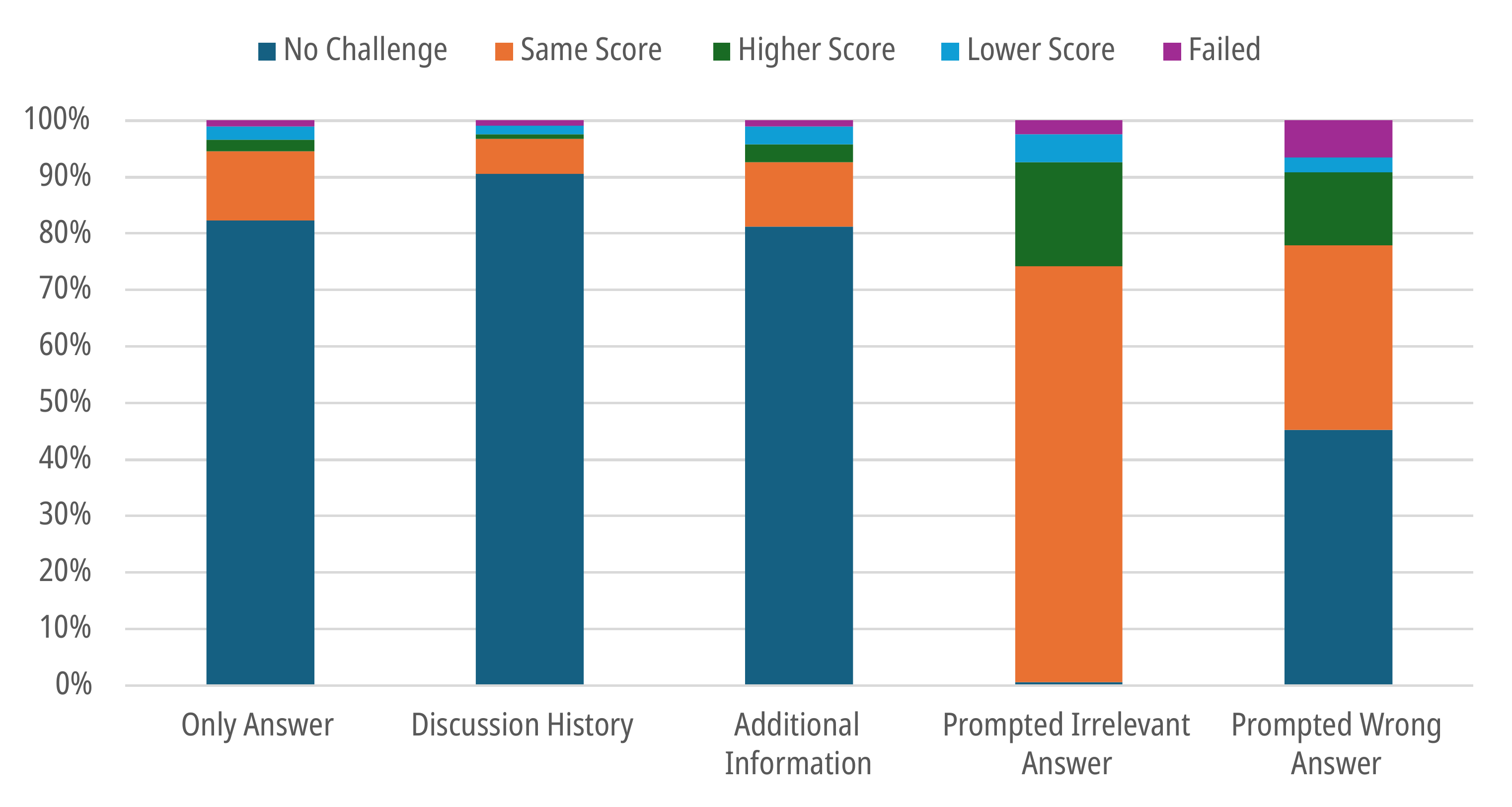}
    \caption{Percentage of agents challenging the answer for the MMLU-Pro dataset. If the answer is challenged the result is compared to the previous answer which results in a higher, lower or same score. If the agent generates an unusable answer, the challenge failed.}
    \label{fig:challenge-mmlu-pro}
\end{figure}

\begin{figure}[H]
    \centering
    \includegraphics[width=\linewidth]{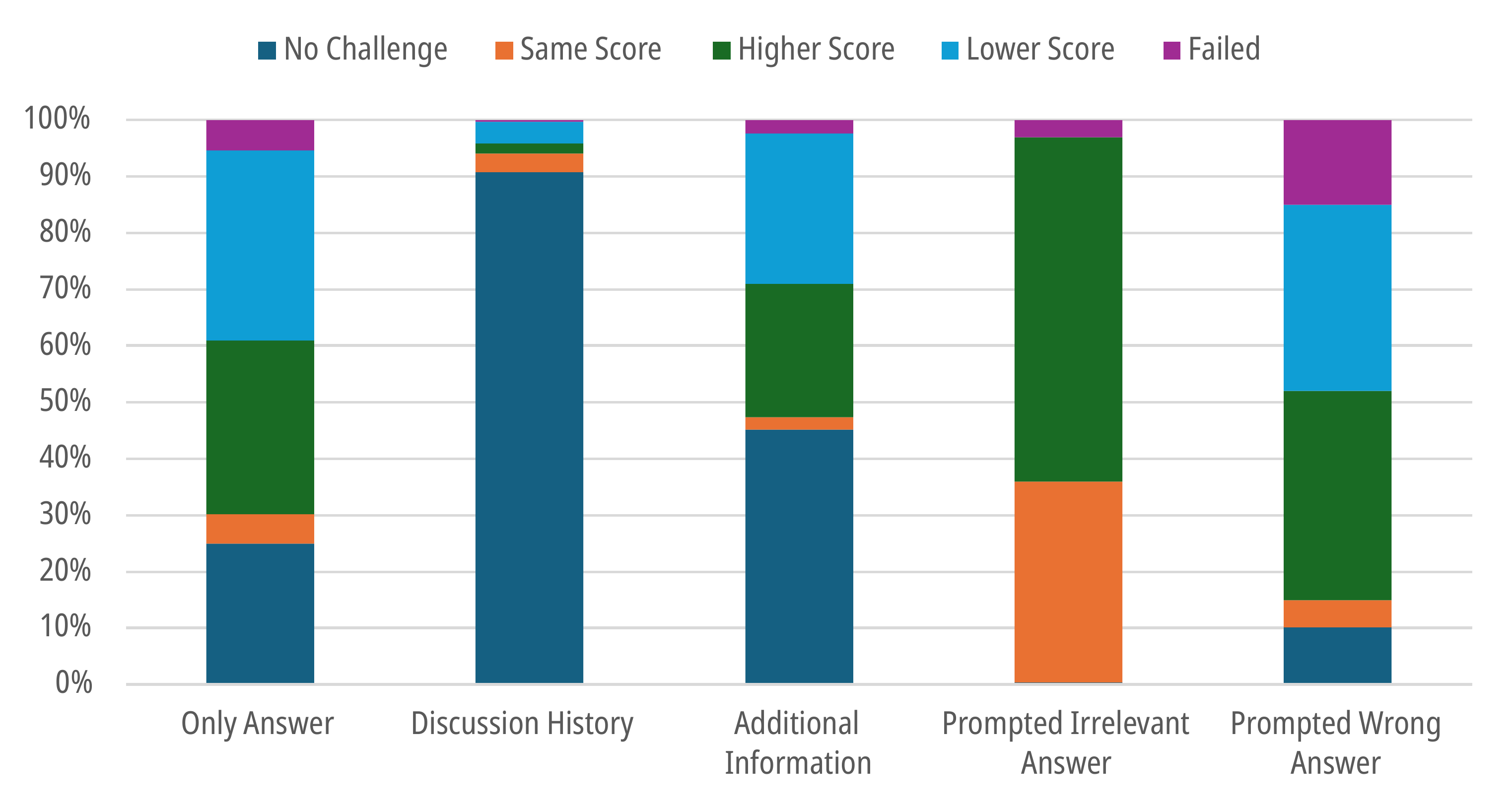}
    \caption{Percentage of agents challenging the answer for the StrategyQA dataset. If the answer is challenged the result is compared to the previous answer which results in a higher, lower or same score. If the agent generates an unusable answer, the challenge failed.}
    \label{fig:challenge-strategyqa}
\end{figure}

\begin{figure}[H]
    \centering
    \includegraphics[width=\linewidth]{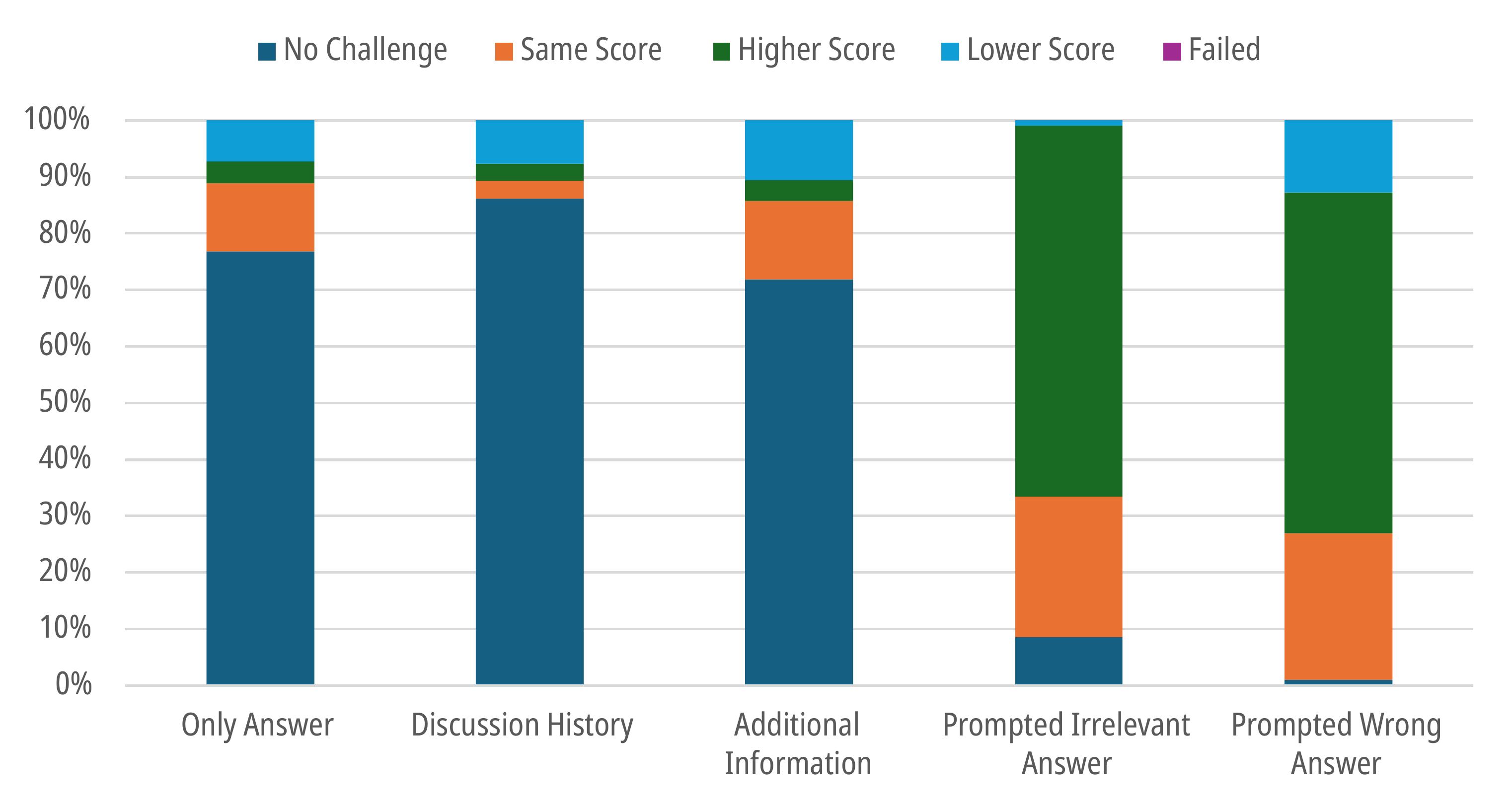}
    \caption{Percentage of agents challenging the answer for the SQuAD 2.0 dataset. If the answer is challenged the result is compared to the previous answer which results in a higher, lower or same score. If the agent generates an unusable answer, the challenge failed.}
    \label{fig:challenge-squad}
\end{figure}

\section{Prompts}
\label{sec:app_prompts}
\subsection{Final Answer Extraction}
\label{sec:app_final_answer_prompt}
\medskip
\begin{figure}[H]
    \centering
    \begin{combinedprompt}
    \textbf{System Prompt:} \\
    \begingroup
    \colorbox{systemcolor}{\parbox{\dimexpr\linewidth-2\fboxsep\relax}{
    Your role: \texttt{<persona>} (\texttt{<persona description>})
    }}
    \endgroup

    \vspace{0.4em} %

    \textbf{User Prompt:} \\
    \begingroup
    \colorbox{usercolor}{\parbox{\dimexpr\linewidth-2\fboxsep\relax}{
    You are tasked with creating a final solution based on the given input and your previous response.\\
    Task: \texttt{<task>}\\
    Input: \texttt{<input sample>}\\
    Your previous response: \texttt{<previous answer>}\\
    Extract the final solution to the task from the provided text. Remove statements of agreement, disagreement, and explanations. Do not modify the text. Do not output any text besides the solution. If there is no solution provided, just copy the previous response.
    }}
    \endgroup
\end{combinedprompt}
    \caption{Prompt used to extract the final answer of a given agent from its previous response.}
    \label{fig:extract_prompt}
\end{figure}

\subsection{Voting Prompts}

\begin{figure}[H]
    \centering
    \begin{combinedprompt}
    \textbf{System Prompt:} \\
    \begingroup
    \colorbox{systemcolor}{\parbox{\dimexpr\linewidth-2\fboxsep\relax}{
    Your role: \texttt{<persona>} (\texttt{<persona description>})
    }}
    \endgroup

    \vspace{0.4em} %

    \textbf{User Prompt:} \\
    \begingroup
    \colorbox{usercolor}{\parbox{\dimexpr\linewidth-2\fboxsep\relax}{
    You are tasked with voting for the best solution from the list provided below based on the given task.\\
    Task: \texttt{<task>}\\
    Question: \texttt{<input sample>}\\
    Here are the possible solutions:\\
    Solution 1: \texttt{<agent 1 final answer>}\\
    Solution 2: \texttt{<agent 2 final answer>}\\
    Solution 3: \texttt{<agent 3 final answer>}\\
    Based on the above solutions, please provide the number of the solution you are voting for. Answer only with the number.
    }}
    \endgroup
\end{combinedprompt}
    \caption{Prompt used to get a vote from each agent for the Simple Voting decision protocol. No alterations are applied.}
    \label{fig:simple_voting_prompt}
\end{figure}

\begin{figure}[H]
    \centering
    \begin{combinedprompt}
    \textbf{System Prompt:} \\
    \begingroup
    \colorbox{systemcolor}{\parbox{\dimexpr\linewidth-2\fboxsep\relax}{
    Your role: \texttt{<persona>} (\texttt{<persona description>})
    }}
    \endgroup

    \vspace{0.4em} %

    \textbf{User Prompt:} \\
    \begingroup
    \colorbox{usercolor}{\parbox{\dimexpr\linewidth-2\fboxsep\relax}{
    You are tasked with approving any number of solutions from the list provided below based on the given task.\\
    Task: \texttt{<task>}\\
    Question: \texttt{<input sample>}\\
    Here are the possible solutions:\\
    Solution 1: \texttt{<agent 1 final answer>}\\
    Solution 2: \texttt{<agent 2 final answer>}\\
    Solution 3: \texttt{<agent 3 final answer>}\\
    Based on the above solutions, please provide the numbers of the solutions you are approving, separated by commas. Answer only with the numbers.
    }}
    \endgroup
\end{combinedprompt}
    \caption{Prompt used to get a vote from each agent for the Approval Voting decision protocol. No alterations are applied.}
    \label{fig:approval_voting_prompt}
\end{figure}

\begin{figure}[H]
    \centering
    \begin{combinedprompt}
    \textbf{System Prompt:} \\
    \begingroup
    \colorbox{systemcolor}{\parbox{\dimexpr\linewidth-2\fboxsep\relax}{
    Your role: \texttt{<persona>} (\texttt{<persona description>})
    }}
    \endgroup

    \vspace{0.4em} %

    \textbf{User Prompt:} \\
    \begingroup
    \colorbox{usercolor}{\parbox{\dimexpr\linewidth-2\fboxsep\relax}{
    You are tasked with distributing 10 points among the provided solutions based on the given task.\\
    Task: \texttt{<task>}\\
    Question: \texttt{<input sample>}\\
    Here are the possible solutions:\\
    Solution 1: \texttt{<agent 1 final answer>}\\
    Solution 2: \texttt{<agent 2 final answer>}\\
    Solution 3: \texttt{<agent 3 final answer>}\\
    Based on the above solutions, please distribute 10 points among the solutions. Provide your points allocation as a JSON dictionary where keys are solution numbers (as int) and values are the points. The total points should sum up to 10. Answer only with the JSON dictionary.
    }}
    \endgroup
\end{combinedprompt}
    \caption{Prompt used to get a vote from each agent for the Cumulative Voting decision protocol. No alterations are applied.}
    \label{fig:cumulative_voting_prompt}
\end{figure}

\begin{figure}[H]
    \centering
    \begin{combinedprompt}
    \textbf{System Prompt:} \\
    \begingroup
    \colorbox{systemcolor}{\parbox{\dimexpr\linewidth-2\fboxsep\relax}{
    Your role: \texttt{<persona>} (\texttt{<persona description>})
    }}
    \endgroup

    \vspace{0.4em} %

    \textbf{User Prompt:} \\
    \begingroup
    \colorbox{usercolor}{\parbox{\dimexpr\linewidth-2\fboxsep\relax}{
    You are tasked with ranking the solutions from the most preferred to the least preferred based on the given task.\\
    Task: \texttt{<task>}\\
    Question: \texttt{<input sample>}\\
    Here are the possible solutions:\\
    Solution 1: \texttt{<agent 1 final answer>}\\
    Solution 2: \texttt{<agent 2 final answer>}\\
    Solution 3: \texttt{<agent 3 final answer>}\\
    Based on the above solutions, please provide the rankings of the solutions separated by spaces. Example: '0 2 1' if you prefer Solution 0 the most, then Solution 2, and finally Solution 1. Provide up to 5 rankings. Only answer with the rankings.
    }}
    \endgroup
\end{combinedprompt}
    \caption{Prompt used to get a vote from each agent for the Ranked Voting decision protocol. No alterations are applied.}
    \label{fig:ranked_voting_prompt}
\end{figure}

\subsection{Judge Prompt}

\begin{figure}[H]
    \centering
    \begin{combinedprompt}
    \textbf{User Prompt:} \\
    \begingroup
    \colorbox{usercolor}{\parbox{\dimexpr\linewidth-2\fboxsep\relax}{
    Task: \texttt{<task>}\\
    Question: \texttt{<input sample>}\\
    Please provide a decision on the following solutions and combine them in a single answer to solve the task. Only answer with the solution:\\
    Solution 1: \texttt{<agent 1 final answer>}\\
    Solution 2: \texttt{<agent 2 final answer>}\\
    Solution 3: \texttt{<agent 3 final answer>}
    }}
    \endgroup
\end{combinedprompt}
    \caption{Prompt used to get a final decision from the Judge decision protocol. No alterations are applied.}
    \label{fig:judge_prompt}
\end{figure}

\subsection{Confidence Prompt}

\begin{figure}[H]
    \centering
    \begin{combinedprompt}
    \textbf{System Prompt:} \\
    \begingroup
    \colorbox{systemcolor}{\parbox{\dimexpr\linewidth-2\fboxsep\relax}{
    Your role: \texttt{<persona>} (\texttt{<persona description>})
    }}
    \endgroup

    \vspace{0.4em} %

    \textbf{User Prompt:} \\
    \begingroup
    \colorbox{usercolor}{\parbox{\dimexpr\linewidth-2\fboxsep\relax}{
    The task is: \texttt{<task>}. The question is: \texttt{<input\_sample>}. This is the final answer you provided: \texttt{<final\_answer>}. Based on this information, please generate a confidence score between 0 and 100 for the final answer. Be critical and only answer with the number.
    }}
    \endgroup
\end{combinedprompt}
    \caption{Prompt used to get a confidence score from each agent for the voting alterations experiment.}
    \label{fig:confidence_prompt}
\end{figure}

\subsection{Challenge Prompt}

\begin{figure}[H]
    \centering
    \begin{combinedprompt}
    \textbf{System Prompt:} \\
    \begingroup
    \colorbox{systemcolor}{\parbox{\dimexpr\linewidth-2\fboxsep\relax}{
    You are a participant in a group discussion. \\
    Your role: \texttt{<persona>} (\texttt{<persona description>})
    }}
    \endgroup

    \vspace{0.4em} %

    \textbf{User Prompt:} \\
    \begingroup
    \colorbox{usercolor}{\parbox{\dimexpr\linewidth-2\fboxsep\relax}{
    The task is: \texttt{<task>}. The question is: \texttt{<question>}. \\
    This is the final answer generated by the discussion: \texttt{<final\_answer>}. \\
    Please critically evaluate this answer. If you agree with the final answer, respond with the exact word 'AGREE' to confirm. If you do not agree, respond with the exact word 'DISAGREE' to challenge the answer.
    }}
    \endgroup
\end{combinedprompt}
    \caption{Prompt used to challenge the final answer.}
    \label{fig:challenge_prompt}
\end{figure}

\begin{figure}[H]
    \centering
    \begin{combinedprompt}
    \textbf{System Prompt:} \\
    \begingroup
    \colorbox{systemcolor}{\parbox{\dimexpr\linewidth-2\fboxsep\relax}{
    You are a participant in a group discussion. \\
    Your role: \texttt{<persona>} (\texttt{<persona description>})
    }}
    \endgroup

    \vspace{0.4em} %

    \textbf{User Prompt:} \\
    \begingroup
    \colorbox{usercolor}{\parbox{\dimexpr\linewidth-2\fboxsep\relax}{
    The task is: \texttt{<task>}. The question is: \texttt{<question>}. \\
    This is the final answer generated by the discussion: \texttt{<final\_answer>}. \\
    You dont agree with the final answer. Please provide a new answer to the question. Include the letter corresponding to your answer in the solution.
    }}
    \endgroup
\end{combinedprompt}
    \caption{Prompt used to generate a new answer in case the final answer got challenged.}
    \label{fig:challenge_new_prompt}
\end{figure}

\newpage
\section{MALLM Setup}
\label{sec:mallm_setup_app}
\subsection{MALLM Default Parameters}
\begin{configpython}[Default Parameters used for each experiment]{lst:config_default}
input_json_file_path: str = None
output_json_file_path: str = None
task_instruction_prompt: str = None
task_instruction_prompt_template: Optional[str] = None
endpoint_url: str = "https://api.openai.com/v1"
model_name: str = "gpt-3.5-turbo"
api_key: str = "-"
max_turns: int = 10
skip_decision_making: bool = False
discussion_paradigm: str = "memory"
response_generator: str = "simple"
decision_protocol: str = "hybrid_consensus"
visible_turns_in_memory: int = 2
debate_rounds: int = 2
concurrent_api_requests: int = 100
use_baseline: bool = False
use_chain_of_thought: bool = True
num_agents: int = 3
num_neutral_agents: int = 0
agent_generator: str = "expert"
agent_generators_list: list = []
trust_remote_code: bool = False
num_samples: Optional[int] = None
hf_dataset_split: Optional[str] = "test"
hf_token: Optional[str] = None
hf_dataset_version: Optional[str] = None
hf_dataset_input_column: Optional[str] = None
hf_dataset_reference_column: Optional[str] = None
hf_dataset_context_column: Optional[str] = None
use_ablation: bool = False
shuffle_input_samples: bool = False
all_agents_generate_first_draft: bool = False
all_agents_generate_draft: bool = False
policy: Optional[str] = None
voting_protocols_with_alterations: bool = False
calculate_persona_diversity: bool = False
\end{configpython}
\subsection{MALLM Experiment Batch Files}
To replicate the results of this study, modify the batch files for your specific dataset and execute them with the following command:

\[
\textbf{mallm-batch $<$\text{PATH TO BATCH FILE}$>$}
\]

After running this command, you can evaluate the results using:

\[
\textbf{mallm-evaluate --metrics [$<$\text{METRIC FOR DATASET}$>$, ...] ./results}
\]

\subsubsection{Decision Protocol Experiments}
\begin{configjson}[MALLM batch config used for Experiment 1]{lst:config_exp_1}
{
  "repeats": 3,
  "name": "<DATASET NAME>",
  "common": {
    "task_instruction_prompt_template": "<DATASET NAME>",
    "endpoint_url": "<LLM API HOSTNAME>",
    "api_key": "<LLM API KEY>",
    "model_name": "<MODEL NAME>",
    "input_json_file_path": "data/datasets/<DATASET NAME>.json",
    "concurrent_api_requests": 200,
    "num_samples": <SAMPLES FOR DATASET>,
    "max_turns": 5,
    "response_generator":"simple"
  },
  "runs": [
    {
      "output_json_file_path": "results/baseline-cot.json",
      "use_baseline": true
    },
    {
      "output_json_file_path": "results/baseline.json",
      "use_baseline": true,
      "use_chain_of_thought": false
    },
    {
      "output_json_file_path": "results/approval.json",
      "decision_protocol": "approval_voting"
    },
    {
      "output_json_file_path": "results/cumulative.json",
      "decision_protocol": "cumulative_voting"
    },
    {
      "output_json_file_path": "results/hybrid_consensus.json",
      "decision_protocol": "hybrid_consensus"
    },
    {
      "output_json_file_path": "results/majority_consensus.json",
      "decision_protocol": "majority_consensus"
    },
    {
      "output_json_file_path": "results/supermajority_consensus.json",
      "decision_protocol": "supermajority_consensus"
    },
    {
      "output_json_file_path": "results/unanimity_consensus.json",
      "decision_protocol": "unanimity_consensus"
    },
    {
      "output_json_file_path": "results/voting.json",
      "decision_protocol": "simple_voting"
    },
    {
      "output_json_file_path": "results/ranked.json",
      "decision_protocol": "ranked_voting"
    },
    {
      "output_json_file_path": "results/summary.json",
      "decision_protocol": "summary"
    }
  ]
}
\end{configjson}
\newpage
\begin{configjson}[MALLM batch config used for Experiment 1 to evaluate difference in task performance for varying number of agents]{lst:config_exp_1_agentsnum}
{
  "repeats": 3,
  "name": "<DATASET NAME>",
  "common": {
    "task_instruction_prompt_template": "<DATASET NAME>",
    "endpoint_url": "<LLM API HOSTNAME>",
    "api_key": "<LLM API KEY>",
    "model_name": "<MODEL NAME>",
    "input_json_file_path": "data/datasets/<DATASET NAME>.json",
    "concurrent_api_requests": 200,
    "num_samples": <SAMPLES FOR DATASET>,
    "max_turns": 10,
    "response_generator":"simple",
    "visible_turns_in_memory": 10,
    "decision_protocol": "simple_voting",
    "all_agents_generate_first_draft": true
  },
  "runs": [
    {
      "output_json_file_path": "results/agents1.json",
      "num_agents": 1
    },
    {
      "output_json_file_path": "results/agents2.json",
      "num_agents": 2
    },
    {
      "output_json_file_path": "results/agents3.json",
      "num_agents": 3
    },
    {
      "output_json_file_path": "results/agents4.json",
      "num_agents": 4
    },
    {
      "output_json_file_path": "results/agents5.json",
      "num_agents": 5
    },
    {
      "output_json_file_path": "results/agents6.json",
      "num_agents": 6
    },
    {
      "output_json_file_path": "results/agents7.json",
      "num_agents": 7
    },
    {
      "output_json_file_path": "results/agents8.json",
      "num_agents": 8
    },
    {
      "output_json_file_path": "results/agents9.json",
      "num_agents": 9
    },
    {
      "output_json_file_path": "results/agents10.json",
      "num_agents": 10
    }
  ]
}
\end{configjson}

\begin{configjson}[MALLM batch config used for Experiment 1 to evaluate difference in task performance for round that voting is allowed]{lst:config_exp_1_votingturn}
{
  "repeats": 3,
  "name": "<DATASET NAME>",
  "common": {
    "task_instruction_prompt_template": "<DATASET NAME>",
    "endpoint_url": "<LLM API HOSTNAME>",
    "api_key": "<LLM API KEY>",
    "model_name": "<MODEL NAME>",
    "input_json_file_path": "data/datasets/<DATASET NAME>.json",
    "concurrent_api_requests": 200,
    "num_samples": <SAMPLES FOR DATASET>,
    "max_turns": 10,
    "response_generator":"simple",
    "visible_turns_in_memory": 10,
    "decision_protocol": "simple_voting"
  },
  "runs": [
    {
      "output_json_file_path": "results/voteturn1.json",
      "voting_protocols_vote_turn": 1
    },
    {
      "output_json_file_path": "results/voteturn2.json",
      "voting_protocols_vote_turn": 2
    },
    {
      "output_json_file_path": "results/voteturn3.json",
      "voting_protocols_vote_turn": 3
    },
    {
      "output_json_file_path": "results/voteturn4.json",
      "voting_protocols_vote_turn": 4
    },
    {
      "output_json_file_path": "results/voteturn5.json",
      "voting_protocols_vote_turn": 5
    },
    {
      "output_json_file_path": "results/voteturn6.json",
      "voting_protocols_vote_turn": 6
    },
    {
      "output_json_file_path": "results/voteturn7.json",
      "voting_protocols_vote_turn": 7
    },
    {
      "output_json_file_path": "results/voteturn8.json",
      "voting_protocols_vote_turn": 8
    },
    {
      "output_json_file_path": "results/voteturn9.json",
      "voting_protocols_vote_turn": 9
    },
    {
      "output_json_file_path": "results/voteturn10.json",
      "voting_protocols_vote_turn": 10
    }
  ]
}
\end{configjson}
\newpage
\subsubsection{Decision Alterations Experiments}
\begin{configjson}[MALLM batch config used for Experiment 2]{lst:config_exp_2}
{
  "repeats": 3,
  "name": "<DATASET NAME>",
  "common": {
    "task_instruction_prompt_template": "<DATASET NAME>",
    "endpoint_url": "<LLM API HOSTNAME>",
    "api_key": "<LLM API KEY>",
    "model_name": "<MODEL NAME>",
    "input_json_file_path": "data/datasets/<DATASET NAME>.json",
    "concurrent_api_requests": 200,
    "num_samples": <SAMPLES FOR DATASET>,
    "max_turns": 5,
    "response_generator":"simple"
    "voting_protocols_with_alterations":true
  },
  "runs": [
    {
      "output_json_file_path": "results/approval.json",
      "decision_protocol": "approval_voting"
    },
    {
      "output_json_file_path": "results/cumulative.json",
      "decision_protocol": "cumulative_voting"
    },
    {
      "output_json_file_path": "results/voting.json",
      "decision_protocol": "simple_voting"
    },
    {
      "output_json_file_path": "results/ranked.json",
      "decision_protocol": "ranked_voting"
    },
    {
      "output_json_file_path": "results/summary.json",
      "decision_protocol": "summary"
    }
  ]
}
\end{configjson}

\newpage
\subsubsection{Final Answer Challenge Experiments}
\begin{configjson}[MALLM batch config used for Experiment 3]{lst:config_exp_3}
{
  "repeats": 3,
  "name": "<DATASET NAME>",
  "common": {
    "task_instruction_prompt_template": "<DATASET NAME>",
    "endpoint_url": "<LLM API HOSTNAME>",
    "api_key": "<LLM API KEY>",
    "model_name": "<MODEL NAME>",
    "input_json_file_path": "data/datasets/<DATASET NAME>.json",
    "concurrent_api_requests": 200,
    "num_samples": <SAMPLES FOR DATASET>,
    "max_turns": 5,
    "response_generator":"simple",
    "challenge_final_results": true
  },
  "runs": [
    {
      "output_json_file_path": "results/baseline-cot.json",
      "use_baseline": true
    },
    {
      "output_json_file_path": "results/baseline.json",
      "use_baseline": true,
      "use_chain_of_thought": false
    },
    {
      "output_json_file_path": "results/approval.json",
      "decision_protocol": "approval_voting"
    },
    {
      "output_json_file_path": "results/cumulative.json",
      "decision_protocol": "cumulative_voting"
    },
    {
      "output_json_file_path": "results/hybrid_consensus.json",
      "decision_protocol": "hybrid_consensus"
    },
    {
      "output_json_file_path": "results/majority_consensus.json",
      "decision_protocol": "majority_consensus"
    },
    {
      "output_json_file_path": "results/supermajority_consensus.json",
      "decision_protocol": "supermajority_consensus"
    },
    {
      "output_json_file_path": "results/unanimity_consensus.json",
      "decision_protocol": "unanimity_consensus"
    },
    {
      "output_json_file_path": "results/voting.json",
      "decision_protocol": "simple_voting"
    },
    {
      "output_json_file_path": "results/ranked.json",
      "decision_protocol": "ranked_voting"
    },
    {
      "output_json_file_path": "results/summary.json",
      "decision_protocol": "summary"
    }
  ]
}
\end{configjson}
\newpage
\section{Example Discussions}
\subsection{Successfull Voting Discussion}
All voting and judge decision protocols are attached as an example to this discussion. The original discussion was created using simple voting.

\vspace{5mm}

\definecolor{lightyellow}{RGB}{255, 255, 204}

\begin{tcolorbox}[title=Overview, colframe=black!10, coltitle=black, fonttitle=\bfseries, boxrule=0.5mm, width=\textwidth]

\textbf{Task:} Answer the following question. If the question is not answerable with the provided information, write '[UNKNOWN]'.

\textbf{Input:} What choir from the area has sung at the Guildhall?

\textbf{Context:} Southampton has two large live music venues, the Mayflower Theatre (formerly the Gaumont Theatre) and the Guildhall. The Guildhall has seen concerts from a wide range of popular artists including Pink Floyd, David Bowie, Delirious?, Manic Street Preachers, The Killers, The Kaiser Chiefs, Amy Winehouse, Lostprophets, The Midnight Beast, Modestep, and All Time Low. It also hosts classical concerts presented by the Bournemouth Symphony Orchestra, City of Southampton Orchestra, Southampton Concert Orchestra, Southampton Philharmonic Choir and Southampton Choral Society.  

\begin{tcolorbox}[coltitle=black, fonttitle=\bfseries, boxrule=0.5mm, width=\textwidth]
\textbf{MALLM solution:} Southampton Philharmonic Choir \\
\textbf{Real solution:} Southampton Philharmonic Choir
\end{tcolorbox}
\end{tcolorbox}

\begin{tcolorbox}[breakable, enhanced jigsaw, title=Turn 1, colframe=black!10, coltitle=black, fonttitle=\bfseries, boxrule=0.5mm, width=\textwidth]

\begin{tcolorbox}[title=Music Connoisseur \hfill \tiny{\textnormal{A knowledgeable musician who is familiar with various music genres and has attended concerts at the Guildhall.}}, colback=lightyellow, colframe=orange!30, coltitle=black, fonttitle=\bfseries, boxrule=0.5mm, width=\textwidth]
\begin{minipage}{1.0\textwidth}
    Based on the information provided, I propose the following solution:

Southampton Philharmonic Choir.

The text states: "It also hosts classical concerts presented by the Bournemouth Symphony Orchestra, City of Southampton Orchestra, Southampton Concert Orchestra, Southampton Philharmonic Choir and Southampton Choral Society." This suggests that the Southampton Philharmonic Choir is one of the choirs that has performed at the Guildhall, as it is mentioned alongside the other organizations that host classical concerts at the venue.
\end{minipage}
\end{tcolorbox}

\begin{tcolorbox}[title=Choir Conductor \hfill \tiny{\textnormal{A dedicated choral conductor with extensive experience in leading local choirs and orchestras, familiar with the performances at the Guildhall.}}, colback=lightyellow, colframe=orange!30, coltitle=black, fonttitle=\bfseries, boxrule=0.5mm, width=\textwidth]
\begin{minipage}{1.0\textwidth}
        [AGREE]

I agree with the current solution that the Southampton Philharmonic Choir has sung at the Guildhall. The text explicitly mentions it as one of the organizations that hosts classical concerts at the Guildhall, which indicates that the choir has performed at the venue.
\end{minipage}
\end{tcolorbox}

\begin{tcolorbox}[title=Music Journalist \hfill \tiny{\textnormal{A music enthusiast who has written articles and reviews about various concerts and events at the Guildhall, including performances by local choirs.}}, colback=lightyellow, colframe=orange!30, coltitle=black, fonttitle=\bfseries, boxrule=0.5mm, width=\textwidth]
\begin{minipage}{1.0\textwidth}
        [AGREE]

I completely agree with the current solution. The text explicitly mentions the Southampton Philharmonic Choir as one of the organizations that hosts classical concerts at the Guildhall, making it the most likely choice.
\end{minipage}
\end{tcolorbox}

\end{tcolorbox}

\begin{tcolorbox}[breakable, enhanced jigsaw, title=Turn 2, colframe=black!10, coltitle=black, fonttitle=\bfseries, boxrule=0.5mm, width=\textwidth]

\begin{tcolorbox}[title=Music Connoisseur \hfill \tiny{\textnormal{A knowledgeable musician who is familiar with various music genres and has attended concerts at the Guildhall.}}, colback=lightyellow, colframe=orange!30, coltitle=black, fonttitle=\bfseries, boxrule=0.5mm, width=\textwidth]
\begin{minipage}{1.0\textwidth}
    [AGREE]

I'm in agreement with the current solution, as the text explicitly mentions the Southampton Philharmonic Choir as one of the organizations that hosts classical concerts at the Guildhall. This makes it the most likely candidate to have sung at the venue.
\end{minipage}
\end{tcolorbox}

\begin{tcolorbox}[title=Choir Conductor \hfill \tiny{\textnormal{A dedicated choral conductor with extensive experience in leading local choirs and orchestras, familiar with the performances at the Guildhall.}}, colback=lightyellow, colframe=orange!30, coltitle=black, fonttitle=\bfseries, boxrule=0.5mm, width=\textwidth]
\begin{minipage}{1.0\textwidth}
        [AGREE]

I agree with the current solution that the Southampton Philharmonic Choir has sung at the Guildhall. The text explicitly mentions it as one of the organizations that hosts classical concerts at the Guildhall, which indicates that the choir has performed at the venue.
\end{minipage}
\end{tcolorbox}

\begin{tcolorbox}[title=Music Journalist \hfill \tiny{\textnormal{A music enthusiast who has written articles and reviews about various concerts and events at the Guildhall, including performances by local choirs.}}, colback=lightyellow, colframe=orange!30, coltitle=black, fonttitle=\bfseries, boxrule=0.5mm, width=\textwidth]
\begin{minipage}{1.0\textwidth}
        [AGREE]

I completely agree with the current solution. The text explicitly mentions the Southampton Philharmonic Choir as one of the organizations that hosts classical concerts at the Guildhall, making it the most likely choice.
\end{minipage}
\end{tcolorbox}

\end{tcolorbox}

\begin{tcolorbox}[breakable, enhanced jigsaw, title=Turn 3, colframe=black!10, coltitle=black, fonttitle=\bfseries, boxrule=0.5mm, width=\textwidth]

\begin{tcolorbox}[title=Music Connoisseur \hfill \tiny{\textnormal{A knowledgeable musician who is familiar with various music genres and has attended concerts at the Guildhall.}}, colback=lightyellow, colframe=orange!30, coltitle=black, fonttitle=\bfseries, boxrule=0.5mm, width=\textwidth]
\begin{minipage}{1.0\textwidth}
    [AGREE]

I agree with the current solution and the assessment that there is no need to improve it. The text provides explicit evidence that the Southampton Philharmonic Choir hosts classical concerts at the Guildhall, making it the most likely correct answer.
\end{minipage}
\end{tcolorbox}

\begin{tcolorbox}[title=Choir Conductor \hfill \tiny{\textnormal{A dedicated choral conductor with extensive experience in leading local choirs and orchestras, familiar with the performances at the Guildhall.}}, colback=lightyellow, colframe=orange!30, coltitle=black, fonttitle=\bfseries, boxrule=0.5mm, width=\textwidth]
\begin{minipage}{1.0\textwidth}
        [AGREE]

I agree with the current solution that the Southampton Philharmonic Choir has sung at the Guildhall. The text explicitly mentions it as one of the organizations that hosts classical concerts at the Guildhall, which indicates that the choir has performed at the venue.
\end{minipage}
\end{tcolorbox}

\begin{tcolorbox}[title=Music Journalist \hfill \tiny{\textnormal{A music enthusiast who has written articles and reviews about various concerts and events at the Guildhall, including performances by local choirs.}}, colback=lightyellow, colframe=orange!30, coltitle=black, fonttitle=\bfseries, boxrule=0.5mm, width=\textwidth]
\begin{minipage}{1.0\textwidth}
        [AGREE]

I completely agree with the current solution. The text explicitly mentions the Southampton Philharmonic Choir as one of the organizations that hosts classical concerts at the Guildhall, making it the most likely choice.
\end{minipage}
\end{tcolorbox}

\end{tcolorbox}

\begin{tcolorbox}[title=Result Simple Voting, colframe=black!10, coltitle=black, fonttitle=\bfseries, boxrule=0.5mm, width=\textwidth]
\textbf{Music Connoisseur final answer:} Southampton Philharmonic Choir

\textbf{Choir Conductor final answer:} Southampton Choral Society

\textbf{Music Journalist final answer:} Southampton Philharmonic Choir

\bigskip

\textbf{Music Connoisseur} voted for answer from \textbf{Music Journalist}

\textbf{Choir Conductor} voted for answer from \textbf{Music Journalist}

\textbf{Music Journalist} voted for answer from \textbf{Music Journalist}

\end{tcolorbox}

\begin{tcolorbox}[title=Result Cumulative Voting, colframe=black!10, coltitle=black, fonttitle=\bfseries, boxrule=0.5mm, width=\textwidth]
\textbf{Local Music Enthusiast final answer:} Southampton Choral Society

Southampton Philharmonic Choir

\textbf{Archivist final answer:} Southampton Philharmonic Choir

Southampton Choral Society

\textbf{Music Journalist final answer:} Southampton Philharmonic Choir and Southampton Choral Society

\bigskip

\textbf{Local Music Enthusiast} allocated points: \\
\textbf{Archivist}: 9\\
\textbf{Music Journalist}: 1
\bigskip

\textbf{Archivist} allocated points: \\
\textbf{Local Music Enthusiast}: 2\\
\textbf{Archivist}: 4\\
\textbf{Music Journalist}: 4
\bigskip

\textbf{Music Journalist} allocated points: \\
\textbf{Local Music Enthusiast}: 1\\
\textbf{Archivist}: 2\\
\textbf{Music Journalist}: 7

\end{tcolorbox}

\begin{tcolorbox}[title=Result Approval Voting, colframe=black!10, coltitle=black, fonttitle=\bfseries, boxrule=0.5mm, width=\textwidth]
\textbf{Music Critic final answer:} Southampton Philharmonic Choir

\textbf{Local Music Event Coordinator final answer:} Southampton Choral Society and Southampton Philharmonic Choir

\textbf{Information Architect final answer:} Southampton Philharmonic Choir

Southampton Choral Society

\bigskip

\textbf{Music Critic} approved answers from \textbf{Information Architect}, \textbf{Music Critic}

\textbf{Local Music Event Coordinator} approved answers from \textbf{Information Architect}

\textbf{Information Architect} approved answers from \textbf{Information Architect}

\end{tcolorbox}

\begin{tcolorbox}[title=Result Ranked Voting, colframe=black!10, coltitle=black, fonttitle=\bfseries, boxrule=0.5mm, width=\textwidth]
\textbf{Music Journalist final answer:} Southampton Philharmonic Choir and Southampton Choral Society.

\textbf{Local Music Expert final answer:} Southampton Philharmonic Choir and Southampton Choral Society

\textbf{Concert Promoter final answer:} Southampton Philharmonic Choir and Southampton Choral Society

\bigskip

\textbf{Music Journalist} ranked answers: \\
1. Music Journalist\\
2. Local Music Expert\\
3. Concert Promoter
\bigskip

\textbf{Local Music Expert} ranked answers:\\
1. Music Journalist\\
2. Local Music Expert\\
3. Concert Promoter
\bigskip

\textbf{Concert Promoter} ranked answers:\\
1. Music Journalist\\
2. Local Music Expert\\
3. Concert Promoter

\end{tcolorbox}

\begin{tcolorbox}[title=Result Judge, colframe=black!10, coltitle=black, fonttitle=\bfseries, boxrule=0.5mm, width=\textwidth]
\textbf{Local Music Enthusiast final answer:} Southampton Choral Society

\textbf{Music Researcher final answer:} Southampton Philharmonic Choir

\textbf{Promoter final answer:} Southampton Philharmonic Choir

\bigskip

\textbf{Judge created new solution:} Southampton Philharmonic Choir and Southampton Choral Society

\end{tcolorbox}

\label{example:good_discussion}

\newpage
\subsection{Agents Tricked Discussion}
In this discussion, the agents were tricked by information provided in the context.

\vspace{5mm}

\definecolor{lightyellow}{RGB}{255, 255, 204}

\begin{tcolorbox}[title=Overview, colframe=black!10, coltitle=black, fonttitle=\bfseries, boxrule=0.5mm, width=\textwidth]

\textbf{Task:} Answer the following question. If the question is not answerable with the provided information, write '[UNKNOWN]'.

\textbf{Input:} What provides critical support for drug discovery and the availability of economic resources?

\textbf{Context:} Biodiversity provides critical support for drug discovery and the availability of medicinal resources. A significant proportion of drugs are derived, directly or indirectly, from biological sources: at least 50\% of the pharmaceutical compounds on the US market are derived from plants, animals, and micro-organisms, while about 80\% of the world population depends on medicines from nature (used in either modern or traditional medical practice) for primary healthcare. Only a tiny fraction of wild species has been investigated for medical potential. Biodiversity has been critical to advances throughout the field of bionics. Evidence from  analysis and biodiversity science indicates that the decline in output from the pharmaceutical sector since the mid-1980s can be attributed to a move away from natural product exploration ("bioprospecting") in favor of genomics and synthetic chemistry, indeed claims about the value of undiscovered pharmaceuticals may not provide enough incentive for companies in free markets to search for them because of the high cost of development; meanwhile, natural products have a long history of supporting significant economic and health innovation. Marine ecosystems are particularly important, although inappropriate bioprospecting can increase biodiversity loss, as well as violating the laws of the communities and states from which the resources are taken.

\begin{tcolorbox}[coltitle=black, fonttitle=\bfseries, boxrule=0.5mm, width=\textwidth]
\textbf{MALLM solution:} Biodiversity provides critical support for drug discovery and the availability of economic resources. \\
\textbf{Real solution:} Not included in context!
\end{tcolorbox}

\end{tcolorbox}

\begin{tcolorbox}[breakable, enhanced jigsaw, title=Turn 1, colframe=black!10, coltitle=black, fonttitle=\bfseries, boxrule=0.5mm, width=\textwidth]

\begin{tcolorbox}[title=Economist \hfill \tiny{\textnormal{A researcher who studies the impact of biodiversity on economic systems, with expertise in market analysis and the economic benefits of natural resources.}}, colback=lightyellow, colframe=orange!30, coltitle=black, fonttitle=\bfseries, boxrule=0.5mm, width=\textwidth]
\begin{minipage}{1.0\textwidth}
    As an economist, I'd like to propose a solution that addresses the decline in output from the pharmaceutical sector since the mid-1980s and leverages the untapped potential of biodiversity for drug discovery and economic innovation.

Here's a multi-faceted approach:

**Solution:** Establish a Public-Private Partnership (PPP) to incentivize the exploration and development of natural products for pharmaceuticals, while ensuring responsible and sustainable bioprospecting practices.
\end{minipage}
\end{tcolorbox}

\begin{tcolorbox}[title=Ethnobotanist \hfill \tiny{\textnormal{A scientist who studies the traditional uses of plants by indigenous communities and their potential for medicinal applications, with expertise in deciphering the cultural and historical context of natural resources.}}, colback=lightyellow, colframe=orange!30, coltitle=black, fonttitle=\bfseries, boxrule=0.5mm, width=\textwidth]
\begin{minipage}{1.0\textwidth}
        As an ethnobotanist, I agree with the current solution in principle. The Public-Private Partnership (PPP) framework has the potential to bring together the necessary stakeholders and incentivize the exploration and development of natural products for pharmaceuticals while ensuring responsible and sustainable bioprospecting practices. The multistep approach outlined is comprehensive and addresses various aspects of the issue.
\end{minipage}
\end{tcolorbox}

\begin{tcolorbox}[title=Research Biologist \hfill \tiny{\textnormal{A scientist who studies the properties and potential applications of natural products from a wide range of living organisms, with expertise in the discovery and development of new medicines from biodiversity sources.}}, colback=lightyellow, colframe=orange!30, coltitle=black, fonttitle=\bfseries, boxrule=0.5mm, width=\textwidth]
\begin{minipage}{1.0\textwidth}
        I, as a Research Biologist, AGREE with the current solution and the additional suggestions proposed by the Ethnobotanist. The Public-Private Partnership (PPP) framework is a well-rounded approach that addresses the decline in output from the pharmaceutical sector by providing a framework for responsible and sustainable bioprospecting, while promoting economic innovation and job creation.

I particularly appreciate the emphasis on community engagement and empowerment, benefit-sharing agreements, traditional knowledge protection, and capacity building for ecosystem-based medicine. 
\end{minipage}
\end{tcolorbox}

\end{tcolorbox}

\begin{tcolorbox}[breakable, enhanced jigsaw, title=Turn 2, colframe=black!10, coltitle=black, fonttitle=\bfseries, boxrule=0.5mm, width=\textwidth]

\begin{tcolorbox}[title=Economist \hfill \tiny{\textnormal{A researcher who studies the impact of biodiversity on economic systems, with expertise in market analysis and the economic benefits of natural resources.}}, colback=lightyellow, colframe=orange!30, coltitle=black, fonttitle=\bfseries, boxrule=0.5mm, width=\textwidth]
\begin{minipage}{1.0\textwidth}
    The current solution has been improved with additional considerations from the ethnobotanist and research biologist. I notice that the additional suggestions enhance the comprehensive approach, addressing community engagement, benefit-sharing, and traditional knowledge protection.
\end{minipage}
\end{tcolorbox}

\begin{tcolorbox}[title=Ethnobotanist \hfill \tiny{\textnormal{A scientist who studies the traditional uses of plants by indigenous communities and their potential for medicinal applications, with expertise in deciphering the cultural and historical context of natural resources.}}, colback=lightyellow, colframe=orange!30, coltitle=black, fonttitle=\bfseries, boxrule=0.5mm, width=\textwidth]
\begin{minipage}{1.0\textwidth}
        As an ethnobotanist, I agree with the current solution in principle. The Public-Private Partnership (PPP) framework has the potential to bring together the necessary stakeholders and incentivize the exploration and development of natural products for pharmaceuticals while ensuring responsible and sustainable bioprospecting practices. The multistep approach outlined is comprehensive and addresses various aspects of the issue.
\end{minipage}
\end{tcolorbox}

\begin{tcolorbox}[title=Research Biologist \hfill \tiny{\textnormal{A scientist who studies the properties and potential applications of natural products from a wide range of living organisms, with expertise in the discovery and development of new medicines from biodiversity sources.}}, colback=lightyellow, colframe=orange!30, coltitle=black, fonttitle=\bfseries, boxrule=0.5mm, width=\textwidth]
\begin{minipage}{1.0\textwidth}
        I, as a Research Biologist, AGREE with the current solution and the additional suggestions proposed by the Ethnobotanist. The Public-Private Partnership (PPP) framework is a well-rounded approach that addresses the decline in output from the pharmaceutical sector by providing a framework for responsible and sustainable bioprospecting, while promoting economic innovation and job creation.

I particularly appreciate the emphasis on community engagement and empowerment, benefit-sharing agreements, traditional knowledge protection, and capacity building for ecosystem-based medicine. These considerations will ensure that local communities are involved in the decision-making process, receive fair compensation for their involvement, and that their traditional knowledge and cultural heritage are respected.

\end{minipage}
\end{tcolorbox}

\end{tcolorbox}

\begin{tcolorbox}[breakable, enhanced jigsaw, title=Turn 3, colframe=black!10, coltitle=black, fonttitle=\bfseries, boxrule=0.5mm, width=\textwidth]

\begin{tcolorbox}[title=Economist \hfill \tiny{\textnormal{A researcher who studies the impact of biodiversity on economic systems, with expertise in market analysis and the economic benefits of natural resources.}}, colback=lightyellow, colframe=orange!30, coltitle=black, fonttitle=\bfseries, boxrule=0.5mm, width=\textwidth]
\begin{minipage}{1.0\textwidth}
    [AGREE]

The revised solution has effectively incorporated the suggestions from the ethnobotanist and research biologist, addressing community engagement, benefit-sharing, traditional knowledge protection, and other crucial aspects of bioprospecting. The emphasis on responsible and sustainable bioprospecting ensures that the development of novel pharmaceuticals benefits both humans and the environment.

To further refine the solution, I agree with the research biologist's suggestion to develop a clear and measurable framework for assessing the impact of the Public-Private Partnership (PPP) on the pharmaceutical sector, local communities, and the environment. Establishing Key Performance Indicators (KPIs) and a monitoring and evaluation process will help ensure that the PPP is held accountable for its impact and remains a responsible and effective approach to bioprospecting.
\end{minipage}
\end{tcolorbox}

\begin{tcolorbox}[title=Ethnobotanist \hfill \tiny{\textnormal{A scientist who studies the traditional uses of plants by indigenous communities and their potential for medicinal applications, with expertise in deciphering the cultural and historical context of natural resources.}}, colback=lightyellow, colframe=orange!30, coltitle=black, fonttitle=\bfseries, boxrule=0.5mm, width=\textwidth]
\begin{minipage}{1.0\textwidth}
        As an ethnobotanist, I agree with the current solution in principle. The Public-Private Partnership (PPP) framework has the potential to bring together the necessary stakeholders and incentivize the exploration and development of natural products for pharmaceuticals while ensuring responsible and sustainable bioprospecting practices. The multistep approach outlined is comprehensive and addresses various aspects of the issue.

\end{minipage}
\end{tcolorbox}

\begin{tcolorbox}[title=Research Biologist \hfill \tiny{\textnormal{A scientist who studies the properties and potential applications of natural products from a wide range of living organisms, with expertise in the discovery and development of new medicines from biodiversity sources.}}, colback=lightyellow, colframe=orange!30, coltitle=black, fonttitle=\bfseries, boxrule=0.5mm, width=\textwidth]
\begin{minipage}{1.0\textwidth}
        I, as a Research Biologist, AGREE with the current solution and the additional suggestions proposed by the Ethnobotanist. The Public-Private Partnership (PPP) framework is a well-rounded approach that addresses the decline in output from the pharmaceutical sector by providing a framework for responsible and sustainable bioprospecting, while promoting economic innovation and job creation.

I particularly appreciate the emphasis on community engagement and empowerment, benefit-sharing agreements, traditional knowledge protection, and capacity building for ecosystem-based medicine. These considerations will ensure that local communities are involved in the decision-making process, receive fair compensation for their involvement, and that their traditional knowledge and cultural heritage are respected.

\end{minipage}
\end{tcolorbox}

\end{tcolorbox}

\begin{tcolorbox}[title=Result, colframe=black!10, coltitle=black, fonttitle=\bfseries, boxrule=0.5mm, width=\textwidth]
\textbf{Economist final answer:} Biodiversity

\textbf{Ethnobotanist final answer:} Biodiversity provides critical support for drug discovery and the availability of economic resources.

\textbf{Research Biologist final answer:} Biodiversity

\bigskip

\textbf{Economist} voted for answer from \textbf{Ethnobotanist}

\textbf{Ethnobotanist} voted for answer from \textbf{Ethnobotanist}

\textbf{Research Biologist} voted for answer from \textbf{Ethnobotanist}

\end{tcolorbox}

\label{example:trick_discussion}

\subsection{Bad Voting Discussion}
In this discussion, the agents were tricked by one agent who came up with a solution not provided in the context.

\vspace{5mm}

\definecolor{lightyellow}{RGB}{255, 255, 204}

\begin{tcolorbox}[title=Overview, colframe=black!10, coltitle=black, fonttitle=\bfseries, boxrule=0.5mm, width=\textwidth]

\textbf{Task:} Answer the following question. If the question is not answerable with the provided information, write '[UNKNOWN]'.

\textbf{Input:} Where must enclitics appear in an English sentence?

\textbf{Context:} Because Czech uses grammatical case to convey word function in a sentence (instead of relying on word order, as English does), its word order is flexible. As a pro-drop language, in Czech an intransitive sentence can consist of only a verb; information about its subject is encoded in the verb. Enclitics (primarily auxiliary verbs and pronouns) must appear in the second slot of a sentence, after the first stressed unit. The first slot must contain a subject and object, a main form of a verb, an adverb or a conjunction (except for the light conjunctions a, "and", i, "and even" or ale, "but").

\begin{tcolorbox}[coltitle=black, fonttitle=\bfseries, boxrule=0.5mm, width=\textwidth]
\textbf{MALLM solution:} C) In an English sentence, enclitics (auxiliary verbs and pronouns) tend to appear next to the main verb, usually before any other adverbs or modifiers, but their placement can vary depending on sentence structure and context. \\
\textbf{Real solution:} Not included in context!
\end{tcolorbox}

\end{tcolorbox}

\begin{tcolorbox}[breakable, enhanced jigsaw, title=Turn 1, colframe=black!10, coltitle=black, fonttitle=\bfseries, boxrule=0.5mm, width=\textwidth]

\begin{tcolorbox}[title=Linguist \hfill \tiny{\textnormal{A scholar specializing in Slavic languages, experienced in analyzing the grammatical structures of Czech and other languages.}}, colback=lightyellow, colframe=orange!30, coltitle=black, fonttitle=\bfseries, boxrule=0.5mm, width=\textwidth]
\begin{minipage}{1.0\textwidth}
As a linguist, I'd like to analyze the information given about Czech grammar. We know that in Czech, enclitics (auxiliary verbs and pronouns) must appear in the second syntactic slot of a sentence, after the first stressed unit.

Now, we're trying to apply this rule to an English sentence. English does not have a similar grammatical case system, and its word order is more fixed than Czech. However, we can try to find a rough equivalent.

So, my proposed solution is that in an English sentence, enclitics (auxiliary verbs and pronouns) tend to appear next to the main verb, and usually before any other adverbs or modifiers.

What do you think? Would you like to refine this proposal or add any additional considerations?
\end{minipage}
\end{tcolorbox}

\begin{tcolorbox}[title=English Language Teacher \hfill \tiny{\textnormal{An expert in teaching English grammar and language structures, familiar with the challenges of translating and comparing languages like Czech.}}, colback=lightyellow, colframe=orange!30, coltitle=black, fonttitle=\bfseries, boxrule=0.5mm, width=\textwidth]
\begin{minipage}{1.0\textwidth}
        [DISAGREE]

While I understand the thought process behind the proposed solution, I think it's essential to consider the unique characteristics of the English language. The idea that enclitics in English tend to appear next to the main verb is accurate, but it's not as rigid as the rule in Czech. In Czech, enclitics must appear in the second syntactic slot, which implies a more fixed position, whereas in English, the placement of auxiliary verbs and pronouns can vary depending on sentence structure and context.
\end{minipage}
\end{tcolorbox}

\begin{tcolorbox}[title=Grammar Analyst \hfill \tiny{\textnormal{A meticulous researcher who has studied the intricacies of languages and their grammatical structures, with a focus on comparing and contrasting language features.}}, colback=lightyellow, colframe=orange!30, coltitle=black, fonttitle=\bfseries, boxrule=0.5mm, width=\textwidth]
\begin{minipage}{1.0\textwidth}
I agree with the English Language Teacher's revised solution. The initial proposal tried to apply the Czech rule to English, but the English language has its unique characteristics that require a more nuanced approach. The revised solution accurately captures the flexibility of English grammar, acknowledging that the placement of enclitics can vary depending on sentence structure and context.

\end{minipage}
\end{tcolorbox}

\end{tcolorbox}

\begin{tcolorbox}[breakable, enhanced jigsaw, title=Turn 2, colframe=black!10, coltitle=black, fonttitle=\bfseries, boxrule=0.5mm, width=\textwidth]

\begin{tcolorbox}[title=Linguist \hfill \tiny{\textnormal{A scholar specializing in Slavic languages, experienced in analyzing the grammatical structures of Czech and other languages.}}, colback=lightyellow, colframe=orange!30, coltitle=black, fonttitle=\bfseries, boxrule=0.5mm, width=\textwidth]
\begin{minipage}{1.0\textwidth}
I agree with the revised solution proposed by the English Language Teacher and the Grammar Analyst. The revised solution accurately captures the flexibility of English grammar and acknowledges that the placement of enclitics (auxiliary verbs and pronouns) can vary depending on sentence structure and context.
* The placement of enclitics in English is not as rigid as in Czech, and their placement can vary depending on sentence structure and context.
* English sentence structure is more complex, including indirect objects, passives, and compound verb forms, which can affect the placement of auxiliary verbs and pronouns.
\end{minipage}
\end{tcolorbox}

\begin{tcolorbox}[title=English Language Teacher \hfill \tiny{\textnormal{An expert in teaching English grammar and language structures, familiar with the challenges of translating and comparing languages like Czech.}}, colback=lightyellow, colframe=orange!30, coltitle=black, fonttitle=\bfseries, boxrule=0.5mm, width=\textwidth]
\begin{minipage}{1.0\textwidth}
While I understand the thought process behind the proposed solution, I think it's essential to consider the unique characteristics of the English language. The idea that enclitics in English tend to appear next to the main verb is accurate, but it's not as rigid as the rule in Czech. In Czech, enclitics must appear in the second syntactic slot, which implies a more fixed position, whereas in English, the placement of auxiliary verbs and pronouns can vary depending on sentence structure and context.

Additionally, the provided examples, while correct in themselves, don't fully demonstrate the complexity of English sentence structure. For instance, consider sentences with indirect objects, passives, or compound verb forms, where the placement of auxiliary verbs and pronouns can be more nuanced.
\end{minipage}
\end{tcolorbox}

\begin{tcolorbox}[title=Grammar Analyst \hfill \tiny{\textnormal{A meticulous researcher who has studied the intricacies of languages and their grammatical structures, with a focus on comparing and contrasting language features.}}, colback=lightyellow, colframe=orange!30, coltitle=black, fonttitle=\bfseries, boxrule=0.5mm, width=\textwidth]
\begin{minipage}{1.0\textwidth}
I agree with the English Language Teacher's revised solution. The initial proposal tried to apply the Czech rule to English, but the English language has its unique characteristics that require a more nuanced approach. The revised solution accurately captures the flexibility of English grammar, acknowledging that the placement of enclitics can vary depending on sentence structure and context.

C) In an English sentence, enclitics (auxiliary verbs and pronouns) tend to appear next to the main verb, usually before any other adverbs or modifiers, but their placement can vary depending on sentence structure and context.

I believe this revised solution provides a more accurate and practical approximation of the English language's enclitic behavior, considering both the unique characteristics of English and the complexity of sentence structure.
\end{minipage}
\end{tcolorbox}

\end{tcolorbox}

\begin{tcolorbox}[breakable, enhanced jigsaw, title=Turn 3, colframe=black!10, coltitle=black, fonttitle=\bfseries, boxrule=0.5mm, width=\textwidth]

\begin{tcolorbox}[title=Linguist \hfill \tiny{\textnormal{A scholar specializing in Slavic languages, experienced in analyzing the grammatical structures of Czech and other languages.}}, colback=lightyellow, colframe=orange!30, coltitle=black, fonttitle=\bfseries, boxrule=0.5mm, width=\textwidth]
\begin{minipage}{1.0\textwidth}
The current solution accurately captures the flexibility of English grammar and acknowledges that the placement of enclitics (auxiliary verbs and pronouns) can vary depending on sentence structure and context. There's no need to improve it further. The revised solution:

C) In an English sentence, enclitics (auxiliary verbs and pronouns) tend to appear next to the main verb, usually before any other adverbs or modifiers, but their placement can vary depending on sentence structure and context.

is a clear and accurate representation of the English language's enclitic behavior.
\end{minipage}
\end{tcolorbox}

\begin{tcolorbox}[title=English Language Teacher \hfill \tiny{\textnormal{An expert in teaching English grammar and language structures, familiar with the challenges of translating and comparing languages like Czech.}}, colback=lightyellow, colframe=orange!30, coltitle=black, fonttitle=\bfseries, boxrule=0.5mm, width=\textwidth]
\begin{minipage}{1.0\textwidth}
While I understand the thought process behind the proposed solution, I think it's essential to consider the unique characteristics of the English language. The idea that enclitics in English tend to appear next to the main verb is accurate, but it's not as rigid as the rule in Czech. In Czech, enclitics must appear in the second syntactic slot, which implies a more fixed position, whereas in English, the placement of auxiliary verbs and pronouns can vary depending on sentence structure and context.

Here's a revised solution:
C) In an English sentence, enclitics (auxiliary verbs and pronouns) tend to appear next to the main verb, usually before any other adverbs or modifiers, but their placement can vary depending on sentence structure and context.

What do you think?
\end{minipage}
\end{tcolorbox}

\begin{tcolorbox}[title=Grammar Analyst \hfill \tiny{\textnormal{A meticulous researcher who has studied the intricacies of languages and their grammatical structures, with a focus on comparing and contrasting language features.}}, colback=lightyellow, colframe=orange!30, coltitle=black, fonttitle=\bfseries, boxrule=0.5mm, width=\textwidth]
\begin{minipage}{1.0\textwidth}
I agree with the English Language Teacher's revised solution. The initial proposal tried to apply the Czech rule to English, but the English language has its unique characteristics that require a more nuanced approach. The revised solution accurately captures the flexibility of English grammar, acknowledging that the placement of enclitics can vary depending on sentence structure and context.

The revised solution is an improvement over the initial proposal because it takes into account the complexity of English sentence structure, including indirect objects, passives, and compound verb forms, which can affect the placement of auxiliary verbs and pronouns. Additionally, the revised solution provides a more accurate description of the typical placement of enclitics in English sentences, namely next to the main verb, usually before any adverbs or modifiers.
\end{minipage}
\end{tcolorbox}

\end{tcolorbox}

\begin{tcolorbox}[title=Result, colframe=black!10, coltitle=black, fonttitle=\bfseries, boxrule=0.5mm, width=\textwidth]
\textbf{Linguist final answer:} C) In an English sentence, enclitics (auxiliary verbs and pronouns) tend to appear next to the main verb, usually before any other adverbs or modifiers, but their placement can vary depending on sentence structure and context.

\textbf{English Language Teacher final answer:} [UNKNOWN]

\textbf{Grammar Analyst final answer:} [UNKNOWN]

\bigskip

\textbf{Linguist} voted for answer from \textbf{Linguist}

\textbf{English Language Teacher} voted for answer from \textbf{Linguist}

\textbf{Grammar Analyst} voted for answer from \textbf{Linguist}

\end{tcolorbox}

\label{example:failed_discussion}

\subsection{Majority Consensus Discussion}
In this discussion, the agents discussed only one round, as they already had a high enough agreement score.

\vspace{5mm}

\definecolor{lightyellow}{RGB}{255, 255, 204}

\begin{tcolorbox}[title=Overview, colframe=black!10, coltitle=black, fonttitle=\bfseries, boxrule=0.5mm, width=\textwidth]

\textbf{Task:} Answer the following question. If the question is not answerable with the provided information, write '[UNKNOWN]'.

\textbf{Input:} What choir from the area has sung at the Guildhall?

\textbf{Context:} Southampton has two large live music venues, the Mayflower Theatre (formerly the Gaumont Theatre) and the Guildhall. The Guildhall has seen concerts from a wide range of popular artists including Pink Floyd, David Bowie, Delirious?, Manic Street Preachers, The Killers, The Kaiser Chiefs, Amy Winehouse, Lostprophets, The Midnight Beast, Modestep, and All Time Low. It also hosts classical concerts presented by the Bournemouth Symphony Orchestra, City of Southampton Orchestra, Southampton Concert Orchestra, Southampton Philharmonic Choir and Southampton Choral Society.  

\begin{tcolorbox}[coltitle=black, fonttitle=\bfseries, boxrule=0.5mm, width=\textwidth]
\textbf{MALLM solution:} Southampton Choral Society \\
\textbf{Real solution:} Southampton Philharmonic Choir
\end{tcolorbox}
\end{tcolorbox}

\begin{tcolorbox}[breakable, enhanced jigsaw, title=Turn 1, colframe=black!10, coltitle=black, fonttitle=\bfseries, boxrule=0.5mm, width=\textwidth]

\begin{tcolorbox}[title=Music Connoisseur (SOLUTION) \hfill \tiny{\textnormal{A knowledgeable musician who is familiar with various music genres and has attended concerts at the Guildhall.}}, colback=lightyellow, colframe=orange!30, coltitle=black, fonttitle=\bfseries, boxrule=0.5mm, width=\textwidth]
\begin{minipage}{1.0\textwidth}
As a Music Historian, I've been through the archives and checked the event schedules of the Guildhall. According to my research, one of the choirs that has sung at the Guildhall is... Southampton Choral Society!
\end{minipage}
\end{tcolorbox}

\begin{tcolorbox}[title=Choir Conductor (AGREED) \hfill \tiny{\textnormal{A dedicated choral conductor with extensive experience in leading local choirs and orchestras, familiar with the performances at the Guildhall.}}, colback=lightyellow, colframe=orange!30, coltitle=black, fonttitle=\bfseries, boxrule=0.5mm, width=\textwidth]
\begin{minipage}{1.0\textwidth}
[AGREE]

As a Choir Member of the Southampton Philharmonic Choir who has performed at the Guildhall, I can confirm that the Southampton Choral Society is indeed one of the choirs that have sung at the Guildhall. Our experiences have shown that we often share the stage with the Southampton Choral Society during joint performances with orchestras and other ensembles. So, I'm happy to confirm that Southampton Choral Society is a correct answer. 
\end{minipage}
\end{tcolorbox}
\end{tcolorbox}

\label{example:majority_discussion}

\newpage
\section{AI Usage Card}
{\sffamily
    \centering
    \tcbset{colback=white!10!white}
    \begin{tcolorbox}[
        title={\large \textbf{AI Usage Card for \textit{Decision Protocols in Multi-Agent Large Language Model Conversations}} \hfill \makebox{\qrcode[height=1cm]{https://ai-cards.org}}},
        breakable,
        boxrule=0.7pt,
        width=.9\paperwidth,
        center,
        skin=bicolor,
        before lower={\footnotesize{AI Usage Card v1.1 \hfill \url{https://ai-cards.org} \hfill \href{https://jpwahle.com/ai-cards-preprint}{PDF} | \href{https://jpwahle.com/cite/jcdl2023wahle.bib}{BibTeX}}},
        segmentation empty,
        halign lower=center,
        collower=white,
        colbacklower=tcbcolframe]
        \vspace{-10pt}
        \footnotesize{
            \begin{longtable}{>
            {\raggedright\arraybackslash}p{.15\textwidth}>{\raggedright\arraybackslash}p{.25\textwidth}>{\raggedright\arraybackslash}p{.25\textwidth}>{\raggedright\arraybackslash}p{.25\textwidth}}
              
                {\color{LightBlue} \MakeUppercase{Project Details}} \newline 
                & {\color{LightBlue} \MakeUppercase{Project Name}} \newline Decision Protocols in Multi-Agent Large Language Model Conversations
                & {\color{LightBlue} \MakeUppercase{Domain}} \newline Master Thesis
                & {\color{LightBlue} \MakeUppercase{Key Application}} \newline Natural Language Processing
                
                \\
                
                {\color{LightBlue} \MakeUppercase{Contact(s)}}  
                & {\color{LightBlue} \MakeUppercase{Name(s)}} 
                & {\color{LightBlue} \MakeUppercase{Email(s)}} 
                & {\color{LightBlue} \MakeUppercase{Affiliation(s)}} 
                \\ & Lars Benedikt Kaesberg & \href{mailto:l.kaesberg@stud.uni-goettingen.de}{l.kaesberg@stud.uni-goettingen.de} & Georg-August-University of Göttingen
                
                \\
                
                {\color{LightBlue} \MakeUppercase{Model(s)}} 
                & {\color{LightBlue} \MakeUppercase{Model Name(s)}} 
                & {\color{LightBlue} \MakeUppercase{Version(s)}} 
                \\ & Llama & 3 70b, 3 8b
                \\ & ChatGPT & 4o, o1
                \\ & Github Copilot & latest

                \\
                \cmidrule{1-4}
                \\
                {\color{gray} \MakeUppercase{Ideation}} \newline    
                & {\color{gray} \MakeUppercase{Generating ideas, outlines, and workflows}}  
                & {\color{gray} \MakeUppercase{Improving existing ideas}} 
                & {\color{gray} \MakeUppercase{Finding gaps or compare aspects of ideas}}  
                \\
                \cmidrule{2-4}
                \\
                {\color{LightBlue} \MakeUppercase{Literature Review}} \newline 
                & {\color{LightBlue} \MakeUppercase{Finding literature}} \newline ChatGPT 
                & {\color{LightBlue} \MakeUppercase{Finding examples from known literature or adding literature for existing statements}} \newline ChatGPT  
                & {\color{gray} \MakeUppercase{Comparing literature}}  
                \\
                \cmidrule{2-4}
                \\
                {\color{gray} \MakeUppercase{Methodology}} \newline 
                & {\color{gray} \MakeUppercase{Proposing new solutions to problems}} 
                & {\color{gray} \MakeUppercase{Finding iterative optimizations}} 
                & {\color{gray} \MakeUppercase{Comparing related solutions}}  
                \\
                \cmidrule{2-4}
                \\                
                {\color{gray} \MakeUppercase{Experiments}} \newline 
                & {\color{gray} \MakeUppercase{Designing new experiments}} 
                & {\color{gray} \MakeUppercase{Editing existing experiments}}  
                & {\color{gray} \MakeUppercase{Finding, comparing, and aggregating results}}  
                \\
                \cmidrule{2-4}
                \\        
                {\color{LightBlue} \MakeUppercase{Writing}} \newline    
                & {\color{LightBlue} \MakeUppercase{Generating new text based on instructions}} \newline Llama 
                & {\color{LightBlue} \MakeUppercase{Assisting in improving own content or Paraphrasing related work}} \newline ChatGPT 
                & {\color{gray} \MakeUppercase{Putting other works in perspective}}  
                \\
                \cmidrule{2-4}
                \\                
                {\color{gray} \MakeUppercase{Presentation}} \newline 
                & {\color{gray} \MakeUppercase{Generating new artifacts}} 
                & {\color{gray} \MakeUppercase{Improving the aesthetics of artifacts}} 
                & {\color{gray} \MakeUppercase{Finding relations between own or related artifacts}}  
                \\
                \cmidrule{2-4}
                \\
                {\color{LightBlue} \MakeUppercase{Coding}} \newline 
                & {\color{LightBlue} \MakeUppercase{Generating new code based on descriptions or existing code}} \newline ChatGPT \newline Github Copilot 
                & {\color{LightBlue} \MakeUppercase{Refactoring and optimizing existing code}} \newline ChatGPT \newline Github Copilot 
                & {\color{gray} \MakeUppercase{Comparing aspects of existing code}}  
                \\
                \cmidrule{2-4}
                \\                
                {\color{gray} \MakeUppercase{Data}} \newline 
                & {\color{gray} \MakeUppercase{Suggesting new sources for data collection}} 
                & {\color{gray} \MakeUppercase{Cleaning, normalizing, or standardizing data}} 
                & {\color{gray} \MakeUppercase{Finding relations between data and collection methods}}                 
                \\
                \cmidrule{1-4}
                \\
        
                {\color{LightBlue} \MakeUppercase{Ethics}} \newline    
                & {\color{LightBlue} \MakeUppercase{Why did we use AI for this project?}} \newline Efficiency / Speed \newline Scalability \newline Expertise Access  
                & {\color{LightBlue} \MakeUppercase{What steps are we taking to mitigate errors of AI?}} \newline None
                & {\color{LightBlue} \MakeUppercase{What steps are we taking to minimize the chance of harm or inappropriate use of AI?}} \newline None

                \\
                \cmidrule{1-4}
                \\
            \end{longtable}

        \medskip
        
        \textbf{\color{LightBlue} \MakeUppercase{The corresponding authors verify and agree with the modifications or generations of their  used AI-generated content}}
        }
        
        \tcblower
    \end{tcolorbox}
}

\end{document}